
\magnification\magstep1
\parindent 0pt
\def\sqr#1#2{{\vcenter{\vbox{\hrule height.#2pt
\hbox{\vrule width.#2pt height#1pt \kern#1pt
\vrule width.#2pt}
\hrule height.#2pt}}}}
\def\sqre{\mathchoice\sqr34\sqr34\sqr{2.1}\sqr{1.5}3}
\def\Bbb{\bf}
\tolerance=1600
\parskip=6pt
\def\sqr#1#2{{\vcenter{\vbox{\hrule height.#2pt
\hbox{\vrule width.#2pt height#1pt \kern#1pt
\vrule width.#2pt}
\hrule height.#2pt}}}}
\def\sqre{\mathchoice\sqr{7}4\sqr{7}4\sqr{6}3\sqr{6}3}

\def\hhalf{{\textstyle{1\over 2}}}

\def\sg14{{\partial\over{\partial\sigma_1^{k_1}}}...
{\partial\over{\partial\sigma_4^{k_4}}}}

\def\ssig12{{\partial\over{\partial\sigma_1^{k_1}}}...
{\partial\over{\partial\sigma_4^{k_4}}}}
\def\quar{{\textstyle{1\over 4}}}
\def\hb#1{\hbox to 4truecm{#1\hfil}}

\def\hhalf{{\textstyle{1\over 2}}}
\def\half{{1\over 2}}

\def\J{{\cal J}}
\def\Jp{{\J^\perp}}
\def\bU{{\bf U}}

\def\sq{{\sqrt 2}}
\def\fred{{\textstyle{1\over{\sqrt 2}}}}
\def\one{{\underline 1}}

\def\Aut{{\rm Aut}}
\def\rank{{\rm rank}\,}

\def\ze{{\Bbb Z}}

\def\Wbar{\overline{W}}


\def\pbar{\overline{\psi}}

\def\phbar{\overline{\phi}}

\def\X{{\cal X}}
\def\ce{{\Bbb C}}

\def\dt{{\cdot}}

\def\H{{\cal H}}
\def\THX{{\widetilde\H}}

\def\C{{\cal C}}
\def\F{{\cal F}}
\def\K{{\cal K}}

\def\ip#1,#2.{\langle #1,\,#2\rangle}
\def\bc#1,#2.{\left({#1\atop #2}\right)}

\def\vac{|0\rangle}
\def\ref#1{$^{[#1]}$}
\def\bra#1.{|#1\rangle}

\def\V{{\cal V}}

\def\pagenumber{\footline={\hss\tenrm\folio\hss}}
\def\bV{{\bf V}}
\def\T{{\cal T}}
\def\h{\hbox}
\def\hba#1{\hbox to 3truecm{#1\hfil}}
\def\hbb#1{\hbox to 5truecm{#1\hfil}}
\def\hbt#1{\hbox to 10truecm{#1\hfil}}
\def\hbp#1#2{\hbox to 4truecm{#1\hfil #2}}
\def\hbap#1#2{\hbox to 3truecm{#1\hfil #2}}
\def\hbbp#1#2{\hbox to 5truecm{#1\hfil #2}}
\nopagenumbers
\rightline{DAMTP 94-80}
\vskip 16pt
\centerline{\bf CONFORMAL FIELD THEORIES, REPRESENTATIONS}
\vskip2pt
\centerline{\bf AND LATTICE CONSTRUCTIONS}
\vskip 40 pt
\centerline {\bf L. Dolan}
\vskip 10 pt
\centerline {\it Department of Physics and Astronomy,}
\centerline {\it University of North Carolina,}
\centerline {\it Chapel Hill, NC 27599, U.S.A.}
\vskip16pt
\centerline {{\bf P. Goddard} {\it and} {\bf P. Montague}}
\vskip10pt
\centerline {\it Department of Applied Mathematics and Theoretical Physics,}
\centerline {\it University of Cambridge, Silver Street,}
\centerline {\it Cambridge, CB3 9EW, U.K.}
\vskip 40 pt
\centerline {\bf ABSTRACT}
\vskip 5 pt

{\rightskip=18 true mm \leftskip=18 true mm  \noindent

An account is given of the structure and representations of chiral bosonic
meromorphic conformal field theories (CFT's), and, in particular, the
conditions under which such a CFT may be extended by a representation to
form a new theory. This general approach is illustrated by considering the
untwisted and $\ze_2$-twisted theories, $\H(\Lambda)$ and $\tilde
\H(\Lambda)$ respectively, which may be constructed from a suitable even
Euclidean lattice $\Lambda$. Similarly, one may construct lattices $\Lambda_\C$
and $\tilde\Lambda_\C$ by analogous constructions from a doubly-even binary
code $\C$. In the case when $\C$ is self-dual, the corresponding lattices are
also. Similarly, $\H(\Lambda)$ and $\tilde
\H(\Lambda)$ are self-dual if and only if $\Lambda$ is. We show that
$\H(\Lambda_\C)$ has a natural ``triality'' structure, which induces an
isomorphism $\H(\tilde\Lambda_\C)\equiv\tilde\H(\Lambda_\C)$ and also a
triality structure on $\tilde\H(\tilde\Lambda_\C)$. For $\C$ the Golay code,
$\tilde\Lambda_\C$ is the Leech lattice, and the triality on
$\tilde\H(\tilde\Lambda_\C)$ is the symmetry which extends the natural action
of (an extension of) Conway's group on this theory to the Monster, so setting
triality and Frenkel, Lepowsky and Meurman's construction of the natural
Monster module in a more general context. The results also serve to shed some
light on the classification of self-dual CFT's. We find that of the 48 theories
$\H(\Lambda)$ and $\tilde\H(\Lambda)$ with central charge 24 that there are 39
distinct ones, and further that all 9 coincidences are accounted for by the
isomorphism detailed above, induced by the existence of a doubly-even self-dual
binary code.
}

\vfill
\eject
\pagenumber

\centerline{\bf 1. Introduction}
\nobreak
\vskip3pt
\nobreak
In this paper we shall provide the details omitted from the summary of our
results given in [1].

The principal result of the paper will be to show how a study of binary linear
codes leads to an understanding of some of the symmetries of conformal field
theories (CFT's). We shall restrict ourselves to self-dual chiral bosonic
theories, which are regarded as trivial by approaches to the CFT classification
problem which rely upon a study of the fusion rules for the representations of
some chiral algebra\ref{2}. (For general reviews of CFT see [3,4].) Hence, a
complete understanding of these ``trivial'' theories would seem to be
essential to obtain,
and further our results show that such theories are not necessarily
without an interesting structure.

Indeed, one such theory, constructed initially by Frenkel, Lepowsky and Meurman
(FLM)\ref{5} from the Leech lattice, possesses only discrete automorphisms,
which close to form the largest of the sporadic simple groups,
the Monster group\ref{8-10}.
Building on previous work generalising the construction of this Monster module
to other lattices\ref{11}, we show that a certain subgroup of discrete
symmetries, known as triality, which is the key to the construction of the
action of the Monster in the FLM theory, can be seen in this more general
context. Triality is seen to occur in some theories as an obvious consequence
of the existence of a corresponding binary code, and can be lifted to provide
an isomorphism between otherwise potentially distinct CFT's and further to the
triality structure of the form exhibited by FLM, though in a more general
setting. Hence, we see that triality and binary codes provide insight into the
classification of bosonic self-dual theories, and a more general framework in
which to view the hitherto mysterious Monster group.

In addition to these investigations of lattice constructions, we provide a
general treatment of the representations of bosonic CFT's. We discuss the
notions of a subconformal field theory and of a hermitian structure on a CFT,
and demonstrate that, under certain conditions, we may extend a CFT by a
representation to form a new CFT. A particular example of this is provided by
the twisted lattice construction, which gives the Monster module in the case of
the Leech lattice. Our treatment is based on the approach of
[12], which was inspired by the work of FLM and Borcherds' general approach to
``vertex operator algebras''\ref{13}. Results in a similar direction have also
been independently described in [14].

The layout of the paper is as follows. Sections 2-4 cover the general aspects
of conformal field theories and their representations. In section 5, we sketch
the straight and $\ze_2$-twisted lattice constructions of CFT's and the
analogous constructions of lattices from binary codes. Further details may be
found in [11]. Section 6 gives the results of these constructions, and
discusses the connection with the Monster provided by the work of FLM, while in
sections 7 and 8 we exhibit the triality structure in this general framework.
Our conclusions are presented in section 9.
\vskip12pt
\centerline{\bf 2. Definitions and elementary properties}
\nobreak
\vskip3pt
\nobreak
In this section, we define what we shall mean by
a ``conformal field theory'',
and review some of the properties which follow from
this definition. [Note that we shall take conformal
field theory to refer to bosonic meromorphic chiral conformal field theories
defined on the Riemann sphere only,
{\it i.e.} they are holomorphic, in the sense that there is only a dependence
on the complex variable $z$ and not its conjugate $z^\ast$, with meromorphic
matrix elements and ``commuting'' vertex operators in the sense of (2.4)].

\noindent {\it {\bf Definition 2.1}}
A {\it conformal field theory} ($\H$, $\F$, $\bV$, $|0\rangle$, $\psi_L$)
consists of a
Hilbert space of states $\H$, a dense subspace $\F$
[typically the Fock space of states of finite occupation number for some set
of harmonic oscillators] and a
set $\bV$ of linear operators called
{\it vertex operators} $V(\psi,z)$ in one-to-one
correspondence with the states $\psi\in\F$.
[We shall use the Dirac notation $|\psi\rangle$, but will write this simply
as $\psi$ where it is notationally convenient.]
There are two special states in
$\F$, the {\it vacuum} $|0\rangle$ and a {\it conformal state} $\psi_L$.
The theory must satisfy the properties P1-6 detailed below, and is said to
be a {\it hermitian conformal field theory} if it satisfies in addition
property P7.

\noindent {\bf P1} We define the
moments of
the vertex operator of $\psi_L$ to be given by
$$V(\psi_L,z)=\sum_{n\in\ze}L_nz^{-n-2}\,,\eqno(2.1)$$
and demand that they
provide a representation of the centrally extended Virasoro algebra
$$[L_m,L_n]=(m-n)L_{m+n}+{c\over 12}m(m^2-1)\delta_{m,-n}\,,\eqno(2.2)$$
with $L_n^\dagger=L_{-n}$ and $L_n|0\rangle
=0$, $n\geq -1$.
[Note that we shall see later that the requirement (2.2) may be weakened
slightly and still hold, in the presence of the remaining axioms.]

\noindent {\bf P2} The vertex operators satisfy
$$V(\psi,z)|0\rangle =e^{zL_{-1}}\psi\,,\eqno(2.3)$$
and also

\noindent {\bf P3} the bosonic ``locality'' relation
$$V(\psi,z)V(\phi,\zeta)=V(\phi,\zeta)V(\psi,z)\,.\eqno(2.4)$$
\noindent
More precisely, we require that the matrix elements of the product $V(\psi,z)
V(\phi,\zeta)$ between states in $\F$ should be defined for $|z|>|\zeta|$
and that the function this defines by analytic continuation be regular
except for possible poles at $z$, $\zeta=0$, $\infty$ and $z=\zeta$. Then
(2.4) should be interpreted to mean that the functions obtained from either
side in such a manner are equal.
(Note that any extension of the definition of the vertex operators $V(\psi,z)$
from $\F$ to the space of generalised coherent states $V(\psi_1,z_1)
V(\psi_2,z_2)\ldots V(\psi_N,z_N)\phi$ for $|z|>|z_1|>|z_2|>\ldots
|z_N|$ and $\phi\in\F$, if it exists, is unique because of the fact that
$\F$ is dense in $\H$.)
(Note also that we could allow a relative minus sign between the two sides
of (2.4), corresponding to fermionic fields.)

The property P2 is equivalent to the conditions
$$[L_{-1},V(\psi,z)]={d\over{dz}}V(\psi,z)\eqno(2.5)$$
and
$$\lim_{z\rightarrow 0}V(\psi,z)|0\rangle =\psi\,.\eqno(2.6)$$
(In particular $\psi_L=L_{-2}|0\rangle$.)
Equivalently (2.5) can be expressed in the form
of the ``translation property''
$$e^{wL_{-1}}V(\psi,z)e^{-wL_{-1}}=V(\psi,z+w)\,.\eqno(2.7)$$
{}From the locality condition (2.4) we can establish a uniqueness property of
the vertex operators.

\noindent
{\it {\bf Proposition 2.2}}
{\it If} $U(z)$ {\it satisfies}
$$U(z)|0\rangle =e^{zL_{-1}}\phi\,,\eqno(2.8)$$
{\it for some} $\phi\in\F$, {\it and is local with respect to the
system of vertex
operators, then} $U(z)=V(\phi,z)$.

\noindent
{\it Proof.}
The proof is straightforward, since
$$\eqalignno{
U(z)e^{\zeta L_{-1}}\psi &=U(z)V(\psi,\zeta)|0\rangle
=V(\psi,\zeta)U(z)|0\rangle\cr
&=V(\psi,\zeta)e^{zL_{-1}}\phi =V(\psi,\zeta)V(\phi,z)|0\rangle\cr
&=V(\phi,z)V(\psi,\zeta)|0\rangle =V(\phi,z)e^{\zeta L_{-1}}\psi\,
. & (2.9)}$$
Thus, taking $\zeta\rightarrow 0$, we deduce $U(z)=V(\phi,z)$. $\sqre$

Thus,
to demonstrate a given operator to be a vertex operator for a particular state
all we have to do is show that it is local with respect to $\bV$ and has the
appropriate action on the vacuum.

We may apply this uniqueness property to (2.5) to deduce that, since
${d\over{dz}}V(\psi,z)$ is local with respect to $\bV$,
$${d\over{dz}}V(\psi,z)=V(L_{-1}\psi,z)\,,\eqno(2.10)$$
since both sides are local with respect to $\bV$ and have the same action
on the vacuum, from (2.3).

Similarly, uniqueness immediately implies that $V(\psi,z)$ is linear in $\psi$
and that $V(|0\rangle ,z)\equiv 1$, again using (2.3) (and the fact that
$L_{-1}|0\rangle =0$).

In addition, we have

\noindent
{\it {\bf Proposition 2.3}}
{\it The ``duality'' relation:}
$$V(\psi,z)V(\phi,\zeta)=V(V(\psi,z-\zeta)\phi ,\zeta )\,,\eqno(2.11)$$
\noindent
{\it Proof.}
Again this is
a consequence of the uniqueness argument (note that the product on the
left hand side
of (2.11) is local with respect to $\bV$, because each of the factors is).
We use (2.3) and the translation property, {\it i.e.}
$$\eqalignno{
V(\psi,z)V(\phi,\zeta)|0\rangle = & V(\psi,z)e^{\zeta L_{-1}}\phi
=e^{\zeta L_{-1}}V(\psi,z-\zeta)\phi\cr
&=V(V(\psi,z-\zeta)\phi,\zeta)|0\rangle\, .\,\,\sqre &(2.12)}$$
These results serve to demonstrate the powerful role played by locality
in conformal field theory.

\noindent
{\it {\bf Proposition 2.4}} {\it Skew-symmetry:}
$$V(\psi,z)\phi =e^{zL_{-1}}V(\phi,-z)\psi\,.\eqno(2.13)$$
\noindent
{\it Proof.}
Using (2.11) together with (2.3) we obtain
$$\eqalignno{
V(\psi,z)V(\phi,\zeta)|0\rangle &=V(V(\psi,z-\zeta)\phi,\zeta)|0\rangle\cr
&=e^{\zeta L_{-1}}V(\psi,z-\zeta)\phi\, . & (2.14)}$$
But using (2.4) first gives
$$V(\psi,z)V(\phi,\zeta)|0\rangle =e^{zL_{-1}}V(\phi,\zeta-z)\psi\,.
\eqno(2.15)$$
Thus, comparing (2.14) and (2.15) one obtains (2.13). $\sqre$

This result will be of use in later chapters when we come to defining what
are called the intertwining operators. Note that it also immediately
implies linearity of $V(\psi,z)$ in the state $\psi$.

Let
us also assume

\noindent {\bf P4}
$x^{L_0}$ acts locally with respect to $\bV$, {\it i.e.}
$x^{L_0}V(\psi,z)x^{-L_0}$ is local with respect to $\bV$.

[Note: we could alternatively {\it assume} that the spectrum of $L_0$ is
integral
(which we deduce in Proposition 2.9 in our present treatment),
and then (2.16), and thus the locality of the action of $L_0$, would follow as
a
consequence of (2.59) and its implications for $L_{-1}$ descendent states.]

Then using
$[L_0,L_{-1}]=L_{-1}$, from the Virasoro algebra (2.2), and the uniqueness
argument (together with $L_0|0\rangle =0$) we deduce that
$$x^{L_0}V(\psi,z)x^{-L_0}=x^{h_\psi}V(\psi,xz)\,,\eqno(2.16)$$
when $L_0\psi=h_\psi\psi$, or equivalently
$$[L_0,V(\psi,z)]=\left( z{d\over{dz}}+h_\psi\right) V(\psi,z)\,.\eqno(2.17)$$
For a general state, {\it i.e.} not necessarily an $L_0$ eigenstate,
we can write (2.16) by linearity as
$$x^{L_0}V(\psi,z)x^{-L_0}=V(x^{L_0}\psi,xz)\,.\eqno(2.18)$$
Later,
by imposing the requirement P7
that the conformal field theory has a hermitian
structure, we will see that the conformal weights $h_\psi$
for the states $\psi$ must always be
non-negative integers.

We begin by writing $\F$ as a direct sum of eigenspaces
of $L_0$, {\it i.e.}
$$\F=\bigoplus_h\F_h\,,\eqno(2.19)$$
where $L_0\psi=h\psi$ for $\psi\in\F_h$.

\noindent
{\it {\bf Definition 2.5}}
A state $\psi$ is said to be an
$su(1,1)$ {\it highest weight state} or a
{\it quasi-primary state} if $L_1\psi=0$. The corresponding vertex operator
is said to be
a {\it quasi-primary field}.

\noindent
(The
elements $L_{\pm 1}$, $L_0$ of the Virasoro algebra generate a subalgebra
isomorphic to $su(1,1)$ [note that for $m$, $n=0$, $\pm 1$ in (2.2) the
central term vanishes].)

Let us also assume

\noindent
{\bf P5} The spectrum of $L_0$ is bounded below.

\noindent
Note that
this assumption is physically reasonable,
since in a conformally invariant quantum field theory we have both a
holomorphic and an anti-holomorphic conformal structure, with Virasoro
generators $L_n$ and $\overline L_n$ respectively. The Hamiltonian $H$ of the
theory is given by $L_0+\overline L_0$, and in any sensible quantum field
theory the Hamiltonian should be bounded below. Since the holomorphic and
anti-holomorphic sectors are independent, then $L_0$ and $\overline L_0$
should be separately bounded from below.

\noindent
{\it {\bf Proposition 2.6} The eigenvalues of} $L_0$ {\it are non-negative.}

\noindent
{\it Proof.}
For $\psi\in\F_h$ a quasi-primary state, using $L_{-1}^\dagger
=L_1$ and the relation $[L_1,L_{-1}]=2L_0$, we obtain
$$||L_{-1}\psi||^2=2h||\psi||^2\,,\eqno(2.20)$$
so that, by positive definiteness of the norm on the Hilbert space of states,
$h\geq 0$.
If $\phi$ is an arbitrary non-zero (not necessarily quasi-primary) state
with negative conformal weight $\Delta$, then the sequence of states
$L_1^N\phi$ for $N=0$, $1$, $2,\ldots$ have conformal weights
$\Delta -N$. If any of these states vanishes, let $N_0$ be the smallest such
value of $N$. Then $L_1^{N_0-1}\phi$ is a quasi-primary state $\psi$
with conformal weight $h=\Delta+1-N_0<0$. The left hand side of (2.20) is
non-negative, but the right hand side (since $\psi\neq 0$ as we chose
$N_0$ to be as small as possible) is negative. This contradiction implies that
the sequence of states $L_1^N\phi$ are all non-vanishing. Hence, if any state
has negative weight, the spectrum of $L_0$ is unbounded below. This contradicts
P5 and hence establishes the result. $\sqre$

If a state $\phi$ has conformal weight zero then $L_1\phi=0$, otherwise we
would have a state with negative conformal weight. Thus, we may apply (2.20)
with $\phi=\psi$ to see that $L_{-1}\phi=0$ also, {\it i.e.} a state has zero
conformal weight if and only if it is $su(1,1)$ invariant. We shall assume

\noindent
{\bf P6} The vacuum is the only $su(1,1)$ invariant state in the theory.

{}From the fact that the conformal weights are bounded below, we see that
$L_1^N\phi=0$ for $N$ sufficiently large, where $\phi$ is an arbitrary state of
some definite conformal weight. So we have

\noindent
{\it {\bf Proposition 2.7}}
$\F$ {\it splits up into a direct sum of}
$su(1,1)$ {\it highest weight representations.}

Each is generated by repeated action of
$L_{-1}$ on an $su(1,1)$ highest weight state (a quasi-primary state). This
fact will be of use later in proving certain locality relations, since, by
(2.10), we only have to consider quasi-primary states, and for these the
hermitian structure takes a particularly simple form. Let us now define this
hermitian structure.

\noindent
{\it {\bf Proposition 2.8} If} $V(e^{z^\ast
L_1}{z^\ast}^{-2L_0}
\psi,1/z^\ast)^\dagger$ {\it is local with respect to} $\bV$
{\it then}
$$V(e^{z^\ast
L_1}{z^\ast}^{-2L_0}
\psi,1/z^\ast)^\dagger=V(\overline\psi,z)\,,\eqno(2.24)$$
{\it where}
$$\overline\psi=\lim_{z\rightarrow 0}V(e^{z^\ast
L_1}{z^\ast}^{-2L_0}\psi,1/z^\ast)
^\dagger|0\rangle\,,\eqno(2.25)$$
{\it and
the conjugation map} $\psi\mapsto\overline\psi$ {\it is antilinear.}

{\it Proof.} All we need do is demonstrate that the left hand side of (2.24)
satisfies the obvious analogue of
(2.5), or equivalently the translation property (2.7). Then, being
local with respect to $\bV$, from Proposition 2.2 we see that it must be the
vertex operator with argument $z$ for a particular state $\overline\psi$, which
must, using (2.6), be given by (2.25). (Note that the limit is seen to exist
from the translation property.) The map $\psi\mapsto\overline\psi$ is clearly
antilinear, from the linearity
of the map from states to the corresponding vertex operators noted above as
a simple consequence of the uniqueness theorem. We now establish that the
translation property
is
satisfied.
$e^{\epsilon L_1} V\left(e^{L_1/\zeta}\psi,\zeta\right)e^{-\epsilon L_1}$
is a local operator (see (2.59)) and so we calculate its effect on the vacuum
and
then use the uniqueness theorem. Now
$$\eqalignno{
e^{\epsilon L_1} V\left(e^{L_1/\zeta}\psi,\zeta\right)&e^{-\epsilon L_1}
|0\rangle = e^{\epsilon L_1} e^{\zeta L_{-1}} e^{L_1/\zeta}\psi&(2.26)\cr
&=\exp\left\{{\zeta\over 1-\epsilon\zeta}L_{-1}\right\}
\left(1-\epsilon\zeta\right)^{-2L_0}
\exp\left\{{1\over \zeta(1-\epsilon\zeta)}L_{1}\right\}\psi&(2.27)\cr
&=\left(1-\epsilon\zeta\right)^{-2 h}
\exp\left\{{\zeta\over 1-\epsilon\zeta}L_{-1}\right\}
\exp\left\{{1-\epsilon\zeta\over\zeta}L_{1}\right\}\psi\,,&(2.28)\cr}$$
where to get from (2.26) to (2.27) we have used
$$e^{\epsilon L_1} e^{\zeta L_{-1}}=
\exp\left\{{\zeta\over 1-\epsilon\zeta}L_{-1}\right\}
\left(1-\epsilon\zeta\right)^{-2L_0}
\exp\left\{{\epsilon\over 1-\epsilon\zeta}L_{1}\right\}\,\eqno(2.29)$$
and, to get from (2.27) to (2.28), $[L_0,\,L_{-1}]= L_{-1}$,
and we have taken $L_0\psi=h\psi$.
Thus it follows that
$$e^{\epsilon L_1} V\left(e^{L_1/\zeta} \psi,\zeta\right)e^{-\epsilon L_1}=
\left(1-\epsilon\zeta\right)^{-2 h}V\left(
\exp\left\{{1-\epsilon\zeta\over\zeta}L_{1}\right\}\psi,
{\zeta\over 1-\epsilon\zeta}\right)\,.\eqno(2.30)$$
Taking $\epsilon$ small gives
$$ \left[L_1, \, V\left(e^{L_1/\zeta} \psi,\zeta\right)\right]
= 2 h \zeta V\left(e^{L_1/\zeta} \psi,\zeta\right) -
{d\over d(1/\zeta)} V\left(e^{L_1/\zeta} \psi,\zeta\right)\,.\eqno(2.31)$$
Therefore
$$\left[L_{-1},\,z^{-2 h}V\left(e^{L_1z^\ast}\psi,1/z^\ast\right)^\dagger
\right] =
{d\over dz} \left\{
z^{-2 h}V\left(e^{L_1z^\ast}\psi,1/z^\ast\right)^\dagger \right\}\,,
\eqno(2.32)$$
so that the translation property is satisfied by
$z^{-2h}V(e^{{z^\ast}L_1}\psi,1/z^\ast)^\dagger$. The required
result follows by linearity of the vertex operators. $\sqre$

So, let us assume

\noindent
{\bf P7} For all $\psi\in\F$
$V(e^{z^\ast
L_1}{z^\ast}^{-2L_0}
\psi,1/z^\ast)^\dagger$ is local with respect to $\bV$.

The conclusions of Proposition 2.8 then follow. The following result details
some of the useful properties which follow from this hermitian structure.

\noindent
{\it {\bf Proposition 2.9} For} $\psi$, $\phi\in\F$

\noindent
{\it (i) if} $L_0\psi=h\psi$ {\it then} $L_0\overline\psi=h\overline\psi\,,$
{\it i.e. the
conjugation operation preserves conformal weights}

\noindent
{\it (ii)} $\overline{L_{-1}\psi}=-L_{-1}\overline\psi$

\noindent
{\it (iii)} $\overline{\overline\psi}=\psi$

\noindent
{\it (iv)} $(f_{\phi_1\phi_2\phi_3})^\ast=(-1)^{h_1+h_2+h_3}f_{\overline\phi_1
\overline\phi_2
\overline\phi_3}\,,$

\noindent
{\it where} $\phi_1\,,\,\phi_2\,,\,\phi_3\in\F$ {\it with}
$L_0\phi_j=h_j\phi_j$ {\it for}
$1\leq j\leq 3$ {\it and}
$$f_{\phi_1\phi_2\phi_3}=\langle\overline\phi_1|V(\phi_2,1)|\phi_3\rangle
\eqno(2.33)$$

\noindent
{\it (v)} $\overline{L_1\psi}=-L_1\overline\psi$

\noindent
{\it (vi)  the spectrum of} $L_0$ {\it is integral, i.e. all states have
integral conformal weight}

\noindent
{\it (vii)} $\langle\overline\psi |\phi\rangle=\langle\overline\phi |\psi
\rangle\,.$

\noindent
{\it Proof.}

\noindent
(i) Consider
$$\eqalignno{
e^{\epsilon L_0} V\left(e^{L_1/\zeta}\psi,\zeta\right)e^{-\epsilon L_0}
|0\rangle &= e^{\epsilon L_0} e^{\zeta L_{-1}} e^{L_1/\zeta}\psi&(2.34)\cr
&=e^{ h\epsilon}\exp\left\{e^\epsilon\zeta L_{-1}\right\}
\exp\left\{e^{-\epsilon}L_{1}/\zeta\right\}\psi\,.&(2.35)}$$
Therefore
$$e^{\epsilon L_0} V\left(e^{L_1/\zeta}\psi,\zeta\right)e^{-\epsilon L_0}
=e^{ h\epsilon}V\left(\exp\left\{e^{-\epsilon}L_{1}/\zeta\right\}\psi,
e^\epsilon\zeta\right)\,,\eqno(2.36)$$
and by taking $\epsilon$ small we see
$$\left[ L_0,\, V\left(e^{L_{1}/\zeta}\psi,\zeta\right)\right]
=\left(\zeta{d\over d\zeta} +  h\right)
V\left(e^{L_{1}/\zeta}\psi,\zeta\right)\,,\eqno(2.37)$$
$$\left[ L_0,\, V(\pbar,z)\right] = \left(z{d\over dz} +  h\right)
V(\pbar,z)\,,\eqno(2.38)$$
showing that
$L_0\pbar= h\pbar$ as required.

\noindent
(ii) If $\phi = L_{-1}\psi$,
$$ V(\phbar,z) = z^{-2 h-2}V\left(e^{L_1z^\ast}L_{-1}
\psi,1/z^\ast\right)^\dagger
\,.\eqno(2.39)$$
Now,
$$\eqalignno{
e^{L_1z^\ast}L_{-1}\psi &= \left(L_{-1}+2z^\ast L_0 + (z^\ast)^2L_1\right)
e^{L_1z^\ast}\psi\cr
&= L_{-1}e^{L_1z^\ast}\psi
+ e^{L_1z^\ast}\left(2z^\ast L_0 - (z^\ast)^2L_1\right)\psi\,,&(2.40)}$$
using the algebra (2.2).  Thus,
$$V(\phbar,z) = -{d\over dz} V(\pbar,z) = - V(L_{-1}\pbar,z)\,,
\eqno(2.41)$$
so
$$\phbar = - L_{-1}\pbar\,.\eqno(2.42)$$

\noindent
(iii)
For $\psi$ a quasi-primary state of weight $h_\psi$, (2.24) gives
$$V(\overline\psi,z)=z^{-2h_\psi}V(\psi,1/z^\ast)^\dagger\,.\eqno(2.43)$$
{}From (i), $\overline\psi$ has the same conformal weight
as $\psi$, so that a second application of (2.7) shows that the vertex
operators for $\overline{\overline\psi}$ and $\psi$ are equal, and so by the
uniqueness
theorem $\overline{\overline\psi}=\psi$. But from (ii),
the action of $L_{-1}$ anticommutes with the barring operation. From
the decomposition of $\F$ into a direct sum of $su(1,1)$ highest weight
representations, we thus see that for a general state $\psi\in\F$ (not
necessarily quasi-primary) $\overline{\overline\psi}=\psi$.

\noindent
(iv) To establish this result, we note that
$$\eqalignno{
f_{\phi_1\phi_2\phi_3} &= \langle\phbar_1|V(\phi_2,1)|\phi_3\rangle\cr
&=\langle\phbar_1|e^{L_{-1}}V(\phi_3,-1)|\phi_2\rangle
&\hb{by (2.13)}\cr
&=\langle\phi_2|V(\phi_3,-1)^\dagger e^{L_1}|\phbar_1\rangle^\ast\cr
&=\langle\phi_2|V(e^{-L_1}\phbar_3,-1) e^{L_1}|\phbar_1\rangle^\ast
&\hb{by (2.24) and $\overline{\phbar}_3=\phi_3$}\cr
&=\langle\phi_2|e^{-L_{-1}}V(e^{L_1}\phbar_1,1)e^{-L_1}|\phbar_3\rangle^\ast
&\hb{by (2.13)}\cr
&=\langle\phi_2|e^{-L_{-1}}V(\phi_1,1)^\dagger e^{-L_1}|\phbar_3\rangle^\ast
&\hb{by (2.24) and $\overline{\phbar}_1=\phi_1$}\cr
&=\langle\phi_1|V(e^{-L_{1}}\phi_2,-1)^\dagger |\phbar_3\rangle^\ast
&\hb{by (2.13)}\cr
&=\langle\phi_1|V(\phbar_2,-1) |\phbar_3\rangle^\ast
&\hb{by (2.24)}\cr
&=(-1)^{h_1+h_2+h_3}\left(f_{\phbar_1\phbar_2\phbar_3}\right)^\ast
\,.&(2.44)}$$

\noindent
(v)
A quasi-primary
state remains quasi-primary under the map $\psi\mapsto\overline\psi$, since
$$\eqalignno{
||L_1\overline\psi||^2&=\langle\overline\psi|L_{-1}L_1|\overline\psi\rangle\cr
&=\langle\overline\psi|L_1L_{-1}|\overline\psi\rangle-2\langle\overline\psi|L_0
|\overline\psi
\rangle\cr
&=\langle\psi|L_1L_{-1}|\psi\rangle-2\langle\psi|L_0|\psi\rangle\,,&(2.45)}$$
using first $[L_1,L_{-1}]=2L_0$ and then the facts that bar anticommutes
with $L_{-1}$ and commutes with $L_0$ and that barring preserves the norm
(a special case of (iv) given by setting $\phi_2=|0\rangle$
and $\phi_3=\overline\phi_1$), so that
$$||L_1\overline\psi||^2=||L_1\psi||^2=0\,,\eqno(2.46)$$
for $\psi$ quasi-primary.
Hence we deduce $\overline{L_1\psi}=-L_1\overline\psi$ for an arbitrary state
$\psi$ by induction on $n$ in considering states of the form ${L_{-1}}^n
\phi$ with $\phi$ quasi-primary (as we saw in
Proposition 2.7, these span $\F$), {\it
i.e.} $L_1$ anticommutes with the barring operation.

\noindent
(vi)
{}From (2.7) and (2.18) we have
$$\langle 0|V(\psi,z)V(\phi,\zeta)|0\rangle=(z-\zeta)^{-h_\psi-h_\phi}K_{
\psi\phi}\,,\eqno(2.47)$$
where $L_0\phi=h_\phi\phi$, $L_0\psi=h_\psi\psi$ and $K_{\psi\phi}=\langle
0|V(\psi,1)|\phi\rangle$.
But, using (2.24) and $\overline{\overline\psi}=\psi$,
$V(\psi,1)=V(e^{L_1}\overline\psi,1)^\dagger$.
So, using (2.3), $\langle 0|V(\psi,1)=\langle\overline\psi |e^{L_{-1}}e^{L_1}$,
{\it i.e.}
$$K_{\psi\phi}=\langle e^{L_1}\overline\psi |e^{L_1}\phi\rangle\,.\eqno(2.48)$$
Taking $\phi=\overline\psi$ and $\psi$ (and hence $\phi$) quasi-primary, $K_{
\psi\overline\psi}=||\overline\psi ||^2$ and $K_{\overline\psi\psi}=||\psi
||^2$ (using
$\overline{\overline\psi}=\psi$), both of which are positive (for $\psi\neq
0$).
The locality relation (2.4) applied to (2.47) then implies that $K_{\psi
\overline\psi}=(-1)^{2h_\psi}K_{\overline\psi\psi}$ (as
$h_\psi=h_{\overline\psi}$),
and so $h_\psi$ must be integral. Hence the conformal weights of all quasi-
primary states and hence all of their $L_{-1}$ descendants ({\it i.e.}
all states) must be integral.

\noindent
(vii)
For general states, locality applied to (2.47) and making
use of (2.48) requires
$\langle e^{L_1}\overline\psi |e^{L_1}(-1)^{L_0}\phi\rangle=
\langle e^{L_1}\overline\phi |e^{L_1}(-1)^{L_0}\psi\rangle$. Replacing $\phi$
and $\psi$ by $e^{L_1}\phi$ and $e^{L_1}\psi$ respectively and using the
facts that $L_1$ anticommutes with the bar and that $(-1)^{L_0}e^{L_1}=
e^{-L_1}(-1)^{L_0}$, we obtain the required result.
(The factors of $(-1)^{L_0}$ cancel on either side, as the inner products
clearly vanish when the conformal weights of $\phi$ and $\psi$ are not
equal.) $\sqre$

An analogue of part (iv) of the above
will be of
importance later in proving one of the locality relations
when we come to consider extending the CFT by a
particular representation to give a new CFT.

\noindent
{\it {\bf Definition 2.10}}
The {\it moments} of the vertex operators are given by
$$V(\psi,z)=\sum_{n\in\ze}V_n(\psi)z^{-n-h_\psi}\,.\eqno(2.49)$$

Then, by (2.3),
$$V_{-h_\psi}(\psi)|0\rangle =\psi\,\,,\,\,V_n(\psi)|0\rangle =0\,\,\,
{\rm for}\,\,\,n>-h_\psi\,.\eqno(2.50)$$
({\it c.f.} (2.1) noting that the state $\psi_L=L_{-2}|0\rangle$
has conformal weight 2). (2.5) and (2.17) may then be rewritten in terms of
modes as
$$\eqalignno{
&[L_0,V_n(\psi)]=-nV_n(\psi)\cr
&[L_{-1},V_n(\psi)]=(1-n-h_\psi)V_{n-1}(\psi)\,,&(2.51)}$$
and the duality relation (2.11) can also be rewritten, giving
$$V(\psi,z)V(\phi,\zeta)=\sum_{n=0}^{\infty}(z-\zeta)^{n-h_\phi-h_\psi}
V(\phi_n,\zeta)\,,\eqno(2.52)$$
where
$$\phi_n=V_{h_\phi-n}(\psi)\phi\,,\eqno(2.53)$$
and $h_\phi$ and $h_\psi$ are the conformal weights of $\phi$ and $\psi$
respectively, the sum being bounded below because $V_{h_\phi-n}(\psi)\phi=0$
for $n<0$ (otherwise we would have a non-zero state with negative
conformal weight).
This is a precise version  of the operator product expansion (OPE),
showing that this important result, often assumed in theories as an axiom,
is simply a consequence of locality, emphasising further the important role
played by the requirement (2.4).

Note also that for $\psi$ quasi-primary, (2.27) gives
$$V_n(\psi)^\dagger=V_{-n}(\overline\psi)\,.\eqno(2.54)$$

A fact which will be of use later
when we come to discuss sub-conformal field theories is

\noindent
{\it {\bf Proposition 2.11} The vacuum is generated in the OPE corresponding to
states} $\phi$ {\it and} $\overline\phi$.

\noindent
{\it Proof.}
Considering the OPE (2.52) for $\psi=\overline\phi$, we see that the leading
term in
the expansion on the right hand side is $(z-\zeta)^{-2h_\phi}V(\phi_0,\zeta)$
where $\phi_0=V_{h_\phi}(\overline\phi)\phi$. $\phi_0$ has conformal weight
zero,
from (2.51), and so must be proportional to the vacuum state $|0\rangle$ by
our assumption P6
about the uniqueness of the $su(1,1)$ invariant state, {\it
i.e.} $\phi_0=k|0\rangle$, and so the leading term is $k(z-\zeta)^{-2h_\phi}$.
Comparing with (2.47), we see that $k=K_{\overline\phi\phi}=||e^{L_1}\phi||^2$
from
(2.48). So $k>0$, and therefore the vertex operator for $|0\rangle$ appears in
the OPE. $\sqre$

Consider now the OPE (2.52) for $\psi_L$ with $\psi_L$. The singular terms in
the expansion are $\phi_0$, $\phi_1$, $\phi_2$ and $\phi_3$. $\phi_0={c'\over
2}|0\rangle$, where $c'=2||\psi_L||^2$, $\phi_1=L_1L_{-2}|0\rangle=3L_{-1}|0
\rangle=0$, $\phi_2=L_0\psi_L=2\psi_L$ and $\phi_3=L_{-1}\psi_L$. Then,
setting $V(\psi_L,z)=L(z)$,
$$
L(z)L(\zeta)={c'\over 2}(z-\zeta)^{-4}+2(z-\zeta)^{-2}L(\zeta)+
(z-\zeta)^{-1}{d\over{d\zeta}}L(\zeta)+O(1)\,,\eqno(2.55)$$
where the $(z-\zeta)^{-1}$ term is rewritten using (2.10) and $O(1)$ stands for
terms regular at $z=\zeta$. From (2.55) we can use the usual contour
manipulation arguments to derive (2.2) (with $c=c'$). That is
$$\eqalignno{
[L_n,L_m]&={1\over{(2\pi i)^2}}\left[\oint_{|z|>|\zeta|}dz\oint d\zeta-
\oint_{|\zeta|>|z|}dz\oint d\zeta\right] z^{n+1}\zeta^{m+1}L(z)L(\zeta)\cr
&={1\over{(2\pi i)^2}}\oint_0d\zeta\oint_\zeta dz\,z^{n+1}\zeta^{m+1}
L(z)L(\zeta)\,,&(2.56)}$$
where the $z$ integral is taken on a contour positively encircling $\zeta$
excluding $z=0$ and the $\zeta$ contour is then taken positively about $
\zeta=0$. Substituting in from (2.55) gives the required result. In other
words, we can deduce the entire Virasoro algebra from the conformal field
theory structure and the few simple properties used immediately
above (2.55) (in
particular, the relation $[L_1,L_{-2}]=3L_{-1}$),
{\it i.e.} we can weaken the requirement P1 slightly and it still holds true in
full,
showing once more the
powerful consequences which follow from the structure of local vertex
operators.

We may similarly deduce the conformal properties of vertex operators. We have
$$\eqalignno{
L(z)V(\phi,\zeta)=\ldots+&(z-\zeta)^{-4}V(L_2\phi,\zeta)+
(z-\zeta)^{-3}V(L_1\phi,\zeta)+\cr
&h_\phi(z-\zeta)^{-2}V(\phi,\zeta)+(z-\zeta)^{-1}{d\over{d\zeta}}
V(\phi,\zeta)+O(1)\,,&(2.57)}$$
for a state $\phi$ of weight $h_\phi$. Hence, if $L_n\phi=0$ for $n=1$, 2 (and
so for all $n\geq 1$ by (2.2)),
$$[L_n,V(\phi,\zeta)]=\zeta^n\left\{ \zeta{d\over{d\zeta}}+(n+1)h_\phi\right\}
V(\phi,\zeta)\,,\eqno(2.58)$$
for all $n$, by the contour manipulation argument.

\noindent
{\it {\bf Definition 2.12}}
A state $\phi$
said to be a {\it (conformal) primary state} if it
is a highest weight state for the Virasoro algebra,
{\it i.e.} if $L_n\phi=0$ for $n=1$, 2.
The corresponding vertex operator is
said to be a {\it (conformal) primary field}.

By using once more the fact that the conformal weights are bounded below, we
see that $\F$ splits up into a direct sum of Virasoro highest weight
representations, each generated by the action of the operators $L_{-n}$ for
$n>0$ on Virasoro highest weight states.
If instead we have $\phi$ only quasi-primary,
then (2.58) holds only for $n=0$, $\pm 1$. The relation (2.58) is a
generalisation of (2.5) and (2.17), which hold for all states.

For a quasi-primary state $\psi$ of weight $h_\psi$, (2.58) for $n=0$, $\pm 1$
is equivalent to the M\"obius transformation property
$$D_\gamma V(\psi,z)D_\gamma^{-1}=\left[{d\gamma(z)\over{dz}}\right]^{h_\psi}
V(\psi,\gamma(z))\,,\eqno(2.59)$$
where
$$D_\gamma=\exp\left\{{b\over d}L_{-1}\right\}\left({\sqrt{ad-bc}\over
d}\right)
^{2L_0}\exp\left\{-{c\over d}L_1\right\}\,,\eqno(2.60)$$
and
$$\gamma(z)={az+b\over{cz+d}}\,.\eqno(2.61)$$
This freedom to perform M\"obius transformations on the variables for
quasi-primary fields means that we can write the three-point
function for quasi-primary states $\phi_1$, $\phi_2$ and $\phi_3$ in the form
$$\eqalignno{
\langle 0|V(\phi_1,z_1)V(\phi_2,z_2)V(\phi_3,z_3)|0\rangle=
(z_1-&z_2)^{h_3-h_1-h_2}(z_1-z_3)^{h_2-h_1-h_3}\cr
&\cdot (z_2-z_3)^{h_1-h_2-h_3}
f_{\phi_1\phi_2\phi_3}\,,&(2.62)}$$
where $f_{\phi_1\phi_2\phi_3}$is defined as above
and $L_0\phi_j=h_j\phi_j$ for $1\leq j\leq 3$. Locality therefore implies
$$
f_{\phi_1\phi_2\phi_3}=f_{\phi_2\phi_3\phi_1}=(-1)^{h_1+h_2+h_3}
f_{\phi_1\phi_3\phi_2}=(-1)^{h_1+h_2+h_3}f_{\phi_3\phi_2\phi_1}\,.\eqno(2.63)
$$

\noindent
{\it {\bf Definition 2.13}}
We shall say that two conformal field theories $\H$ and $\H'$, with dense
subspaces $\F$ and $\F'$ respectively and corresponding vertex operators
$V(\psi,z)$ and $V'(\psi',z)$, are {\it isomorphic}
if there exists a unitary map
$u:\H\rightarrow\H'$ such that
$$V'(u\psi,z)=uV(\psi,z)u^{-1}\,,\eqno(2.64)$$
for $\psi\in\F$.

\noindent
{\it {\bf Proposition 2.14} If} $u:\H\rightarrow\H'$ {\it is an isomorphism of
conformal field theories} $\H$, $\H'$ {\it then} $u|0\rangle=|0'\rangle$,
$u\psi_L=\psi'_L$ {\it where} $|0\rangle$, $|0'\rangle$ {\it are the vacuum
states and} $\psi_L$, $\psi'_L$ {\it are the conformal states in} $\H$ {\it
and} $\H'$ {\it respectively.}

\noindent
{\it Proof.}
Taking $\psi=|0\rangle$, the vacuum in $\F$, we have
$V'(u|0\rangle,z)\equiv 1$, so by uniqueness $u|0\rangle=|0'\rangle$, the
vacuum state in $\F'$. Also, we can show that $u$ must map the other special
state in $\F$, the conformal state $\psi_L$, into the corresponding state
$\psi'_L$ in $\F'$. From the action of the vertex operators on the vacuum,
(2.3), together with (2.64) we see that $uL_{-1}u^{-1}=L'_{-1}$. Set $\tilde
L_n=uL_nu^{-1}$ for all $n$. Then $\tilde L_{-1}=L'_{-1}$. By conjugation we
have $\tilde L_1=L'_1$, and by $[L_1,L_{-1}]=2L_0$ we see that $\tilde
L_0=L'_0$.

\noindent
{}From the OPE (2.52), we have
$$V'(\psi_L',z)V'(u\psi_L,\zeta)=\sum_{n=0}^\infty(z-\zeta)^{n-4}V'(\phi_n,
\zeta)\,,\eqno(2.65)$$
where
$$\phi_n=L'_{2-n}u\psi_L\,.\eqno(2.66)$$
(Note that the conformal weight of $u\psi_L$ is still 2, as $\tilde L_0=L_0'$
implies that $u$ preserves conformal weights.) $L'_2u\psi_L$ has zero conformal
weight, and so is ${k\over 2}|0'\rangle$, by the uniqueness assumption made
above, for some $k\in\ce$. $L'_1u\psi_L=uL_1\psi_L=0$, $L'_0u\psi_L=2u\psi_L$
(see comment above) and $V'(L_{-1}'u\psi_L,\zeta)={d\over{d\zeta}}
V'(u\psi_L,\zeta)$ ({\it c.f.} (2.10)), so that (2.65) becomes
$$V'(\psi_L',z)V'(u\psi_L,\zeta)={k\over 2}(z-\zeta)^{-4}+2(z-\zeta)^{-2}
V'(u\psi_L,\zeta)+(z-\zeta)^{-1}{d\over{d\zeta}}V'(u\psi_L,\zeta)+O(1)\,,
\eqno(2.67)$$
Hence, comparing with (2.55), we see that the contour manipulation argument
gives
$$[L'_m,\tilde L_n]=(m-n)\tilde L_{m+n}+{k\over{12}}m(m^2-1)\delta_{m,-n}\,,
\eqno(2.68)$$
(the modes of $V'(u\psi_L,z)$ are $\tilde L_n$ by (2.64)).

\noindent
On the other hand, we find similarly
$$V'(u\psi_L,z)V'(\psi_L',\zeta)={k'\over 2}(z-\zeta)^{-4}+2(z-\zeta)^{-2}
V'(\psi_L',\zeta)+(z-\zeta)^{-1}{d\over{d\zeta}}V'(\psi_L',\zeta)+O(1)\,,
\eqno(2.69)$$
giving
$$[\tilde L_m,L_n']=(m-n)L'_{m+n}+{k'\over{12}}m(m^2-1)\delta_{m,-n}\,.
\eqno(2.70)$$
Comparing (2.68) and (2.70) shows $\tilde L_n=L_n'$ for $n\neq 0$, and so
$\psi_L'=u\psi_L$ as required (we already have the case $n=0$ from the above
discussion). $\sqre$

Finally in this section, we discuss the notion of a sub-conformal field theory.
This concept is not particularly exploited in the following sections, but it
does
provide an interesting example of the techniques and structures discussed
above.

\noindent
{\it {\bf Definition 2.15}}
A {\it sub-conformal field theory} of a conformal field theory
$\H$ is defined to be a subspace $\J$ of $\H$
such that

\leftline{(i) $\,\,\J$ is an invariant subspace for each $V(\phi,z)$, $\phi\in
\F_\J\equiv\J\cap\F$}

\leftline{(ii) $\J$ is invariant under the $su(1,1)$ algebra $L_{\pm 1}$,
$L_0$}

\leftline{(iii) $\overline{\F_\J}\equiv\{\overline\phi:\phi\in\F_\J\}=\F_\J$.}

We noted earlier that the vertex operator for $|0\rangle$ appears in the OPE of
$\phi$ and $\overline\phi$. From (i) and (iii), this immediately implies that
$|0\rangle\in\J$. (This automatically gives invariance under $L_{-1}$ from
$V(\psi,z)|0\rangle=e^{zL_{-1}}\psi$ for $\psi\in\J$ together with (i).) Let us
denote the orthogonal complement of $\J$ as $\Jp$, {\it i.e.}
$\Jp=\{\psi\in\H:\langle\psi|\phi\rangle=0\,\forall\,\phi\in\J\}$. If $\J$ is
invariant under $L_n$ for some $n$, then for $\phi\in\J$ and $\psi\in\Jp$
$0=\langle\psi|L_n|\phi\rangle=\langle\phi|L_{-n}|\psi\rangle^\ast$, {\it i.e.}
$\Jp$ is invariant under $L_{-n}$. Hence (ii) is equivalent to saying that
$\Jp$ is $su(1,1)$ invariant. Also, by (vii)
of Proposition 2.9, (iii) could equally well be
stated for $\Jp$.

\noindent
{\it {\bf Proposition 2.16} A sub-conformal field theory of a (hermitian)
conformal
field theory
is itself a (hermitian) conformal field
theory.}

\noindent
{\it Proof.} Suppose $\J$ is a sub-conformal field theory of $\H$.
Let $P^\J$ be the orthogonal projection onto $\J$, and set
$\psi_L^\J=P^\J\psi_L$. Write $L^\J(z)=V(\psi_L^\J,z)$ and $K(z)=V(\psi_K,z)$,
where $\psi_K=\psi_L-\psi_L^\J\in\Jp$. To evaluate the OPE $L(z)L^\J(\zeta)$,
we require the actions of $L_2$, $L_1$ and $L_0$ on $\psi_L^\J$.
$L_0\psi_L=2\psi_L$ gives $L_0(\psi_L^\J+\psi_K)=2(\psi_L^\J+\psi_K)$. But $\J$
and $\Jp$ are $L_0$ invariant. So, comparing both sides, we see that
$L_0\psi_L^\J=2\psi_L^\J$ (and $L_0\psi_K=2\psi_K$), by uniqueness of the
decomposition of a state into a sum of a state in $\J$ and a state in $\Jp$.
$L_2\psi_L^\J$ has conformal weight zero, and so is ${c^\J\over 2}|0\rangle$,
for some $c^\J\in\ce$. So
$$
{c^\J\over 2}=\langle 0|L_2|\psi_L^\J\rangle
=\langle\psi_L|\psi_L^\J\rangle
=\langle\psi_L|P^\J|\psi_L^\J\rangle
=||\psi_L^\J||^2\,,\eqno(2.71)$$
since $P^\J={P^\J}^\dagger$. Finally, $L_1\psi_L=0=L_1\psi_L^\J+L_1\psi_K$. But
$\J$ and $\Jp$ are $L_1$ invariant. So $L_1\psi_L^\J=-L_1\psi_K\in\J\cap
\Jp=\{0\}$. From this, it follows by the usual contour manipulation argument
(or comparing with (2.55)) that
$$[L_m,L_n^\J]=(m-n)L_{m+n}^\J+{c^\J\over{12}}m(m^2-1)\delta_{m,-n}\,,
\eqno(2.72)$$
where $L_n^\J$ are the modes of $L^\J(z)$. To deduce from (2.72) that the
$L_n^\J$ satisfy the (centrally extended) Virasoro algebra, we need to show
that $[K_m,L_n^\J]=0$, where $K_n$ are the modes of $K(z)$ ($L_n=L_n^\J+
K_n$). To do this, we consider the OPE $K(z)L^\J(\zeta)$. We need to prove that
it contains no singular terms (at $z=\zeta$), so that the commutator vanishes
on
applying the contour manipulation argument, {\it i.e.} we need $K_n\psi_L^\J=0$
for $-1\leq n\leq 2$.

\noindent
To show this, let us look at the action of vertex operators for states in $\Jp$
on states in $\J$ and vice versa. Let $\times$ denote the operation of taking
the operator product and identifying states with the corresponding vertex
operators. Then (i) becomes $\J\times\J\subset\J$. Therefore, for $\chi\in\Jp$
and $\phi$, $\psi\in\J$, $\langle\chi|V(e^{z^\ast L_1}{z^\ast}^{-2L_0}\overline
\phi,{1\over{z^\ast}})|\psi\rangle=0$, the state forming the argument of the
vertex operator being in $\J$ by (ii) and (iii). So, conjugating and using
(2.24), $\langle\psi|V(\phi,z)|\chi\rangle=0$, {\it i.e.}
$\J\times\Jp\subset\Jp$. But, from (2.13), $V(\chi,z)\phi=e^{zL_{-1}}V(\phi,
-z)\chi$. So, since $\Jp$ is invariant under $L_{-1}$, this result also implies
$\Jp\times\J\subset\Jp$.
Thus, $K_n\psi_L^\J\in\Jp$ and $L_n^\J\psi_L^\J\in\J$. But, from the above
calculation of the OPE (2.72), we see that $L_n\psi_L^\J\in\J$ for $-1\leq
n\leq 2$. So $K_n\psi_L^\J=0$ for $-1\leq n\leq 2$, as required.

\noindent
This gives us a type of generalised coset construction\ref{15},
with $L_n^\J$ and $K_n$
satisfying commuting Virasoro algebras with $L_n=L_n^\J+K_n$,
$$[L_m^\J,L_n^\J]=(m-n)L_{m+n}^\J+{c^\J\over{12}}m(m^2-1)\delta_{m,-n}
\eqno(2.73)$$
$$[K_m,K_n]=(m-n)K_{m+n}+{c^K\over{12}}m(m^2-1)\delta_{m,-n}\,,\eqno(2.74)$$
where $c^K=c-c^\J=2||\psi_K||^2$.

\noindent
Since $L_n\J\subset\J$ for $n=0$, $\pm 1$ by (ii) and $L_n^\J\J\subset\J$, we
see that $K_n\J=0$ for $n=0$, $\pm 1$, {\it i.e.} we can replace $L_0$, $L_{\pm
1}$ by $L_0^\J$, $L^\J_{\pm 1}$ respectively when acting on $\J$. Hence, $\J$
becomes a conformal field theory, with vertex operators $V(\psi,z)$ for
$\psi\in\J$ restricted to $\J$, vacuum state $|0\rangle$ and conformal state
$\psi_L^\J$. It also retains the hermitian structure possessed by $\H$. (Note
also that the zero conformal weight state remains unique, since $L_0=L_0^\J$ on
$\J$, and similarly the spectrum of $L_0^\J$ is bounded below (and hence
non-negative).) $\sqre$

Set $\J^0=\{\psi\in\H:L_0^\J\psi=0\}$. We have

\noindent
{\it {\bf Proposition 2.17}}

\noindent
\line{
{\it (i)} $\psi\in\J^0\iff[L_n^\J,V(\psi,z)]=0\,\forall\,n\in\ze$
\hfill (2.75)}

\noindent
{\it (ii)} $\J^0$ {\it is a sub-conformal field theory of} $\H$.

\noindent
{\it (iii) The vertex operators corresponding to} $\J$ {\it and} $\J^0$ {\it
commute.}

\noindent
{\it (iv) $\J\subset{(\J^0)}^0$. In fact, $(\J^0)^0$ is the largest
sub-conformal field theory in $\H$
containing $\J$ and sharing the same conformal structure.}

\noindent
{\it Proof.}

\noindent
(i) If $\psi\in\J^0$, then $L_n^\J\psi=0$
for $n>0$, since the spectrum of $L_0^\J$ is non-negative. Also
$L_{-1}^\J\psi=0$, since $||L_{-1}^\J\psi||^2=||L_1^\J\psi||^2=0$, using
$[L_1^\J,L_{-1}^\J]=2L_0^\J$.
If the commutator vanishes, applying it to the vacuum for $n\geq -1$ (for which
$L_n^\J|0\rangle=0$, since $L_n|0\rangle=0$ and so
$L_n^\J|0\rangle=-K_n|0\rangle\in\J\cup\Jp=\{0\}$) gives
$$L_n^\J V(\psi,z)|0\rangle=0\,,\eqno(2.76)$$
so that (2.6) gives the left hand side of (2.75). Conversely, if $\psi\in\J^0$,
we look at the OPE $L^\J(z)V(\psi,\zeta)$. The singular terms involve the
vertex operators for the states $L_n^\J\psi$ for $n\geq -1$, and so vanish.
Therefore, the right hand side of (2.75) follows by contour integration.

\noindent
(ii)
$\J^0$ is invariant under $L_{\pm 1}$, $L_0$, since if $\psi\in\J^0$ then
$L_0^\J L_m\psi=L_mL_0^\J\psi-[L_m,L_0^\J]\psi$. The first term vanishes by
definition of $\J^0$, and the second term, from (2.72), is $-mL_m^\J\psi$ for
$-1\leq m\leq 1$, which vanishes by the argument given in (i). Also,
conjugation of the right hand side of (2.75) implies that $\J^0$ is invariant
under the bar operation. Also, for $\phi$, $\psi\in\J^0$,
$L_0^\J V(\phi,z)\psi=V(\phi,z)L_0^\J\psi$, by (2.75), which vanishes
by definition of $\J^0$, {\it i.e.} $\J^0$ is invariant under $V(\phi,z)$ for
$\phi\in\J^0$.
Thus, we see that $\J^0$ is a sub-conformal field theory of $\H$.

\noindent
(iii)
For $\phi\in\J^0$, $\langle\psi_L^\J|\phi\rangle=\langle 0|L_2^\J|\phi
\rangle=0$, {\it i.e.} $\psi_L^\J\in{\J^0}^\perp$. Also, from (2.75) and the
fact that $[L_m^\J,K_n]=0$, we see that $\psi_K\in\J^0$. Hence, $P^{\J^0}
\psi_L=\psi_K$, or $L_n^{\J^0}=K_n$. Thus, the Virasoro algebras of $\J$ and
$\J^0$ are complementary, (that is, they commute, and add to give the Virasoro
algebra for $\H$). More generally,
the vertex operators corresponding to $\J$ and
$\J^0$ commute, since the singular terms in the OPE $V(\psi,z)V(\phi,\zeta)$
for $\psi\in\J$ and $\phi\in\J^0$ vanish. To see this, we note by positivity of
the $L_0$ eigenspace that $V_n(\psi)\phi=0$ for $n>h_\phi$. Then from
$L_{-1}^\J V_n(\psi)\phi=-(h_\psi+n-1)V_{n-1}(\psi)\phi$ (since
$L_{-1}^\J\phi=0$
as $\phi\in\J^0$, so we can replace $L_{-1}^\J V_n(\psi)$ by its commutator,
and
as $\psi\in\J$ $L_{-1}^\J$ can be replaced by $L_{-1}$ and we then use (2.51))
we can deduce recursively that $V_n(\psi)\phi=0$ for $n>-h_\psi$, as required.

\noindent
(iv)
The above result gives,
in particular,
$[K_n,V(\psi,z)]=0$ for all $n$ and all $\psi\in\J$,
so that by (2.75) we see that $\J\subset(\J^0)^0$. Also the conformal state for
$(\J^0)^0$ is complementary to that for $\J^0$, and so coincides with that for
$\J$. $\sqre$
\vskip12pt
\centerline{\bf 3. Representations of conformal field theories}
\nobreak
\vskip3pt
\nobreak

\noindent
{\it {\bf Definition 3.1}}
A {\it representation} $(\bU,\K)$ of the conformal field theory $\H$ is
a Hilbert space $\K$ and a set of linear operators $U(\psi,z):\K\rightarrow\K$
linear in $\psi$ for $\psi\in\F$ such that
$$U(\psi,z)U(\phi,\zeta)=U(V(\psi,z-\zeta)\phi,\zeta)\,,\eqno(3.1)$$
({\it c.f.} the duality relation (2.11)), with $U(|0\rangle,z)\equiv 1$
(otherwise we could have $U(\psi,z)$ vanishing on some subspace of $\K$).

Equivalently, using the mode expansion of $V$,
$$U(\psi,z)U(\phi,\zeta)=\sum_{n=0}^\infty(z-\zeta)^{n-h_\phi-h_\psi}U(\phi_n,
\zeta)\,,\eqno(3.2)$$
where $h_\phi$ and $h_\psi$ are the conformal weights of $\phi$ and $\psi$
respectively and the $\phi_n$ are as in (2.53).
The representations which we consider will be meromorphic, that is matrix
elements of operators in $\bU$ will be meromorphic functions of the
complex arguments of the operators.

As a simple consequence of this definition, we have

\noindent
{\it {\bf Proposition 3.2}}

\noindent
{\it (i) The operators in} $\bU$ {\it are local}

\noindent
{\it (ii) The modes of} $U(\psi_L,z)$ {\it satisfy the Virasoro algebra (2.2)}

\noindent
{\it (iii)} $U(\psi,z)$ {\it possesses the
analogous translation property to (2.5)}

\noindent
{\it Proof.}

\noindent
(i)
First, note that by taking $\phi=|0\rangle$ in
(3.1) and using (2.3) we obtain
$$U(\psi,z)=U(e^{(z-\zeta)L_{-1}}\psi,\zeta)\,.\eqno(3.3)$$
Hence
$$\eqalignno{
U(\psi,z)U(\phi,\zeta)&=U(e^{(z-\zeta)L_{-1}}V(\phi,\zeta-z)\psi,\zeta)
&\hb{by (3.1) and (2.13)}\cr
&=U(V(\phi,\zeta-z)\psi,z)&\hb{by (3.3)}\cr
&=U(\phi,\zeta)U(\psi,z)&\hb{by (3.1)}}$$
{\it i.e.}
$$U(\psi,z)U(\phi,\zeta)=U(\phi,\zeta)U(\psi,z)\,,\eqno(3.4)$$
again in the sense of analytic continuation of matrix elements of either side.

\noindent
(ii)
Set $U(\psi_L,z)=L(z)=\sum_nL_nz^{-n-2}$ (using the same notation as for
$V(\psi_L,z)$, but the distinction will always be obvious from the context).
Then (3.2) and the usual contour manipulation argument show that the $L_n$
satisfy the Virasoro algebra with the same central term as for the conformal
field theory $\H$.

\noindent
(iii)
(3.3) implies that
$${d\over{dz}}U(\psi,z)=U(L_{-1}\psi,z)\,.\eqno(3.5)$$
But from (3.2) with $\psi=\psi_L$ we see that
$$[L_{-1},U(\psi,z)]=U(L_{-1}\psi,z)\,,\eqno(3.6)$$
and so the result follows. $\sqre$

\noindent
{\it {\bf Example 3.3}}
If $\J$ is a sub-conformal field
theory of the conformal field theory $\H$, as defined at the end of section
2,
then $\Jp$ forms a representation of $\J$, with $U(\phi,z)=V(\phi,z)$
restricted to $\Jp$ for $\phi\in\J$.

\noindent
{\it {\bf Proposition 3.4}
The existence of the representation given in Definition 3.1 is equivalent
to the existence of ``intertwining'' operators}
$W(\chi,z):\H\rightarrow\K$ {\it for}
$\chi\in\K$ {\it (or rather a dense subspace of $\K$) such that}
$$U(\psi,z)W(\chi,\zeta)=W(\chi,\zeta)V(\psi,z)\eqno(3.7)$$
{\it and} $W(\chi,\zeta)|0\rangle\rightarrow\chi$
{\it as} $\zeta\rightarrow 0$.

The
locality relation (3.7), interpreted in the usual sense, is referred to as the
{\it intertwining relation}.

\noindent
{\it Proof.}
Given a representation as above we define $W(\chi,z)$ by
$$W(\chi,z)\phi=e^{zL_{-1}}U(\phi,-z)\chi\,,\eqno(3.8)$$
({\it c.f.} the relation (2.13). We shall ultimately combine the representation
and the CFT to give a new CFT, and the relation (2.13) which must hold for this
CFT requires (3.8) to hold.). Given this definition, we have
$$\eqalignno{
U(\psi,z)W(\chi,\zeta)\phi&=U(\psi,z)e^{\zeta L_{-1}}U(\phi,-\zeta)\chi
&\hb{by (3.8)}\cr
&=e^{\zeta L_{-1}}U(\psi,z-\zeta)U(\phi,-\zeta)\chi\cr
&=e^{\zeta L_{-1}}U(V(\psi,z)\phi,-\zeta)\chi&\hb{by (3.1)}\cr
&=W(\chi,\zeta)V(\psi,z)\phi&\hbp{by (3.8),}{(3.9)}}$$
and taking $\phi=|0\rangle$ in (3.8) and letting $z\rightarrow 0$ we obtain
$W(\chi,z)|0\rangle\rightarrow\chi$ as $z\rightarrow 0$, as required.

\noindent
Conversely, if we are given the intertwining operators $W(\chi,z)$ satisfying
the intertwining relation (3.7) for some operators $U(\psi,z)$ and also the
limiting relation as $z\rightarrow 0$ on $W(\chi,z)|0\rangle$, we have
$$\eqalignno{
U(\psi,z)U(\phi,\zeta)W(\chi,w)&=W(\chi,w)V(\psi,z)V(\phi,\zeta)&\hba{by
(3.7)}\cr
&=W(\chi,w)V(V(\psi,z-\zeta)\phi,\zeta)&\hba{by (2.11)}\cr
&=U(V(\psi,z-\zeta)\phi,\zeta)W(\chi,w)&\hbap{by (3.1).}{(3.10)}}$$
Apply (3.10) to $|0\rangle$ and let $w\rightarrow 0$ to give, since $\chi$ is
arbitrary, the required locality relation on $\bU$. In addition,
$U(|0\rangle,z)W(\chi,\zeta)=W(\chi,\zeta)$ from the intertwining relation, and
again acting on $|0\rangle$ and letting $\zeta\rightarrow 0$ we obtain
$U(|0\rangle,z)\equiv 1$ as required. $\sqre$

Note that defining $W$ by (3.8) from the representation $\bU$ gives
$$W(\chi,z)|0\rangle=e^{zL_{-1}}\chi\,,\eqno(3.11)$$
({\it c.f.} (2.3)). Conversely, given operators $W$ satisfying the intertwining
relation (3.7) and also (3.11) (which is consistent with the required limit as
$z\rightarrow 0$ in the definition of $W$) we obtain, by applying (3.7) to
$|0\rangle$ and using (2.3),
$$U(\psi,z)e^{\zeta L_{-1}}\chi=W(\chi,\zeta)e^{zL_{-1}}\psi\,.\eqno(3.12)$$
Taking $z\rightarrow 0$, the translation property for $U$ then gives the
relation (3.8). Hence, if we impose the stronger condition (3.11) rather than
just the limiting condition in the definition of the intertwining operators, we
always have the relation (3.8) between $W$ and $U$.

{}From (3.8) we have
$${d\over{dz}}W(\chi,z)\phi=e^{zL_{-1}}L_{-1}U(\phi,-z)\chi-e^{zL_{-1}}
[L_{-1},U(\phi,-z)]\chi\,\eqno(3.13)$$
using the translation property for U. Hence
$$\eqalignno{
{d\over{dz}}W(\chi,z)\phi&=e^{zL_{-1}}U(\phi,-z)L_{-1}\chi\cr
&=W(L_{-1}\chi,z)\phi\,,&(3.14)}$$
{\it i.e.}
$${d\over{dz}}W(\chi,z)=W(L_{-1}\chi,z)\,.\eqno(3.15)$$
Note also that the fact that the $U$'s are linear operators implies, from
(3.8), that $W(\chi,z)$ is linear in $\chi$, and since $U(\phi,z)$ is linear in
$\phi$ then the $W$'s are linear operators.

\noindent
{\it {\bf Definition 3.5}}
If $\H$ and $\H'$ are two isomorphic conformal field theories with isomorphism
$u:\H\rightarrow\H'$, as above, and $(\bU,\K)$ and $(\bU',\K')$
are representations of $\H$ and $\H'$ respectively, they are said to
be {\it equivalent}
if there is a unitary map $\rho:\K\rightarrow\K'$ such that
$$\rho U(\psi,z)\rho^{-1}=U'(u\psi,z)\,,\eqno(3.16)$$
for all $\psi\in\H$.

\noindent
{\it {\bf Proposition 3.6}
If}
$(\bU,\K)$
{\it is an irreducible representation,}
i.e. {\it if it has no proper subspaces invariant under all the}
$U(\psi,z)$
{\it for} $\psi\in\F$, {\it then}
$\rho$
{\it is unique up to multiplication by a constant.}

\noindent
{\it Proof.}
If
$\rho_1$ and $\rho_2$ are two suitable maps satisfying (3.16), then
$u=\rho_2^{-1}\rho_1$ is a unitary map commuting with $U(\psi,z)$ for all
$\psi\in\F$. By Schur's lemma ({\it i.e.} that any eigenspace of $u$ is an
invariant subspace, and so must be the whole space for an irreducible
representation), we deduce that $u$ is a multiple, $\kappa$ say,
$\kappa\in\ce$,
of the identity
map, {\it i.e.} $\rho_1=\kappa\rho_2$. $\sqre$
\vskip12pt
\centerline{\bf 4. Extension of a conformal field theory by a real hermitian
representation}
\nobreak
\vskip3pt
\nobreak
\noindent
{\it {\bf Definition 4.1}}
The representation $\bU$ described in the preceding section
is said to be {\it hermitian}
if
$$U(\overline\phi,z)=z^{-2h_\phi}U(e^{z^\ast L_1}\phi,1/z^\ast)^\dagger\,,
\eqno(4.1)$$
where $\phi$ is a state with conformal weight $h_\phi$, ({\it c.f.} (2.24)).

\noindent
Further, it is said to be {\it real} if there is an antilinear map
$\chi\mapsto\overline\chi$ on $\K$ such that
$\overline{\overline\chi}=\chi$, $L_{-1}\overline\chi=-\overline{L_{-1}\chi}$
and
$$(f_{\chi_1\phi\chi_2})^\ast=(-1)^{h_1+h_\phi+h_2}f_{\overline\chi_1
\overline\phi
\overline\chi_2}\,,\eqno(4.2)$$
({\it c.f.} (v) of Proposition
(2.9)) where $L_0\chi_j=h_j\chi_j$ for $j=1$, $2$,
$L_0\phi=h_\phi\phi$ and
$f_{\chi_1\phi\chi_2}=\langle\overline\chi_1|U(\phi,1)|\chi_2\rangle$. Also,
we require that if $L_0\chi=h_\chi\chi$ then
$L_0\overline\chi=h_\chi\overline\chi$.

\noindent
{\it {\bf Definition 4.2}} Set
$$\Wbar(\overline\chi,z)=z^{-2h_\chi}W(e^{z^\ast L_1}\chi,1/z^\ast)^\dagger
\,,\eqno(4.3)$$
where $h_\chi$ is the conformal weight of $\chi\in\K$ ({\it i.e.} eigenvalue of
the zero mode $L_0$ of $U(\psi_L,z)$), and $W$ is given by (3.8).

Note that this
definition extends to all states by linearity of $\Wbar(\chi,z)$ in $\chi$,
which follows from linearity of $W(\chi,z)$ in $\chi$ and the antilinearity of
the map $\chi\mapsto\overline\chi$. Its definition is inspired by (2.24) as we
wish it to be part of a vertex operator in some extended conformal field
theory.
The operator $\Wbar$ is a map from $\K$ to $\H$ which intertwines the conformal
field theory and the representation in the opposite sense to that in which $W$
does in (3.7), {\it i.e.}
$$\Wbar(\chi,\zeta)U(\psi,z)=V(\psi,z)\Wbar(\chi,\zeta)\,.\eqno(4.4)$$
This follows simply by conjugating (3.7) and using (4.3), (4.1) and (2.24).

\noindent
{\it {\bf Proposition 4.3} We have the locality relation}
$$\Wbar(\chi_1,z)W(\chi_2,\zeta)=\Wbar(\chi_2,\zeta)W(\chi_1,z)\,.\eqno(4.5)$$

\noindent
{\it Proof.}
First, consider the action of the left hand side on an arbitrary untwisted
state $\phi$, {\it i.e.}
$$\eqalignno{
\Wbar(\chi_1,z)W(\chi_2,\zeta)\phi&=\Wbar(\chi_1,z)e^{\zeta
L_{-1}}U(\phi,-\zeta)\chi_2&\hb{by (3.8)}\cr
&=e^{\zeta L_{-1}}\Wbar(\chi_1,z-\zeta)U(\phi,-\zeta)\chi_2\,,&(4.6)}$$
where the translation property for $\Wbar(\chi,z)$ used in the last line
follows by the same arguments as in Proposition 2.8 (trivially checking
that the appropriate
assumptions which went into that proposition remain valid in this context).
Hence, using (4.4),
$$\Wbar(\chi_1,z)W(\chi_2,\zeta)\phi=e^{\zeta
L_{-1}}V(\phi,-\zeta)\Wbar(\chi_1,z-\zeta)\chi_2\,.\eqno(4.7)$$
Therefore, the locality relation (4.5) is equivalent to
$$\eqalignno{
V(\phi,-\zeta)\Wbar(\chi_1,z-\zeta)\chi_2&=e^{(z-\zeta)L_{-1}}
V(\phi,-z)\Wbar(\chi_2,\zeta-z)\chi_1\cr
&=V(\phi,-\zeta)e^{(z-\zeta)L_{-1}}\Wbar(\chi_2,\zeta-z)\chi_1\,,&(4.8)}$$
and so it remains to prove
$$\Wbar(\chi_1,z)\chi_2=e^{zL_{-1}}\Wbar(\chi_2,-z)\chi_1\,.\eqno(4.9)$$
Note that this is analogous to the relation (2.13). Conjugating (4.9) by using
(4.3), with $h_1$ and $h_2$ the conformal weights of $\chi_1$ and $\chi_2$
respectively, we have to show the equality of
$$z^{2h_1}\langle\chi_2|W(e^{{1\over z}L_1}\overline\chi_1,z)\qquad{\rm and}
\qquad
z^{2h_2}\langle\chi_1|W(e^{-{1\over z}L_1}\overline\chi_2,-z)e^{{1\over
z}L_1}\,.
\eqno(4.10)$$
Acting on an arbitrary untwisted state $\phi$ again, and using (3.8), we have
to verify
$$z^{2h_1}\langle\chi_2|e^{zL_{-1}}U(\phi,-z)e^{{1\over z}L_1}|\overline
\chi_1\rangle=z^{2h_2}\langle\chi_1|e^{-zL_{-1}}U(e^{{1\over z}L_1}
\phi,z)e^{-{1\over z}L_1}|\overline\chi_2\rangle\,.\eqno(4.11)$$
{}From the fact that the representation is hermitian, {\it i.e.} using (4.1),
we
see that this is equivalent to
$$z^{2h_1}\langle\chi_2|e^{zL_{-1}}U(\phi,-z)e^{{1\over z}L_1}|\overline
\chi_1\rangle=z^{2h_2-2h_\phi}\langle\chi_1|e^{-zL_{-1}}U(\overline\phi
,1/z^\ast)^\dagger e^{{1\over z}L_1}|\overline\chi_2\rangle\,.
\eqno(4.12)$$
So, we see that for $\chi_1$ and $\chi_2$ quasi-primary states, we have to
verify
$$\eqalignno{
z^{2h_1}\langle\chi_2|U(\phi,-z)|\overline\chi_1\rangle&=
z^{2h_2-2h_\phi}\langle\chi_1|U(\overline\phi,1/z^\ast)^\dagger
|\overline\chi_2\rangle\cr
&=z^{2h_2-2h_\phi}\langle\overline\chi_2|U(\overline\phi,
1/z^\ast)|\chi_1\rangle^\ast\,.&(4.13)}$$
Using (2.18), we may remove the $z$ dependence from $U$, and find that the
relation which we have to verify reduces to (4.2), {\it i.e.} for a real
hermitian representation, the locality relation (4.5) holds for quasi-primary
states.

\noindent
To deduce the result in general, we make use of (3.15) and an analogous result
for $\Wbar(\chi,z)$ which we derive below. This enables us, by differentiation,
to infer locality for all $L_{-1}$ descendents of quasi-primary states, which
is sufficient by Proposition 2.7
and linearity of $\Wbar(\chi,z)$ in
$\chi$. From (2.40) together with (4.3), we see that
$${d\over{dz}}\Wbar(\overline\chi,z)=-\Wbar(\overline{L_{-1}\chi},z)\,,
\eqno(4.14)$$
so that, using $\overline{\overline\chi}=\chi$ and $L_{-1}\overline\chi=-
\overline{L_{-1}\chi}$,
$${d\over{dz}}\Wbar(\chi,z)=\Wbar(L_{-1}\chi,z)\,,\eqno(4.15)$$
as required. $\sqre$

Our main result is

\noindent
{\it {\bf Proposition 4.4}
If we also have the locality relation}
$$W(\chi_1,z)\Wbar(\chi_2,\zeta)=W(\chi_2,\zeta)\Wbar(\chi_1,z)\,,\eqno(4.16)$$
{\it and the spectrum of $L_0$ in the representation is strictly
positive, then we may extend the conformal field theory} $\H$ {\it to a
(hermitian) conformal field theory}
$\THX=\H\oplus\K$, {\it with vertex operators defined by}
$$\tilde V(\psi,z)=\pmatrix{V(\psi,z)&0\cr 0&U(\psi,z)}\,,\qquad
\tilde V(\chi,z)=\pmatrix{0&\Wbar(\chi,z)\cr W(\chi,z)&0}\,,\eqno(4.17)$$
{\it where we use the notation}
$\psi$ {\it for} $(\psi,0)$ {\it with} $\psi\in\H$ {\it and similarly}
$\chi$ {\it for} $(0,\chi)$ {\it with} $\chi\in\K$. {\it
The vacuum and conformal states are}
$(|0\rangle,0)$ {\it and} $(\psi_L,0)$
{\it respectively, which are written} $|0\rangle$
{\it and} $\psi_L$ {\it by this convention.}

\noindent
{\it Proof.}
The vertex operators have the required action on the vacuum, from the actions
of $V$ and $W$ on $|0\rangle\in\H$.
Since the modes of $U(\psi_L,z)$ satisfy the same Virasoro algebra as the modes
of $V(\psi_L,z)$, then we have the required Virasoro structure. (Also note that
we have uniqueness of the $su(1,1)$ invariant state and a spectrum of $L_0$
which is bounded below, properties which we assumed for $\H$ and which carry
over into this new theory.)
The locality relations necessary for this
to be a conformal field theory reduce to the six relations
$$V(\psi,z)V(\phi,\zeta)=V(\phi,\zeta)V(\psi,z)\eqno(4.18)$$
$$W(\chi,z)V(\phi,\zeta)=U(\phi,\zeta)W(\chi,z)\eqno(4.19)$$
$$U(\psi,z)U(\phi,\zeta)=U(\phi,\zeta)U(\psi,z)\eqno(4.20)$$
$$V(\psi,z)\Wbar(\chi,\zeta)=\Wbar(\chi,\zeta)U(\psi,z)\eqno(4.21)$$
$$\Wbar(\chi_1,z)W(\chi_2,\zeta)=\Wbar(\chi_2,\zeta)W(\chi_1,z)\eqno(4.22)$$
$$W(\chi_1,z)\Wbar(\chi_2,\zeta)=W(\chi_2,\zeta)\Wbar(\chi_1,z)\,,\eqno(4.23)$$
which we already have. Note also that the hermitian structure on the $U$'s and
the $V$'s together with reality of the representation
gives a hermitian structure on the new vertex operators, with
$\overline{(\psi,\chi)}=(\overline\psi,\overline\chi)$. $\sqre$

$\H$ is a sub-conformal field theory of $\THX$. We have $\psi_L\in\H$ and
$\psi_K=0$ in the previous notation, and there is a symmetric space structure
$$\eqalignno{
\H\times\H\subset\H\,,\quad&\H\times\K\subset\K\,,\quad\K\times\H\subset\K\,,
\quad
\K\times\K\subset\H\,.&(4.24)}$$
$\THX$ has an automorphism $\iota$ which acts as 1 on $\H$ and $-1$ on $\K$.

\noindent
{\it {\bf Proposition 4.5}
If we have another definition of reality on the space $\K$, with the
conjugation map $\chi\mapsto\widehat\chi$, then,
if $\THX'$ is the conformal field theory obtained by using $\widehat\chi$ in
place of $\overline\chi$, $\THX'$ is isomorphic to $\THX$.}

\noindent
{\it Proof.}
We see from
$$f_{\overline\chi_1\overline\phi\chi_2}=
f_{\widehat\chi_1\overline\phi\chi_2}=
\langle\chi_1|U(\overline\phi,1)|\chi_2\rangle\eqno(4.25)$$
and (4.2) that
$$\langle\overline\chi_1|U(\phi,1)|\overline\chi_2\rangle=
\langle\widehat\chi_1|U(\phi,1)|\widehat\chi_2\rangle\,,\eqno(4.26)$$
noting that conjugation preserves the conformal weight.
Let $\widehat\chi=u\chi$. Then (4.26) implies that
$$u^\dagger U(\phi,1)u=U(\phi,1)\,.\eqno(4.27)$$
Setting $\phi=|0\rangle$, we find that the map $u$ is unitary. So (4.27)
becomes
$$U(\phi,1)u=uU(\phi,1)\,.\eqno(4.28)$$
Since $L_0$ commutes with both conjugation operations (by definition), it
commutes with $u$. So we may use (2.18) to deduce from (4.28) that
$$U(\phi,z)u=uU(\phi,z)\,.\eqno(4.29)$$
So, if our representation is irreducible, then $u$ must be a multiple of the
identity, by Schur's lemma, {\it i.e.} $\widehat\chi=w^2\overline\chi$ for some
$w\in\ce$. Since $\overline{\overline\chi}=\widehat{\widehat\chi}=\chi$ and
conjugation
is antilinear, we must have $|w|=1$, since
$$\eqalignno{
\chi=\widehat{\widehat\chi}=\widehat{w^2\overline\chi}&=w^2\overline{w^2\overline\chi}\cr
&=w^2{w^\ast}^2\overline{\overline\chi}\cr
&=|w|^4\chi\,.&(4.30)}$$
Thus
$$\tilde V'(v\varrho,z)=v\tilde V(\varrho,z)v^{-1}\,,\eqno(4.31)$$
where
$$v=\pmatrix{1&0\cr 0&w}\eqno(4.32)$$
is the (unitary) isomorphism. $\sqre$

\noindent
{\it {\bf Proposition 4.6}
Let $\H$ and
$\H'$ be
two isomorphic hermitian
conformal field theories  with isomorphism $u:\H\rightarrow\H'$.
Suppose $(\bU,\K)$ and
$(\bU',\K')$ are equivalent representations of $\H$ and $\H'$
respectively, with a unitary map $\rho$ satisfying (3.16). Then

\noindent
(i) $u$ preserves the hermitian structure,} i.e. {\it $u\overline\psi=
\overline{u\psi}$ $\forall$ $\psi\in\H$

\noindent
(ii) $\rho W(\chi,z)u^{-1}=W'(\rho\chi,z)$ $\forall$ $\chi\in\K$

\noindent
(iii) If both representations are real, with conjugation denoted by barring,
$\overline{\rho\chi}=\rho\overline\chi$ $\forall$ $\chi\in\K$ (rescaling
$\rho$ if necessary)

\noindent
(iv) If, in addition, the final locality relation (4.23) holds in both theories
(and the representations $\bU$ and $\bU'$ are hermitian),
we may
extend $\H$ and $\H'$ to conformal field theories $\THX$ and $\THX'$
respectively, as in (4.17), (note that (4.23) is not affected by the
redefinition $\rho\mapsto w^{-1}\rho$) and
$u\oplus\rho$
defines an isomorphism $\THX\rightarrow\THX'$.
}

\noindent
{\it Proof.}

\noindent (i) This follows simply by conjugating (2.64)
and comparing with (2.24).

\noindent (ii)
For $\phi\in\H'$ and $\chi\in\K$,
$$\eqalignno{
\rho W(\chi,z)u^{-1}\phi&=\rho e^{zL_{-1}}U(u^{-1}\phi,-z)\chi&\hb{by
(3.8)}\cr
&=\rho e^{zL_{-1}}\rho^{-1}U'(\phi,-z)\rho\chi&\hb{by (3.16)}\cr
&=e^{zL_{-1}'}U'(\phi,-z)\rho\chi\cr
&=W'(\rho\chi,z)\phi&\hbp{by (3.8),}{(4.33)}}$$
where $L_{-1}'=\rho L_{-1}\rho^{-1}$ is the appropriate moment of the vertex
operator for the conformal state in $\K'$, from (3.16) and the fact that
$u\psi_L=\psi_L'$. Thus the result follows.

\noindent (iii)
We can
define a second conjugation operation on $\K'$ by
$\widehat{\chi'}=\rho\overline
{\rho^{-1}\chi'}$ for $\chi'\in\K'$. With respect to this conjugation,
$(\bU',\K')$ is still real, since
$$\eqalignno{
\widehat{\widehat{\chi'}}&=\rho\overline{\rho^{-1}\widehat{\chi'}}
=\rho\overline{\rho^{-1}\rho\overline{\rho^{-1}\chi'}}\cr
&=\rho\overline{\overline{\rho^{-1}\chi'}}=\rho\rho^{-1}\chi'\cr
&=\chi'\,,
&(4.34)}$$
$$\eqalignno{
\widehat{L_{-1}'\chi'}&=\rho\overline{\rho^{-1}L_{-1}'\chi'}
=\rho\overline{L_{-1}\rho^{-1}\chi'}\cr
&=-\rho L_{-1}\overline{\rho^{-1}\chi'}
=-L_{-1}'\rho\overline{\rho^{-1}\chi'}\cr
&=-L_{-1}'\widehat{\chi'}\,,&(4.35)}$$
and if $L_0'\chi'=h_{\chi'}\chi'$, then
$$\eqalignno{L_0'\widehat{\chi'}&=L_0'\rho\overline{\rho^{-1}\chi'}
=\rho L_0\overline{\rho^{-1}\chi'}\cr
&=\rho\overline{L_0\rho^{-1}\chi'}
=\rho\overline{\rho^{-1}L_0'\chi'}\cr
&=h_{\chi'}\rho\overline{\rho^{-1}\chi'}
=h_{\chi'}\widehat{\chi'}\,.&(4.36)}$$
Also, if $\widehat
f_{\chi_1'\phi'\chi_2'}=\langle\widehat{\chi_1'}|U'(\phi',1)|\chi_2'
\rangle$, for $\chi_1'$, $\chi_2'\in\K'$ and $\phi'\in\H'$, then
$$\eqalignno{
\widehat f_{\chi_1'\phi'\chi_2'}&=\langle\overline{\rho^{-1}\chi_1'}|\rho^{-1}
U'(\phi',1)|\chi_2'\rangle\cr
&=\langle\overline{\rho^{-1}\chi_1'}|U(u^{-1}\phi',1)|\rho^{-1}\chi_2'\rangle
\cr
&=f_{\rho^{-1}\chi_1'u^{-1}\phi'\rho^{-1}\chi_2'}\,,&(4.37)}$$
and the reality condition (4.2) follows from that in $\H$, noting that $\rho$
and $u$ preserve the conformal weights (since $L_0'=uL_0u^{-1}$ on $\H'$ and
$\rho L_0\rho^{-1}$ on $\K'$).
By the previous argument leading to (4.32), we see that
$\widehat{\chi'}=w^2\overline{\chi'}$ for some $w\in\ce$ with $|w|=1$, {\it
i.e.}
we just replace $\rho$ by $w^{-1}\rho$ to obtain
$\overline{\rho\chi}=\rho\overline\chi$ for $\chi\in\K$.

\noindent
(iv)
We have, for $\chi\in\K$,
$$\eqalignno{
u\Wbar(\chi,z)\rho^{-1}&=\left(\rho W(e^{z^\ast L_1}{z^\ast}^{-2L_0}\overline
\chi,1/z^\ast)u^{-1}\right)^\dagger\qquad{\rm by}\,\,(4.3)\,\,{\rm and}\,
\,u\,,\,\rho\,\,{\rm unitary}\cr
&=W'(\rho e^{z^\ast L_1}{z^\ast}^{-2L_0}\overline\chi,1/z^\ast)^\dagger
&\hb{by (ii)}\cr
&=W'(e^{z^\ast L_1'}{z^\ast}^{-2L_0'}\rho\overline\chi,1/z^\ast)
^\dagger\cr
&=W'(e^{z^\ast L_1'}{z^\ast}^{-2L_0'}\overline{\rho\chi},1/z^\ast)
^\dagger\cr
&=\Wbar'(\rho\chi,z)&\hbp{by (4.4).}{(4.38)}}$$
Hence, together with (ii), (3.16) and (2.64), we obtain the result.
Note that the requirement
$\overline{\rho\chi}=\rho\overline\chi$ determines $\rho$ up to a sign, which
corresponds to the automorphism $\iota$ of $\THX$ which was noted earlier.
$\sqre$
\vskip12pt
\centerline{\bf 5. Lattice constructions}
\nobreak
\vskip3pt
\nobreak
The general theory described in sections 2-4 is illustrated in the case of the
straight and $\ze_2$-twisted constructions of a conformal field theory from a
lattice, which we define below. For full details of these constructions and
proofs of their consistencies as meromorphic conformal field theories see
[11].
They can be regarded as being analogues of constructions of lattices from
binary codes, and it is for this reason that we begin this section with a
discussion of codes and lattices. In section 6, we shall demonstrate that the
connection with codes is more fundamental than at first apparent, and in fact
provides a more general framework in which to consider Frenkel, Lepowsky and
Meurmans' construction of the natural module for the Monster
group\ref{7}.
\vskip 12pt
\leftline{\bf 5.1 Codes and lattices}
\nobreak
\vskip 3pt
\nobreak
Let us begin with some definitions and simple facts.

A {\it binary linear code}
is a linear subspace $\C$ of the vector space ${{\Bbb
F}_2}^n$ over the two element field ${\Bbb F}_2=\{0,1\}$. $n$ is referred to as
the {\it length} of the code, and $\dim\C$ is its {\it dimension}.
Elements of $\C$ are
known as {\it codewords}, and the {\it weight} of a codeword $c\in\C$,
${\rm wt}(c)$, is
the number of non-zero coordinates of $c$, {\it i.e.} $c=(c_1,\ldots,c_n)$ with
$c_j=0$ or 1 and ${\rm wt}(c)=c^2=\one\cdot c$, where $\one=(1,1,\ldots,1)$ and
we use the inner product $x\cdot y=\sum_{j=1}^nx_jy_j$, with
$x=(x_1,\ldots,x_n)$ and $y=(y_1,\ldots,y_n)$ [and the arithmetic here is
not performed modulo 2!] The {\it dual} of the code $\C$ is the orthogonal
space
$\C^\ast=\{x\in{{\Bbb F}_2}^n:x\cdot y\equiv {0\bmod 2}\,\forall\,y\in\C\}$,
and is
also clearly a binary linear code. So we have $\dim\C^\ast=n-\dim\C$. A code
$\C$ is said to be {\it self-dual} if $\C=\C^\ast$
(so that $\dim\C=\half n$, {\it
i.e.} its length must be even). Clearly $\C$ is self-dual if and only if
$\C\subset\C^\ast$ and $\dim\C=\dim\C^\ast$. $\C$ is said to be
{\it even} if $c^2$
is even for all $c\in\C$. $\C$ is said to
be {\it doubly-even}
if $c^2$ is a multiple of 4 for all $c\in\C$. The length of any
doubly-even self-dual code has to be a multiple of 8.

An {\it $n$-dimensional Euclidean lattice}
$\Lambda$ is a subset of $n$-dimensional
Euclidean space which has integral coordinates in some basis $e_j$, $1\leq
j\leq n$, {\it i.e.} $\Lambda=\{\sum_{j=1}^nn_je_j:n_j\in\ze\}$ is the integral
span of a set of $n$ linearly independent $n$-dimensional Euclidean vectors.
(The definition can clearly be extended to the non-Euclidean case by dropping
the requirement that the inner product be positive definite.) The {\it length}
of a
vector $x\in\Lambda$ is $x^2$. $\Lambda$ is said to be
{\it integral} if $x\cdot y
\in\ze$ for all $x$, $y\in\Lambda$ and
{\it unimodular} if $||\Lambda||^2\equiv
\det (e_i\cdot e_j)=1$. The {\it dual lattice}
$\Lambda^\ast=\{y:x\cdot y\in\ze
\,\forall\,x\in\Lambda\}$ (which is obviously a lattice). Clearly $\Lambda$ is
integral if and only if $\Lambda\subset\Lambda^\ast$. Also, we see that
$\Lambda$ is {\it self-dual}, {\it i.e.} $\Lambda=\Lambda^\ast$, if and only if
$\Lambda$ is both integral and unimodular, since $||\Lambda^\ast||=||\Lambda||^
{-1}$. The lattice $\Lambda$ is said to be {\it even} if $x^2$ is even for all
$x\in\Lambda$. The dimension of an even self-dual lattice has to be a multiple
of 8.
\vskip10pt
\hbox to\hsize{\hfil
\vbox{\offinterlineskip
\halign{\vrule#&
\strut\quad\hfil # \hfil\quad&\vrule#&
\strut\quad\hfil # \hfil\quad&\vrule#&
\strut\quad\hfil # \hfil\quad&\vrule#\cr
\noalign{\hrule}
height4pt&\omit&&\omit&&\omit&\cr
&Codes&&Lattices&&Conformal field
&\cr
&\omit&&\omit&&theories&\cr
height4pt&\omit&&\omit&&\omit&\cr
\noalign{\hrule}
height4pt&\omit&&\omit&&\omit&\cr
&length&&dimension&&$c$&\cr
&weight&&(half) length&&conformal weight&\cr
&$\C\subset\C^\ast$&&integral or
$\Lambda\subset\Lambda^\ast$&&meromorphic&\cr
&$\dim\C=\dim\C^\ast$&&unimodular&&?&\cr
&self-dual&&self-dual&&self-dual&\cr
&doubly-even&&even&&bosonic&\cr
&Euclidean&&Euclidean&&?&\cr
&$W_\C(p)$&&$\Theta_\Lambda(\tau)$&&$\chi_\H(\tau)$&\cr
height6pt&\omit&&\omit&&\omit&\cr
\noalign{\hrule\vskip8pt}
\multispan{6} \hfil{Table 1. Comparison between codes, lattices and
conformal field
theories}
\hfil\cr
}}\hfil}
\vskip6pt
We can define a construction of a lattice
from a code,
known as the straight construction\ref{12}.
We start from a binary linear code $\C$ of length
$d$ and define a lattice $\Lambda_\C$ by
$$\Lambda_\C={\textstyle{1\over{\sqrt 2}}}\C+{\sqrt 2}\ze^d\,.\eqno(5.1)$$
We see that $\Lambda_\C$ is integral if and only if $\C\subset\C^\ast$, and
that
$\Lambda_\C$ is even if and only if $\C$ is doubly-even. Also, the length of
the vector ${1\over{\sqrt 2}}c\in\Lambda_\C$ corresponding to the codeword
$c\in\C$ is half the weight of $c$, while the dimension of $\Lambda_\C$ is
clearly just the length of $\C$, and the Euclidean structure is preserved. The
theta function
$$\Theta_\Lambda(\tau)=\sum_{x\in\Lambda}q^{\half x^2}\,,\quad q=e^{2\pi i
\tau}\,,\eqno(5.2)$$
of the lattice is given in terms of the weight enumerator
$$W_\C(p)=\sum_{c\in\C}p^{{\rm wt}(c)}\,,\eqno(5.3)$$
of the code as
$$\Theta_{\Lambda_\C}(\tau)=\Theta_3(\tau)^dW_\C(\Theta_2(\tau)/\Theta_3(\tau))
\,,\eqno(5.4)$$
where
$$\Theta_2(\tau)=\sum_{r\in\ze+\half}q^{r^2}\,,\qquad\Theta_3(\tau)=\sum_{m\in\ze
}q^{m^2}\,.\eqno(5.5)$$
It is clear that $(\Lambda_\C)^\ast=\Lambda_{\C^\ast}$, so that $\Lambda_\C$ is
self-dual if and only if $\C$ is self-dual. Also, $\Lambda_\C$ is unimodular
(which is equivalent to saying that it has one point per unit volume) if and
only if $\dim\C=\half d$ ($=\dim\C^\ast$). Therefore, this construction
implies
the correspondence between the properties listed in
the first two columns of table 1.
In section 5.2, we shall give a corresponding construction of a conformal field
theory from a lattice which justifies the correspondence
between the second and third columns of the table.
[We will discuss the notion of self-duality later. Bosonic
corresponds to evenness, since we have seen that the bosonic locality relation
(2.4) when combined with the hermitian condition requires all conformal
weights to be integral.
The function
$\chi_\H(\tau)$, the character or partition function, for the
conformal field theory $\H$ is defined to be
$$\eqalignno{\chi_\H(\tau)&=q^{-c/24}{\rm tr}(q^{L_0})\cr
&=q^{-c/24}\sum_h\dim\F_hq^h\,,&(5.6)}$$
where $q=e^{2\pi i\tau}$ as before, using the decomposition (2.19).]

We can divide the lattice $\Lambda_\C$ into two cosets, by defining
$$\Lambda_0(\C)=\fred\C+{\sqrt 2}\ze_+^d\eqno(5.7)$$
$$\Lambda_1(\C)=\fred\C+{\sqrt 2}\ze_-^d\,,\eqno(5.8)$$
where
$$\ze_+^d=\{x\in\ze^d:x^2\in 2\ze\}\eqno(5.9)$$
$$\ze_-^d=\{x\in\ze^d:x^2\in 2\ze +1\}\,,\eqno(5.10)$$
so that $\Lambda_\C=\Lambda_0(\C)\cup\Lambda_1(\C)$.

A second
construction of an even self-dual
lattice from a doubly-even self-dual code
can be
obtained by first defining
$$\Lambda_2(\C)=\fred\C+{\textstyle{1\over{2\sqrt 2}}}\one+{\sqrt
2}\ze^d_{(-)^{n+1}}\eqno(5.11)$$
$$\Lambda_3(\C)=\fred\C+{\textstyle{1\over{2\sqrt 2}}}\one+{\sqrt
2}\ze^d_{(-)^n}\,,\eqno(5.12)$$
where $d=8n$ ($d$ must be a multiple of 8 as noted above) and setting
$\widetilde\Lambda_\C=\Lambda_0(\C)\cup\Lambda_3(\C)$. This is known as the
twisted
construction. It is easily seen to be even. Thus, it must be integral, and
self-duality will follow if we can show that $\widetilde\Lambda_\C$ is
unimodular,
{\it i.e.} has one point per unit volume. This is clear, since $\C$ is
self-dual.

The classification of doubly-even self-dual codes and even self-dual lattices
of length (dimension) 8, 16 and 24 is known\ref{9,16}. If $\C$ is a code of
length $n$, and $\pi$ is a permutation of the $n$ coordinates of $\C$, then
application of $\pi$ to $\C$ produces a code $\C^\pi$ (which clearly shares all
the same properties as $\C$). $\C$ and $\C^\pi$ are said to be {\it equivalent}
codes. If two codes are not related in this way, they are said to be
{\it inequivalent}. There is one doubly-even self-dual linear binary code up to
equivalence of length 8, 2 of length 16 and 9 of length 24 (and 85 of length
32). The one of length 8 is called the Hamming code, and is denoted by $e_8$
(for a reason given below). At length 16, we can take two copies of $e_8$ to
give $e_8\oplus e_8$. The other code is written as $d_{16}$. Among the 9 length
24 codes there is one which has no codewords of weight 4. This is the Golay
code, $g_{24}$, and so can be characterised by the fact that it is the unique
doubly-even self-dual binary linear code of smallest length containing no
codewords of weight 4. Its symmetry group ({\it i.e.} the group of permutations
$\pi$ leaving it invariant) is one of the sporadic simple groups, the Mathieu
group, $M_{24}$ (see section 5.1 for a brief discussion of the classification
of
the finite simple groups). For the even self-dual lattices, the result is that
there is one in 8 dimensions, 2 in 16 dimensions and 24 in 24 dimensions. The
8-dimensional lattice is the root lattice of $E_8$, written as $E_8$, while the
16-dimensional ones are two copies of $E_8$, {\it i.e.} $E_8\oplus E_8$, and
$D_{16}$, which is the union of the root lattice of $D_{16}$ with one of the
spinor cosets of the dual of the root lattice (the weight lattice). The
24-dimensional even self-dual lattices were classified by Niemeier in 1968.
Each such lattice $\Lambda$ is uniquely determined by its set of minimal
vectors, $\Lambda(2)=\{\lambda\in\Lambda:\lambda^2=2\}$, which form root
systems of the following types:
\vskip6pt
\leftline{$\emptyset$, ${A_1}^{24}$, ${A_2}^{12}$, ${A_3}^8$, ${A_4}^6$,
${A_6}^4$,
${A_8}^3$, ${A_{12}}^2$, $A_{24}$, ${D_4}^6$, ${D_6}^4$, ${D_8}^3$,
${D_{12}}^2$, $D_{24}$, ${E_6}^4$, ${E_8}^3$,}
\leftline{${A_5}^4D_4$, ${A_7}^2{D_5}^2$,
${A_9}^2D_6$, $A_{15}D_9$, $E_8D_{16}$, ${E_7}^2D_{10}$, $E_7A_{17}$,
$E_6D_7A_{11}$,\hfill (5.13)}
\vskip4pt
\noindent
where the lattice corresponding to the empty root system is the Leech lattice,
$\Lambda_{24}$. It was shown by Venkov\ref{17} that $|\Lambda(2)|=24h$, where
$|\Lambda(2)|$ is the number of elements in $\Lambda(2)$ and $h$ is the common
dual Coxeter number of the irreducible components of the corresponding root
system, and that the rank of the root system was either 0 or 24. Since the
algebra must be simply laced, we can then derive the above list of
possibilities. We shall denote the lattices corresponding to the non-empty root
systems simply by the root system itself. The Leech lattice can, similarly to
the Golay code, be characterised as the unique even self-dual lattice of
smallest dimension containing no points of length 2. We also similarly obtain a
sporadic simple group, Conway's group $Co_1$, given by $\Aut(\Lambda_{24})/\{
\pm 1\}$, where $\Aut(\Lambda_{24})$ is the group of automorphisms of the Leech
lattice.

Clearly, we must have $\Lambda_{e_8}=\widetilde\Lambda_{e_8}=E_8$. In 16
dimensions, we have $\Lambda_{e_8\oplus e_8}=E_8\oplus E_8$ and
$\Lambda_{d_{16}}=D_{16}$. The twisted construction interchanges the two
lattices, giving $\widetilde\Lambda_{e_8\oplus e_8}=D_{16}$ and
$\widetilde\Lambda_{d_{16}}=E_8\oplus E_8$.

For the length 24 codes, we look at the points of length 2 in the lattice,
and use the results of Venkov.
We find $|\Lambda_\C(2)|=48+16|\C_4|$, where $|\C_4|$ is the
number of codewords of weight 4,
and
$|\widetilde\Lambda_\C(2)|=8|\C_4|$. This, together with a computation of the
number of orthogonal components into which $\Lambda(2)$ decomposes, is
sufficient to identify the lattice.

The results of the two constructions in 24 dimensions are summarised in figures
1, 2 and 3, where we have the codes on the left, with the values of $|\C_4|$
noted, and the lattices on the right, with straight arrows denoting the
straight construction and wavy arrows the twisted construction.

Since there are 24 lattices and only 9 codes, the two constructions can produce
at most 18 of the lattices, and in fact are found to produce only
12. This is
due to some overlap between the two constructions, which enables us to exhibit
the results in the form of figures 1-3. Note that the two exceptional
structures which we discussed earlier, {\it i.e.} the Golay code and the Leech
lattice, are connected by the twisted construction.
\vskip6pt
\hbox to\hsize{\hfil
\vbox{\openup2\jot
\halign{
\strut\quad\quad\hfil $# $\hfil\qquad\qquad&
\strut\quad\quad\hfil $# $\hfil\quad\quad&
\strut\quad\quad\hfil $# $\hfil\quad\quad\cr
|\C_4|&\hbox{code}&\hbox{lattice}\cr
&&D_{10}{E_7}^2\cr
24&d_{10}{e_7}^2&\cr
&&{D_5}^2{A_7}^2\cr
&{\rm Fig.\,\,1}&\cr
\cr
&&D_{24}\cr
66&d_{24}&\cr
&&{D_{12}}^2\cr
30&{d_{12}}^2&\cr
&&{D_6}^4\cr
12&{d_6}^4&\cr
&&{A_3}^8\cr
&{\rm Fig.\,\,2}&\cr
&&{E_8}^3\cr
42&{e_8}^3&D_{16}E_8\cr
42&d_{16}e_8&\cr
&&{D_8}^3\cr
18&{d_8}^3&\cr
&&{D_4}^6\cr
6&{d_4}^6&\cr
&&{A_1}^{24}\cr
0&g_{24}&\cr
&&\Lambda_{24}\cr
&{\rm Fig.\,\,3}&\cr
}}\hfil}
\vskip6pt
We shall construct in sections 5.2 and 5.3
corresponding straight and twisted constructions of conformal field theories
from lattices. Comparing with the results for codes and lattices, we would
expect these constructions to give all of the self-dual theories for $c=8$ and
16, but not for $c=24$. Also, we would expect to obtain, by analogy with the
Leech lattice and the Golay code, some $c=24$ theory, say $V^\natural$, which
will be the unique theory with smallest value of $c$ and no states of conformal
weight one, and we would expect the automorphism group of this theory to be in
some way connected to one of the sporadic finite simple groups. These
conjectures are examined in section 6, where we look at the results of
the two constructions and also at the Monster group.
\vskip12pt
\leftline{\bf 5.2 Untwisted construction of a conformal field theory}
\nobreak
\vskip3pt
\nobreak
In this section, we discuss the straight or untwisted construction of a
conformal field theory from a lattice, the construction being analogous to the
straight construction (5.1) of a lattice from a binary linear code described
in the previous section. We shall define the space of states and the
corresponding vertex operators, as well as the vacuum and conformal states, and
discuss the hermitian structure of the theory. Finally we shall
consider the concept of self-dual theories. The proofs that this is consistent
as a conformal field theory are given in [11].

We start with a Euclidean lattice $\Lambda$ of dimension $d$, and introduce
orthonormal states $|\lambda\rangle\equiv\Psi_\lambda$, $\lambda\in\Lambda$,
$\langle\lambda|\lambda'\rangle=\delta_{\lambda\lambda'}$, in Dirac's notation,
and oscillators $a_n^j$, $n\in\ze$, $1\leq j\leq d$, satisfying the commutation
relations
$$[a^i_m,a^j_n]=m\delta^{ij}\delta_{m,-n}\,,\eqno(5.14)$$
and ${a_n^j}^\dagger=a^j_{-n}$, $a^j_n|\lambda\rangle=0$, $n>0$,
$p^j|\lambda\rangle=\lambda^j|\lambda\rangle$, where $p^j\equiv a_0^j$. The
space of states $\H(\Lambda)$ is then defined to be generated by the action of
the oscillators $a^j_{-n}$, $n>0$, on the momentum states $|\lambda\rangle$,
$\lambda\in\Lambda$. $\H(\Lambda)$ has a basis consisting of states of the form
$$\psi=\left( \prod_{a=1}^Ma^{j_a}_{-m_a}\right)
|\lambda\rangle\,,\eqno(5.15)$$
where $\lambda\in\Lambda$ and the $m_a$ and $j_a$ are positive integers, $1\leq
j_a\leq d$.

We define the position operator $q$, with $q^\dagger=q$, which is a
$d$-dimensional vector and only appears in the form $e^{i\lambda\cdot q}$, by
$$e^{i\lambda\cdot q}|\mu\rangle=|\lambda+\mu\rangle\,,\eqno(5.16)$$
and define the field
$$X^j(z)=q^j-ip^j\ln z+i\sum_{n\neq 0}{a^j_n\over n}z^{-n}\,,\eqno(5.17)$$
which similarly only appears in an exponential or as a derivative (so that the
arbitrariness in the definition of $\ln z$ is irrelevant). The vertex operator
corresponding to the state (5.15) is then defined to be
$$V(\psi,z)=:\left(\prod_{a=1}^M{i\over{(m_a-1)!}}{d^{m_a}X^{j_a}\over
{dz^{m_a}}}(z)\right)\exp\{i\lambda\cdot X(z)\}:\sigma_\lambda\,,\eqno(5.18)$$
where the normal ordering denoted by the colons indicates that $q^j$ is written
to the left of $p^j$ as well as the creation operators being written to the
left of the annihilation operators, {\it i.e.} if
$$X^j_{{>\atop <}}(z)=i\sum_{n{>\atop <}0}{a^j_n\over n}z^{-n}\,,\eqno(5.19)$$
then
$$:\exp\{i\lambda\cdot X(z)\}:=\exp\{i\lambda\cdot X_<(z)\}e^{i\lambda\cdot
q}z^{\lambda\cdot p}\exp\{i\lambda\cdot X_>(z)\}\,.\eqno(5.20)$$
The operator $\sigma_\lambda$ is a cocycle operator, such that
$$\hat\sigma_\lambda\hat\sigma_\mu=(-1)^{\lambda\cdot\mu}\hat\sigma_\mu
\hat\sigma_\lambda\,,\eqno(5.21)$$
where $\hat\sigma_\lambda=e^{i\lambda\cdot q}\sigma_\lambda$, a property which
is necessary for the bosonic locality relation (2.4).
demonstrated below. The construction and properties of cocycle
operators for both this construction and the twisted construction are discussed
in [11].

The vacuum is the state $|0\rangle$ and the conformal state is take to be the
state
$$\psi_L=\half a_{-1}\cdot a_{-1}|0\rangle\,.\eqno(5.22)$$

We have\ref{11}

\noindent
{\it {\bf Theorem 5.1} If $\Lambda$ is even, $\H(\Lambda)$, with the structure
defined above, forms a conformal field theory with central charge $c=d$ and
hermitian structure given by
$$\overline\phi=(-1)^{L_0}\theta\phi\,,\eqno(5.23)$$
where $\phi$ is a real linear combination of states of the form (5.15) (the
result can clearly be extended to include complex combinations, since we know
that the map $\phi\mapsto\overline\phi$ is antilinear), and $\theta$ is given
by}
$$\theta a_n^j\theta^{-1}=-a_n^j\,\qquad\theta\Psi_\lambda
=\Psi_{-\lambda}\,.\eqno(5.24)$$

Thus, we have a construction of a chiral bosonic conformal field theory with a
hermitian structure from an even lattice. This is analogous to the construction
(5.1) of a lattice from a doubly-even binary linear code. Here, the lattice
plays the role of the code, and the $d$-dimensional Heisenberg algebra plays an
analogous role to the cubic lattice $\ze^d$. The construction provides
justification for the correspondences postulated in table 1 between properties
for lattices and conformal field theories. In particular, the dimension of the
lattice becomes the value of the central charge $c$ in the conformal field
theory, while the length of a point in the lattice is related to the conformal
weight of a state, {\it i.e.} the state (5.15) has conformal weight given by
$$h_\psi={1\over 2}\lambda^2+\sum_{a=1}^Mm_a\,.\eqno(5.25)$$
We have also seen that the bosonic locality
relation holds due to the fact that the lattice is even, and we may write the
partition function for the conformal field theory $\H(\Lambda)$ in terms of the
theta function of the corresponding lattice $\Lambda$ as
$$\chi_{\H(\Lambda)}(\tau)=\Theta_\Lambda(\tau)/\eta(\tau)^d\,,\eqno(5.26)$$
where
$$\eta(\tau)=q^{1/24}\prod_{n=1}^\infty(1-q^n)\,.\eqno(5.27)$$

We consider the two transformations $S$ and $T$ of the parameter $\tau$,
defined by
$$S(\tau)=-1/\tau\,,\qquad T(\tau)=\tau+1\,.\eqno(5.28)$$
These generate the modular group $\Gamma=PSL(2,\ze)$. The lattice is clearly
even if and only if $\Theta_\Lambda(\tau+1)=\Theta_\Lambda(\tau)$. Then
$$\chi_{\H(\Lambda)}(\tau+1)=\chi_{\H(\Lambda)}(\tau)e^{i\pi d/12}\,.\eqno(
5.29)$$
Also
$$\Theta_\Lambda(-1/\tau)=(-i\tau)^{\half d}||\Lambda^\ast||
\Theta_{\Lambda^\ast}(\tau)\eqno(5.30)$$
together with $\eta(-1/\tau)=(-i\tau)^\half\eta(\tau)$ gives
$$\chi_{\H(\Lambda)}(-1/\tau)=||\Lambda^\ast||\chi_{\H(\Lambda^\ast)}(\tau)
\,.\eqno(5.31)$$
Hence, we see that the partition function for the conformal field theory is
invariant under the modular group $\Gamma$, {\it i.e.} under the
transformations of (5.28), if the lattice is not only even but, in addition,
self-dual with dimension a multiple of 24. It is invariant
under the transformation $S$ and changes by a phase under $T$ if we only
require $\Lambda$ to be even and self-dual ({\it i.e.} allow $d$ to be an
arbitrary
multiple of 8).
[Note that the full partition function, including the
anti-holomorphic factor, remains modular invariant since the phases cancel.]

\noindent
{\it {\bf Definition 5.2}}
A conformal field theory is said to be {\it self-dual} if its
character is invariant under the transformation $S$.

[Note that this
definition is
consistent with the identification made in table 1.]
\vskip12pt
\leftline{\bf 5.3 $\ze_2$-twisted construction of a conformal field theory}
\nobreak
\vskip3pt
\nobreak
In this section, we shall define a representation of a sub-conformal field
theory of the lattice conformal field theory $\H(\Lambda)$. This
satisfies the necessary properties, as discussed in section 4, to extend the
sub-conformal field theory to a new theory, which we shall
call $\THX(\Lambda)$, the twisted conformal field theory, provided that
$\Lambda$ is even (necessary for $\H(\Lambda)$ to be a conformal field theory),
that the dimension of $\Lambda$ is a multiple of 8 and that $\sqrt
2\Lambda^\ast$ is even. The requirement that $\sqrt 2\Lambda^\ast$
be even comes from the verification of the final locality relation (4.23),
which involves the lattice in a non-trivial way as described in [11].
Note that this is almost a modular invariance condition, {\it i.e.}
almost requires self-duality of $\Lambda$, but arises from a consideration
of locality on the Riemann sphere alone.
This construction is the analogue of the twisted
construction of a lattice from a binary code described in section 5.1.
Again, the relevant proofs of our results may be found in [11].

Let $\Lambda$ be an even Euclidean lattice of dimension $d$. As noted in
[11], the map $\theta$ defined by (5.24) is an involution (automorphism of
order 2) of the conformal field theory $\H(\Lambda)$. We write
$$\H^\pm(\Lambda)=\{\psi\in\H(\Lambda):\theta\psi=\pm\psi\}\,,\eqno(5.32)$$
so that $\H(\Lambda)=\H^+(\Lambda)\oplus\H^-(\Lambda)$.

\noindent
{\it {\bf Lemma 5.3} $\H^+(\Lambda)$ is a sub-conformal field theory of
$\H(\Lambda)$.}

Hence $\H^+(\Lambda)$ is a conformal field theory, from Proposition 2.16.
It has vacuum $|0\rangle$, conformal state $\psi_L$ and vertex operators those
of
$\H(\Lambda)$ restricted to $\H^+(\Lambda)$ ( which we shall still write as
$V(\psi,z)$ for $\psi\in\H^+(\Lambda)$).
$\H^+(\Lambda)$ consists of
states of the form $|\lambda\rangle+|-\lambda\rangle$ acted on by an even
number of creation operators and $|\lambda\rangle-|-\lambda\rangle$ acted on by
an odd number of creation operators. We have observed that the $d$-dimensional
Heisenberg algebra plays an analogous role to the lattice $\ze^d$ in the
construction of the lattice $\Lambda_\C$ from a binary code $\C$, and we see
that picking out $\H^+(\Lambda)$ from $\H(\Lambda)$ corresponds to selecting
out the coset $\Lambda_0(\C)$ (in which we restrict $\ze^d$ to $\ze^d_+$)
from $\Lambda_\C$.

To obtain the lattice $\widetilde\Lambda_\C$, we then added in the lattice
$\Lambda_3(\C)$ which is obtained by shifting $\Lambda_0(\C)$ by
${1\over{2\sqrt 2}}\one$ if $d$ is an even multiple of 8 and
${1\over{2\sqrt 2}}\one+\sqrt 2e_j$ for any $j$ with $1\leq j\leq d$ if $d$ is
an odd multiple of 8. So, firstly, we suspect that the corresponding
construction we produce here will only make sense if $d$ is a multiple of 8.
Analogous to the shifting by ${1\over{2\sqrt 2}}\one$, we define a Hilbert
space $\H_T(\Lambda)$ created by the action of half-integrally moded
oscillators $c_r^j$, $r\in\ze +\half$, $1\leq j\leq d$, satisfying
$$[c^i_r,c^j_s]=r\delta_{r,-s}\delta^{ij}\eqno(5.33)$$
and ${c_r^j}^\dagger=c_{-r}^j$, on an irreducible representation space
$\X(\Lambda)$ of the gamma matrix algebra
$\Gamma(\Lambda)=\{\pm\gamma_\lambda:\lambda\in\Lambda\}$ with
$$\gamma_\lambda\gamma_\mu=(-1)^{\lambda\cdot\mu}\gamma_\mu\gamma_\lambda=
\epsilon(\lambda,\mu)\gamma_{\lambda+\mu}\,,\qquad{\gamma_\lambda}^2=(-1)^{
\half\lambda^2}\,,
\eqno(5.34)$$
where $\lambda$, $\mu\in\Lambda$ and $\epsilon(\lambda,\mu)=\pm 1$ (see
[11]), with $c^j_r\chi_0=0$ for $r>0$ and
$\chi_0\in\X(\Lambda)$. The introduction of the space $\X(\Lambda)$ is
necessary, since we have no zero-moded oscillators, and so no momentum space
on which to represent the cocycles,
and we introduce the algebra by analogy
with the cocycle operators in the straight
construction.

The operators
$$L_n=\half\sum_{r=-\infty\atop{r\in\ze+\half}}^\infty:c_r\cdot
c_{n-r}:+{d\over{16}}\delta_{n0}
\eqno(5.35)$$
($r,s,\ldots$ will usually denote elements of $\ze+\half$ in this chapter)
satisfy the Virasoro
algebra (2.2) with $c=d$, and these
turn out to be the moments of the operator corresponding to the conformal state
$\psi_L$ when we define the representation.
Thus, the ground state sector $\X(\Lambda)$ has conformal weight
$d/16$, and so we see that if we wish to extend $\H^+(\Lambda)$ by some
subspace of $\H_T(\Lambda)$ to give a new conformal field theory as described
in section 4, then we must have $d$ a multiple of 8 as postulated above, since
the conformal weights must be integral, and the conformal weight of the state
$$\chi=\left(\prod_{a=1}^Mc_{-r_a}^{j_a}\right)\chi_0\,,\eqno(5.36)$$
where $\chi_0\in\X(\Lambda)$, $1\leq j_a\leq d$ and $r_a=m_a+\half$ with
the $m_a$ positive integers, is
$$h_\chi={d\over{16}}+\sum_{a=1}^Mr_a\,.\eqno(5.37)$$
This also tells us that we must consider only those states with $\theta=1$,
where
$$\theta c_r^j\theta^{-1}=-c_r^j\,,\qquad\theta\chi_0=(-1)^{d/8}\chi_0\,,
\eqno(5.38)$$
gives us an extension of $\theta$ defined in (5.24) from $\H(\Lambda)$ to
$\H(\Lambda)\oplus\H_T(\Lambda)$. Define
$$\H_T^\pm(\Lambda)=\{\chi\in\H_T(\Lambda):\theta\chi=\pm\chi\}\,,\eqno(5.39)$$
({\it c.f.} (5.32)). $\H_T^+(\Lambda)$ is the subspace of $\H_T(\Lambda)$
consisting of states with integral conformal weight. It is seen to be analogous
to $\Lambda_3(\C)$, in a similar way to the correspondence which we have noted
between $\H^+(\Lambda)$ and $\Lambda_0(\C)$. So, we would expect to obtain an
analogue of the twisted construction of a lattice from a binary code by
adjoining the twisted sector
$\H_T^+(\Lambda)$ to $\H^+(\Lambda)$. This is the
extension which was discussed in section
4. We saw that to do so we must have $\H_T^+(\Lambda)$ a real, hermitian
representation of $\H^+(\Lambda)$, satisfying an additional locality relation.

Define the field
$$R^j(z)=i\sum_{r=-\infty}^\infty{c^j_r\over r}z^{-r}\,,\eqno(5.40)$$
by analogy with (5.17) (note here we sum over $r\in\ze+\half$). Then,
corresponding to the state $\psi$ given by (5.15), we define, by analogy with
the definition (5.18) of the vertex operators in the straight theory,
$$V_T^0(\psi,z)=:\left(\prod_{a=1}^M{i\over{(m_a-1)!}}{d^{m_a}R^{j_a}\over
{dz^{m_a}}}(z)\right)\exp\{i\lambda\cdot R(z)\}:\gamma_\lambda\,,\eqno(5.41)$$
where we use the obvious normal ordering, {\it i.e.}
$$:\exp\{i\lambda\cdot R(z)\}:=\exp\{i\lambda\cdot R_<(z)\}\exp\{
i\lambda\cdot R_>(z)\}\,,\eqno(5.42)$$
where
$$R^j_{{>\atop <}}(z)=i\sum_{r{>\atop <}0}{c^j_r\over r}z^{-r}\,.\eqno(5.43)$$
Then set
$$V_T(\psi,z)=V_T^0(e^{A(-z)}\psi,z)\,,\eqno(5.44)$$
where
$$
A(z)
=\hhalf\sum_{n,m\geq 0\atop{m+n>0}}\left({-\half\atop m}\right)
\left({-\half\atop n}
\right) {(-z)^{-m-n}\over{m+n}}a_m\cdot a_n-\hhalf a_0\cdot a_0\ln (-4z)\,.
\eqno(5.45)$$

[Note that the operators $L_n$ which we wrote down in (5.35) are actually the
modes of the operator $V_T(\psi_L,z)$, so that these {\it will} be the Virasoro
generators in the twisted sector, and the conformal weights are as stated. The
normal ordered sum in (5.35)
arises from $V_T^0(\psi_L,z)$ as in the untwisted case.
However, $e^{A(-z)}\psi_L=\psi_L+{d\over{16}}z^{-2}|0\rangle$, so we have the
extra term $V_T^0({d\over{16}}z^{-2}|0\rangle,z)={d\over{16}}z^{-2}$, which
accounts for the shift in $L_0$.]

Let $M$ be a symmetric, unitary matrix satisfying
$$M{\gamma_\lambda}^\ast=\gamma_\lambda M\,.\eqno(5.46)$$
Then the main result of [11] is

\noindent
{\it {\bf Theorem 5.4} The operators $V_T(\psi,z)$ define a real, hermitian
representation of the conformal field theory $\H^+(\Lambda)$, with the
conjugation map on the twisted sector $\H_T^+(\Lambda)$ given by
$$\overline\chi_u=(-1)^{L_0}\theta\chi_{Mu^\ast}\,,\eqno(5.47)$$
for
$$\chi_u=\left(\prod_{a=1}^Mc^{j_a}_{-r_a}\right) u\,,\eqno(5.48)$$
the extension to all twisted states following by antilinearity. Further,
if $\sqrt 2\Lambda^\ast$ is even, $\THX(\Lambda)=\H^+(\Lambda)\oplus
\H_T^+(\Lambda)$ is a hermitian conformal field theory, which is self-dual if
$\Lambda$ is self-dual.}

\noindent
[Note that it has been shown in [23] that the condition that $\sqrt
2\Lambda^\ast$ be even is also {\it necessary} for consistency of the
conformal field theory.]

\vskip12pt
\centerline{\bf 6. Results of the constructions
and connections with the Monster}
\vskip 12pt
\leftline{\bf 6.1 The Monster module}
\nobreak
\vskip3pt
\nobreak
The result of the recently completed classification of finite simple
groups\ref{10}
is that there are 16 infinite families of groups of Lie type,
the alternating
groups on $n$ elements for $n\geq 5$ and 26 so-called sporadic simple
groups, which do not fit into any systematic classification.
One of the sporadic groups is the Mathieu group,
$M_{24}$, the symmetry group of the Golay code, as discussed in section 5.1.
Conway's group $Co_0$ is the automorphism group of the Leech lattice. It
involves, as quotients of subgroups, 12 sporadic simple groups including the
Mathieu group. In 1973, Fischer and Griess predicted independently the
existence of what would turn out to be the largest of the sporadic groups, the
Monster, $F_1$, (which turns out to involve 19 of the other sporadic
groups) which would have order $2^{46}3^{20}5^97^711^213^317.19.23.
29.31.41.47.59.71\approx 8.10^{53}$. It was observed by
Griess, Conway and Norton that the smallest non-trivial irreducible
representation of the Monster would have dimension $d_1\geq 196883$. Norton
showed that this representation would have the structure of a real commutative
non-associative algebra, and Griess\ref{8} explicitly constructed such an
algebra of dimension 196883 and verified enough of its symmetries to prove the
existence of the Monster (and also that $d_1=196883$). (Tits subsequently
showed that the Monster is the full automorphism group of the Griess algebra.)
However, the construction of Griess is inelegant. From the point of view which
we
have been pursuing in this work, {\it i.e.} the analogies between codes,
lattices and conformal field theories, we would expect, as was stated towards
the end of
section 5.1, the automorphism group of the conformal field theory
obtained by the
twisted construction from the Leech lattice to be somehow related to a
sporadic simple group, since the Golay code was related to the Mathieu group
and yielded the Leech lattice under the twisted construction, which was related
to Conway's group $Co_1$.

Further
evidence for this point of view is
provided by the theory of modular functions.
The modular group
$$\Gamma=PSL(2,\ze)=SL(2,\ze)/<\pm 1>\eqno(6.1)$$
has an action on the
upper half complex plane $H$ given by
$$g\cdot\tau={a\tau+b\over{c\tau+d}}\,\,\,\,{\rm for}\,\,g=\pm\pmatrix{a&b\cr
c&d}\in\Gamma\,,\,\,\tau\in H\,.\eqno(6.2)$$
Dedekind and, independently, Klein produced a function $j(\tau)$ on $H$
invariant under $\Gamma$. Set $q=e^{2\pi i\tau}$. Then
$$j(\tau)={\Theta_{{E_8}^3}(\tau)\over{\eta(\tau)^{24}}}\,.\eqno(6.3)$$

\noindent
{\it {\bf Lemma 6.1}
The modular functions (the meromorphic modular-invariant functions on $H\cup\{
i\infty\}$) are given by the field of rational functions of $j(\tau)$. Up to an
additive constant, $j(\tau)$ is the unique such function having a simple pole
at $i\infty$ with residue 1 in $q$.}

We see from (5.26) that $j(\tau)$ is the partition function for the conformal
field theory $\H({E_8}^3)$. We showed there that $\chi_{\H(\Lambda)}(\tau)$ was
modular invariant for $\Lambda$ an even self-dual lattice of dimension a
multiple of 24. So, by the above lemma, taking another even self-dual
24-dimensional lattice in place of ${E_8}^3$ in (6.3) will give $j(\tau)$ up to
an additive constant. The constant term in the expansion of the character as a
power series in $q$ is the number of states of conformal weight 1 in the
theory, which is at least 24 since we always have the states
$a_{-1}^j|0\rangle$ for $1\leq j\leq 24$. For the Leech lattice, these are the
only such states, but for other lattices we have the states $|\lambda\rangle$
for $\lambda$ a lattice point of length 2. We set $J(\tau)$ to be $j(\tau)$
with zero constant term, giving
$$\eqalignno{
J(\tau)&=q^{-1}+0+196884q+21493760q^2+\ldots\cr
&\equiv\sum_na_nq^n\,,&(6.4)}$$
where $a_n\geq 0$ for all $n\in\ze$.
It was noticed by McKay that
$$a_1=d_0+d_1\,,\eqno(6.5)$$
where $d_0=1$ can be interpreted as the dimension of the trivial representation
of the Monster. McKay and Thompson\ref{18} soon noticed
$$a_2=d_0+d_1+d_2\,,\eqno(6.6)$$
where $d_2$ is the dimension of the next largest irreducible Monster module,
and similarly for other terms. It was conjectured from this that there exists a
natural {\it infinite-dimensional}
representation of the {\it finite-dimensional} Monster
$$V^\natural=V_{-1}\oplus V_1\oplus V_2\oplus\ldots\,,\eqno(6.7)$$
such that $\dim V_n=a_n$, $n=-1$, 1, $2,\ldots$. From (6.3), it would be
imagined that the natural choice for $V^\natural$ would be the conformal field
theory $\H(\Lambda)$ associated by the straight construction with a
24-dimensional even self-dual lattice $\Lambda$. However, as mentioned above,
none of these provide zero constant term in their characters. Inspired by the
analogies to codes and lattices as we have mentioned, the obvious thing to
consider is the twisted construction $\THX(\Lambda_{24})$ from the Leech
lattice. The weight one states in the twisted theory $\THX(\Lambda)$ are
given in 24 dimensions by $|\lambda\rangle+|-\lambda\rangle$ for
$\lambda\in\Lambda$ a vector of length 2, since the states $a_{-1}^j|0\rangle$
are projected out by taking the $\theta=1$ subspace. (There is no contribution
from the twisted sector, since in 24 dimensions the twisted ground state has
conformal weight ${3\over 2}$, and so the smallest weight of a twisted state is
2 for the states $c_{-\half}^j\chi_0$, $\chi_0\in\X(\Lambda)$, $1\leq j\leq
24$.) So $\THX(\Lambda_{24})$ has no weight one states and the character is
$J(\tau)$ as required, {\it i.e.} it is conjectured that
$\THX(\Lambda_{24})$ provides the natural module $V^\natural$ for the
Monster, with $V_n$ the space of states in the conformal field theory of weight
$n+1$.

This conjecture was proved by Frenkel, Lepowsky and Meurman\ref{6,7}. The basic
idea of their work is to construct an involution $\sigma$ known as a triality
operator of the conformal field theory which extends the natural action of
Aut($\Lambda_{24}$) on the theory to the Monster. What we shall show in
sections 6 and 7 is that this triality can be understood in a more general
context than the specific case considered for the Monster, for which properties
special to this case were used. In the remainder of this section we shall
discuss in more detail how the Monster arises as the automorphism group of
$\THX(\Lambda_{24})$. In section 6.2, we discuss the results of the
straight and twisted constructions of conformal field theories, and in
particular the coincidences between the straight and twisted theories.

$\Aut(\Lambda)$ is the group of automorphisms of the $d$-dimensional
even self-dual lattice
$\Lambda$, {\it i.e.}
$$\Aut(\Lambda)=\{R\in
SO(d):R\lambda\in\Lambda\,\forall\,\lambda\in\Lambda\}\,.
\eqno(6.8)$$
Its centre is $\ze_P\equiv\{\pm 1\}\cong\ze_2$. As already remarked, in the
case of the Leech lattice the automorphism group is $Co_0$, and the
group $\Aut(\Lambda_{24})/\ze_P$ is Conway's group $Co_1$. $Co_1$ is a simple
group of order $2^{21}3^95^47^211.13.23\approx 8.10^{18}$. We have a
representation of $\Aut(\Lambda)$ on the Hilbert space $\H(\Lambda)$ given by
$R\mapsto u_R$ with
$$u_Ra_n^ju_R^{-1}=R_{ij}a_n^i\,,\qquad u_R|\lambda\rangle=|R\lambda\rangle
\,,\eqno(6.9)$$
for $R\in\Aut(\Lambda)$. However, this is not a group of automorphisms of the
conformal field theory because we must consider the cocycle operators.

We have from [19] that if $\gamma_\lambda$ and ${\gamma'}_\lambda$ are
irreducible representations of gamma matrices with the same symmetry factor
($(-1)^{\lambda\cdot\mu}$ in this case) then there exists a unitary
transformation $S$ such that
$$S\gamma_\lambda S^{-1}=v(\lambda){\gamma'}_\lambda\,,\eqno(6.10)$$
where $v(\lambda)=\pm 1$ and
$$\epsilon'(\lambda,\mu)={v(\lambda+\mu)\over{v(\lambda)v(\mu)}}\epsilon
(\lambda,\mu)\,.\eqno(6.11)$$
Take ${\gamma'}_\lambda=\gamma_{
R\lambda}$ for some $R\in\Aut(\Lambda)$. Let
$$\hat C(\Lambda)=\{(R,S):R\in\Aut(\Lambda),\,S\gamma_\lambda S^{-1}=v_{R,S}
(\lambda)\gamma_{R\lambda}\}\,,\eqno(6.12)$$
for some $v_{R,S}(\lambda)=\pm 1$. Note that in the trivial case
${\gamma'}_\lambda=\gamma_\lambda$, $v(\lambda+\mu)=v(\lambda)v(\mu)$ and
$S\in\Gamma(\Lambda)$. Thus, the kernel of the homomorphism $(R,S)\mapsto R$ is
$\Gamma(\Lambda)$, and so we have the exact sequence
$$1\rightarrow\Gamma(\Lambda)\rightarrow\hat C(\Lambda)\rightarrow
\Aut(\Lambda)\rightarrow 1\,.\eqno(6.13)$$
This provides a group of automorphisms of $\H(\Lambda)\oplus\H_T(\Lambda)$
given by
$$\eqalignno{
&u_{R,S}a_n^ju_{R,S}^{-1}=R_{ij}a_n^i\,,\qquad u_{R,S}c_r^ju_{R,S}^{-1}=
R_{ij}c_r^i\,,\cr
&u_{R,S}|\lambda\rangle=v_{R,S}(\lambda)|R\lambda\rangle\,,\qquad
u_{R,S}\chi=S\chi\,.&(6.14)}$$
Since we have $\gamma_\lambda=\gamma_{-\lambda}$, then $\iota_P=(-1,1)\in
\hat C(\Lambda)$ acts trivially on $\THX(\Lambda)$. Thus, we have a group
of automorphisms $C(\Lambda)\cong\hat C(\Lambda)/\hat\ze_P$, where
$\hat\ze_P\equiv\{(\pm 1,1)\}\cong\ze_2$, and a homomorphism $(\pm R,S)\mapsto
R$ of $C(\Lambda)\rightarrow\Aut(\Lambda)/\ze_P$ with kernel $\Gamma(\Lambda)$,
giving us the exact sequence
$$1\rightarrow\Gamma(\Lambda)\rightarrow C(\Lambda)\rightarrow
\Aut(\Lambda)/\ze_P\rightarrow 1\,.\eqno(6.15)$$
In the case of the Leech lattice, we obtain
$$1\rightarrow\Gamma(\Lambda_{24})\rightarrow C(\Lambda_{24})\rightarrow
Co_1\rightarrow 1\,,\eqno(6.16)$$
and $C(\Lambda_{24})$ together with the triality operator (involution) $\sigma$
generates the Monster as a group of automorphisms of $\THX(\Lambda_{24})$.
$C(\Lambda_{24})$ is the centraliser of the involution $\iota=(1,-1)\in\Gamma$
inside the Monster, {\it i.e.} it preserves the ``fermion number'', or in other
words it maps states in the straight sector into the straight sector and
similarly for the twisted sector, while the triality operator mixes the
straight and twisted sectors.

The group $\Gamma(\Lambda)$ is an extra-special 2-group, {\it i.e.}
$|\Gamma(\Lambda)|=2^{d+1}$ with centre $\ze_2$ and $\Gamma(\Lambda)/\ze_2=
\Lambda/2\Lambda\cong{\ze_2}^d$.

On the other hand, it is $\iota$ which has trivial action when $\hat
C(\Lambda)$ acts on $\H(\Lambda)$. So we have a representation of $Co(\Lambda)=
\hat C(\Lambda)/\ze_\iota$, where $\ze_\iota=\{1,\iota\}\cong\ze_2$. The
homomorphism $(R,\pm S)\mapsto R$ leads to the exact sequence
$$1\rightarrow\Lambda/2\Lambda\rightarrow Co(\Lambda)\rightarrow\Aut(\Lambda)
\rightarrow 1\,.\eqno(6.17)$$
\vskip12pt
\leftline{\bf 6.2 Results of the straight
and twisted lattice constructions}
\nobreak
\vskip3pt
\nobreak
In this section, we shall describe the results of the straight and twisted
constructions of a conformal field theory from an even self-dual lattice in 8,
16 and 24 dimensions.

Let us begin with a few standard results which are relevant.

\noindent
{\it {\bf Proposition 6.2}
Let $\psi^a$, $1\leq
a\leq N$, be an orthogonal real basis for the weight one states
of a conformal field theory $\H$,
{\it i.e.} $\overline{\psi^a}=\psi^a$ and $\langle\psi^a|\psi^b
\rangle=k\delta^{ab}$.
Set
$$V(\psi^a,z)=T^a(z)\equiv\sum_nT^a_nz^{-n-1}\,.\eqno(6.18)$$
Then these modes obey the affine Kac-Moody algebra
$$[T_m^a,T_n^b]=if^{abc}T^c_{m+n}+km\delta^{ab}\delta_{m,-n}\,.\eqno(6.19)$$}

\noindent
{\it Proof.}
The operator product expansion (2.52) gives
$$T^a(z)T^b(\zeta)=k\delta^{ab}(z-\zeta)^{-2}+if^{abc}T^c(\zeta)(z-\zeta)^{-1}
+O(1)\,,\eqno(6.20)$$
where $T_0^a\psi^b=if^{abc}\psi^c$, since this must be a state of weight one,
and also $T_1^a\psi^b=\lambda|0\rangle$ similarly, where $\lambda=\langle 0|
T^a_1|\psi^b\rangle$, {\it i.e.} $\lambda^\ast=\langle\psi^b|V_{-1}(\psi^a)|0
\rangle$ by (2.54) if the states $\psi^a$ are quasi-primary. So $\lambda^\ast=
\langle\psi^b|\psi^a\rangle$ by (2.50), and so $\lambda=k\delta^{ab}$. We have
that
the weight one states $\psi^a$ {\it are} quasi-primary , since
$L_1\psi^a=\lambda^a|0\rangle$ say, so $\lambda^a=\langle 0|L_1|\psi^a\rangle$,
{\it i.e.} ${\lambda^a}^\ast=\langle\psi^a|L_{-1}|0\rangle=0$, since
$|0\rangle$ is $su(1,1)$ invariant. The usual
contour manipulation argument then shows that
(6.20) is equivalent to (6.19). $\sqre$

The zero modes define a compact Lie algebra with structure constants $f^{abc}$,
{\it i.e.} we have a continuous group of automorphisms of the conformal field
theory. We shall denote this Lie algebra by $g_\H$.

\noindent
{\it {\bf Proposition 6.3} For $\Lambda$ an even lattice, the affine algebra of
$\H(\Lambda)$ in
Proposition 6.2 is the affinization $\hat g_\Lambda$ of the Lie algebra
$g_\Lambda$ with root system $\Lambda(2)$. In particular,
$g_{\H(\Lambda)}=g_\Lambda$.}

\noindent
{\it Proof.}
The weight one
states are given by $a_{-1}^j|0\rangle$ and $|\lambda\rangle$ for
$\lambda\in\Lambda$ a vector of length 2. The appropriate operator products may
easily be evaluated, for example
$$
V(\lambda,z)V(\mu,\zeta)=e^{i\lambda\cdot X_<(z)}e^{i\lambda\cdot q}z^{\lambda
\cdot p}e^{i\lambda\cdot X_>(z)}\sigma_\lambda
\cdot e^{i\mu\cdot X_<(\zeta)}e^{i\mu\cdot q}\zeta^{\mu\cdot p}
e^{i\mu\cdot X_>(\zeta)}\sigma_\mu\,.\eqno(6.21)$$
The $\sigma_\lambda$ commutes through to the right and we use
$\sigma_\lambda\sigma_\mu=\epsilon(\lambda,\mu)\sigma_{\lambda+\mu}$.
Moving $z^{\lambda\cdot p}$ past $e^{i\mu\cdot q}$ produces a
factor $z^{\lambda\cdot\mu}$, while
$$e^{i\lambda\cdot X_>(z)}e^{i\mu\cdot X_<(\zeta)}=e^{i\mu\cdot X_<(\zeta)}
e^{i\lambda\cdot X_>(z)}e^{[i\lambda\cdot X_>(z),i\mu\cdot X_<(\zeta)]}\,,
\eqno(6.22)$$
where the commutator is given by
$$\eqalignno{[i\lambda\cdot X_>(z),i\mu\cdot X_<(\zeta)]&=-\sum_{n>0}
\lambda\cdot\mu{1\over n}\left({\zeta\over z}\right)^n\cr
&=\lambda\cdot\mu\ln\left( 1-{\zeta\over z}\right)\,,&(6.23)}$$
from (5.19) and (5.14).
Hence we obtain
$$\eqalignno{
V(\lambda,z)V(\mu,\zeta)= (z - \zeta)^{\lambda\cdot\mu}
\epsilon(\lambda,\mu)&e^{i\lambda\cdot X_<(z)+i\mu\cdot X_<(\zeta)
}e^{i(\lambda+\mu)\cdot q}\cr
&\cdot z^{\lambda
\cdot p}\zeta^{\mu\cdot p}e^{i\lambda\cdot X_>(z)+i\mu\cdot X_>(\zeta)
}\sigma_{\lambda+\mu}
\,.&(6.24)}$$
Since $\lambda$, $\mu$ have length 2, $\lambda\cdot\mu=\pm 2$, $\pm 1$, 0. So
we only obtain singular terms in the operator product for $\lambda\cdot\mu=-1$
and $\lambda\cdot\mu=-2$ ({\it i.e.} $\mu=-\lambda$). In the first case,
$$V(\lambda,z)V(\mu,\zeta)={\epsilon(\lambda,\mu)\over{z-\zeta}}V(\lambda+
\mu,\zeta)+O(1)\,,\eqno(6.25)$$
while in the second case
$$V(\lambda,z)V(-\lambda,\zeta)={\epsilon(\lambda,-\lambda)\over{(z-\zeta)^2}}
\left[
1+(z-\zeta)i\lambda\cdot{d\over{d\zeta}}X(\zeta)\right]+O(1)\,,\eqno(6.26)$$
Taylor expanding the right hand side of (6.24) about $z=\zeta$.
The vertex operator for the state $\epsilon\cdot a_{-1}|0\rangle$ is $i\epsilon
\cdot{d\over{d\zeta}}X(\zeta)$, which has modes $a_n$. Similarly to the above,
we obtain
$$i\epsilon
\cdot{d\over{dz}}X(z)V(\lambda,\zeta)={\epsilon\cdot\lambda\over{z-\zeta}}
V(\lambda,\zeta)+O(1)\,.\eqno(6.27)$$
In commutator form, the algebra of the weight one states is, in the gauge in
which $\epsilon(\lambda,-\lambda)=1$,
$$\eqalignno{
&[V_n(\lambda),V_m(\mu)]=\cases{\epsilon(\lambda,\mu)V_{m+n}(\lambda+
\mu)&$\lambda\cdot\mu=-1$\cr
\lambda\cdot a_{n+m}+n\delta_{n,-m}&$\lambda\cdot\mu=-2$\cr
0&$\lambda\cdot\mu\geq 0$}\cr
&[\epsilon\cdot a_n,V_m(\lambda)]=\epsilon\cdot\lambda V_{n+m}(\lambda)\cr
&[\epsilon\cdot a_n,\eta\cdot a_m]=\epsilon\cdot\eta\delta_{n,-m}\,.
&(6.28)}$$
This is recognisable as $\hat g_\Lambda$ as required. $\sqre$

If we denote\ref{20} the space of conformal fields corresponding to states of
conformal
weight at most $n$ by
$W_n$, then we have a map
$$W_n\times W_m\rightarrow W_{n+m-1}\,,\eqno(6.29)$$
given by the singular part of the
operator product expansion.
Thus for $n=m=1$, the operator algebra, for the modes of the vertex operators
given by taking moments with the usual contour manipulation argument,
as we
have seen above, closes. In the case $n=m=2$, we see that it does not close
in general,
but we can define a new product
such that this {\it cross-bracket} algebra does close on $W_2$. We remove
the term corresponding to a state of conformal weight 3 in the operator
product expansion for two states of conformal weight 2 by multiplying by
$(z-\zeta)$ to obtain
$$(z-\zeta)V(\psi,z)V(\phi,\zeta)=\sum_{n=0}^2(z-\zeta)^{n-3}V\left(V(\psi)_{2-n}\phi,\zeta\right)+
O(1)\,.\eqno(6.30a)$$
For the modes of the corresponding vertex operators, this is
$$\psi_m\times\phi_n=[\psi_{m+1},\phi_n]-[\psi_m,\phi_{n+1}]\,,\eqno(6.30b)
$$
{\it i.e.} it is composed of two brackets which ``cross".

In the case of the theory
$\THX(\Lambda_{24})$, there are no weight one states, and so we do not
have the continuous group of automorphisms corresponding to these fields,
nor indeed any continuous automorphisms\ref{7} (see also [27]),
but
only discrete automorphisms which, as we have stated in section 6.1, close to
form the Monster group. We have an algebra on the space of states $V_1$ of
conformal weight 2 ({\it c.f.} the decomposition (6.7)) given by
$\psi\times\phi=V_0(\psi)\phi$, for $\psi$, $\phi\in V_1$, of which
(6.30) is the commutative affinization\ref{7}. $\half\psi_L$ acts
as the identity element. The algebra is commutative and non-associative.
(Commutativity follows from the mode expansion of (2.13), remembering that
there are no states of conformal weight one.) We may call this the Griess
algebra. In fact, it is a slight modification of the algebra originally defined
by Griess\ref{8}, incorporating a natural identity element\ref{7}. Note that
Tits' proof\ref{21} that the Monster is the full automorphism group of the
Griess algebra, together with
the observation that the modes of the vertex operators in $W_2$
generate $V^\natural$,
implies that the Monster is in fact the full
automorphism group of the conformal field theory.

This
completes the necessary review of the standard concepts which are relevant to
this section.

If the rank of $g_\H$, the Lie algebra corresponding to the affine algebra
generated by the weight one fields in a conformal field theory $\H$, is equal
to the value of the central charge $c$ of the theory, then we have $c$ weight
one fields $P^j(z)$ corresponding to states $\psi^j$ ({\it i.e.} $P^j(z)=
V(\psi^j,z)$) corresponding to a Cartan subalgebra of $g_\H$. We choose the
states $\psi^j$ to be real and
orthonormal, and the moments of $P^j(z)$, {\it i.e.}
$$P^j(z)=\sum_na_n^jz^{-n-1}\,,\eqno(6.31)$$
satisfy
$$[a_m^i,a_n^j]=m\delta_{m,-n}\delta^{ij}\,,\eqno(6.32)$$
(which we see immediately from (6.19)).

\noindent
{\it {\bf Proposition 6.4} The
simultaneous eigenvalues of the $p^j\equiv a_0^j$ form an even lattice
$\Lambda$, and if $\dim\Lambda=c$ then $\H\cong\H(\Lambda)$.}

\noindent
{\it Proof.}
We have that the modes $L_n'$ of
$$L'(z)=\half:P(z)\cdot P(z):=\sum_nL_n'z^{-n-2}\,,\eqno(6.33)$$
where the normal ordering is defined as in section 5.2, satisfy the Virasoro
algebra (2.2) with central charge $c$, {\it i.e.}
$$[L_m',L_n']=(m-n)L_{m+n}'+{c\over{12}}m(m^2-1)\delta_{m,-n}\,.
\eqno(6.34)$$
Also, $P^j(z)|0\rangle=e^{zL_{-1}}a_{-1}^j|0\rangle$ from (2.3) implies that
$a_n^j|0\rangle=0$ for $n\geq 0$. This gives
$$L'(z)|0\rangle\rightarrow
L_{-2}'|0\rangle=\half a_{-1}\cdot
a_{-1}|0\rangle\equiv\psi_L'\,,\eqno(6.35)$$
as
$z\rightarrow 0$. Also note $L_n'|0\rangle=0$ for $n\geq -1$ and
${L_n'}^\dagger=L_{-n}'$ (since ${a_n^j}^\dagger=a_{-n}^j$, as the states
corresponding to the $P^j(z)$ were chosen to be real and we use (2.54)).

The weight one states are Virasoro primary states, {\it i.e.} $L_n\phi=0$ for
$n>0$ if $\phi$ is a state of conformal weight one (this is obvious for $n\geq
2$ as the conformal weights are non-negative. For $n=1$, the result follows by
the argument used in Proposition 6.1 to show that the states $\psi_a$
are annihilated by $L_1$). Thus (2.58) holds for all $n\in\ze$, and we may
deduce for $V(\phi,z)=P^j(z)$ by taking modes that
$$[L_m,a_n^j]=-na_{m+n}^j\,.\eqno(6.36)$$
We may thus deduce that
$$[L_{-1},L'(z)]={d\over{dz}}L'(z)\,.\eqno(6.37)$$
(6.37) and (6.34), together with the fact that $L'(z)$ is clearly local with
respect to the vertex operators (since the $P^j(z)$ are), shows that, by the
uniqueness theorem, $L'(z)=V(\psi_L',z)$.

(6.36) also implies that
$$[L_m,L_n']=(m-n)L_{m+n}'+{c\over{12}}m(m^2-1)\delta_{m,-n}\,,
\eqno(6.38)$$
so that, setting $\tilde L_n=L_n-L_n'$, (6.34) and (6.38) imply
$$[\tilde L_m,\tilde L_n]=(m-n)\tilde L_{m+n}\,.\eqno(6.39)$$
Set $\tilde L(z)=\sum_n\tilde L_nz^{-n-2}=L(z)-L'(z)=V(\psi_L-
\psi_L',z)$. But $||\psi_L-
\psi_L'||^2=\langle 0|\tilde L_2\tilde L_{-2}|0\rangle=\langle 0|4\tilde
L_0|0\rangle+\langle 0|\tilde L_{-2}\tilde L_2|0\rangle$ by (6.39), {\it i.e.}
$||\psi_L-
\psi_L'||^2=0$ by $L_n|0\rangle=L_n'|0\rangle=0$ for $n\geq -1$. Thus
$\psi_L=\psi_L'$ and so $L_n'=L_n$.

Since the action of $a_n^j$ on a state decreases the $L_0$ eigenvalue by $n$,
we can deduce from the non-negative spectrum of $L_0$ that the space may be
built up from states $\Psi^k_K$ satisfying
$$p^j\Psi^k_K=K^j\Psi^k_K\,,\quad\langle\Psi^k_K|\Psi^{k'}_K\rangle=
\delta^{kk'}\,,\quad a_n^j\Psi^k_K=0\,\,{\rm for}\,\,n>0\,,\eqno(6.40)$$
by the action of $a_n^j$ for $n<0$. (The states have been decomposed into
simultaneous eigenstates of the commuting hermitian operators $p^j\equiv
a_0^j$.) The $k$ on $\Psi^k_K$ is a degeneracy label, which will be shown below
to be
unnecessary. The space $\H$ decomposes into a direct sum of spaces
$\H^k_K$ for $K\neq 0$ (generated from $\Psi^k_K$ by the action of the
creation operators $a_n^j$) and $\H_0$ generated by the creation operators
from the vacuum (there is no degeneracy label due to  our uniqueness
assumption about the state of conformal weight zero).

{}From the OPE (2.52) (noting that $\Psi^k_K$ has conformal weight $\half K^2$)
$$\eqalignno{
P^j(z)V(\Psi^k_K,\zeta)&=\sum_{n=0}^\infty V(a_{\half K^2-n}^j\Psi^k_K,\zeta)
(z-\zeta)^{n-\half K^2-1}\cr
&=\sum_{n=0}^\infty V(a_{-n}^j\Psi^k_K,\zeta)(z-\zeta)^{n-1}\cr
&=(z-\zeta)^{-1}K^jV(\Psi^k_K,\zeta)+O(1)\,,&(6.41)}$$
so that
$$[p^j,V(\Psi^k_K,\zeta)]=K^jV(\Psi^k_K,\zeta)\,,\eqno(6.42)$$
{\it i.e.} acting with $V(\Psi^k_K,\zeta)$ on a state with $p^j$ eigenvalue
$\lambda^j$ maps it to a state with $p^j$ eigenvalue $\lambda^j+K^j$ (if the
state is non-zero).

Now, since $\Psi^k_K$ is quasi-primary and has weight $\half K^2$,
$V(\overline{\Psi^k_K},z)=z^{-K^2}V(\Psi^k_K,1/z^\ast)^\dagger$. (6.42) then
implies that $V(\overline{\Psi^k_K},z)$ lowers the eigenvalues of $p^j$ by
$K^j$. So $\overline{\Psi^k_K}$, given by the action of
$V(\overline{\Psi^k_K},z)$ on $|0\rangle$, must be a state with $p^j$
eigenvalue $-K^j$. Also, since from Proposition 2.9
it has the same conformal weight as $\Psi^k_K$,
it must be a state of the form $\sum_{k'}a_{k'}
\Psi^{k'}_{-K}$, for some $a_{k'}\in\ce$. We choose the degeneracy labels such
that $\overline{\Psi^k_K}=\Psi^k_{-K}$. (Note that the map $\psi\mapsto
\overline\psi$ is invertible (as it preserves norms, from
Proposition 2.9) and so
there are the same number of degeneracy labels in the $-K$ eigenspace as in the
$K$ eigenspace.)

Now, from the OPE,
$$V(\overline{\Psi^k_K},z)V(\Psi^{k'}_K,\zeta)\sim\lambda(z-\zeta)^{-K^2}
\,,\eqno(6.43)$$
where $\lambda=\delta^{kk'}$ from (2.48).

Consider
$$V(\Psi^j_K,z)V(\Psi^k_{-K},\zeta)|\Psi^l_K\rangle\,.\eqno(6.44)$$
We see from (6.42) that $V(\Psi^k_{-K},\zeta)|\Psi^l_K\rangle\in\H_0$,
and so $V(\Psi^j_K,z)V(\Psi^k_{-K},\zeta)|\Psi^l_K\rangle\in\H^j_K$.
However, (6.43) implies that, as $z\rightarrow\zeta$,
(6.44) goes like $(z-\zeta)^{-K^2}\delta^{jk}\Psi^l_K$. Thus, we see that
there can be only one degeneracy label, and we drop them from now on.

The momenta $K$ form a lattice also, since the 4-point function
$$\langle 0|V(\Psi_{-K},z_1)V(\Psi_{-L},z_2)V(\Psi_L,z_3)
V(\Psi_K,z_4)|0\rangle$$ must be non-zero
(see below for the details),
implying that $V(\Psi_L,z)V(\Psi_K,\zeta)|0\rangle$ is non-zero,
a momentum eigenstate with momentum $K+L$. The requirement of integral
conformal weights fixes this lattice $\Lambda$
to be even, and so we have the desired
Fock space structure. However, we must still verify that we have the
isomorphism $\H\cong\H(\Lambda)$ of conformal field theories. In particular,
we need to consider the cocycle structure.

{}From the additivity of momenta, we see that we must have the OPE
$$V(\Psi_\lambda,z)V(\Psi_\mu,\zeta)=\epsilon(\lambda,\mu)(z-\zeta)^{
\lambda\cdot\mu}
V(\Psi_{\lambda+\mu},
\zeta)+\ldots\,,\eqno(6.45)$$
where $\epsilon(\lambda,\mu)$ is not necessarily non-zero. (We shall show
below that it is in fact of unit modulus.) Consider the 4-point
function
$$\eqalignno{
\langle 0|V(\Psi_{-\mu},z_1)&V(\Psi_{-\lambda},z_2)V(\Psi_\lambda,z_3)
V(\Psi_\mu,z_4)|0\rangle=(z_1-z_4)^{-\mu^2}(z_2-z_3)^{-\lambda^2}\cr
&\cdot (z_1-z_2)^{\lambda\cdot\mu}
(z_1-z_3)^{-\lambda\cdot\mu}
(z_2-z_4)^{-\lambda\cdot\mu}
(z_3-z_4)^{\lambda\cdot\mu}G(x)
\,,
&(6.46)}$$
using $su(1,1)$ invariance, with
$$x\equiv{(z_1-z_2)(z_3-z_4)\over{(z_1-z_4)(z_3-z_2)}}\eqno(6.47)$$
and $G(x)$ having potential poles at $x=0$, $1$ and $\infty$.
We consider the limits $z_1\rightarrow z_4$,
$z_2\rightarrow z_3$; $z_1\rightarrow z_2$, $z_3\rightarrow z_4$;
$z_1\rightarrow z_3$, $z_2\rightarrow z_4$.
The first of these gives, on
using (6.43),
$$\langle 0|V(\Psi_{-\mu},z_1)V(\Psi_{-\lambda},z_2)V(\Psi_\lambda,z_3)
V(\Psi_\mu,z_4)|0\rangle\sim (z_1-z_4)^{-\lambda^2}
(z_2-z_3)^{-\mu^2}\,,\eqno(6.48)$$
and we deduce that $G(x)$ is regular at infinity.

Considering the second limit and using (6.45) gives
$$\eqalignno{
\langle 0|V(\Psi_{-\mu},z_1)&V(\Psi_{-\lambda},z_2)V(\Psi_\lambda,z_3)
V(\Psi_\mu,z_4)|0\rangle\sim\epsilon(-\mu,-\lambda)\epsilon(
\lambda,\mu)\cr
&\cdot
(z_1-z_2)^{\lambda\cdot\mu}(z_3-z_4)^{\lambda\cdot\mu}
\langle 0|V(\Psi_{-\lambda-\mu},z_2)V(\Psi_{\lambda+\mu},z_4)
|0\rangle\,.&(6.49)}$$
But
$$\langle 0|V(\Psi_{-\lambda-\mu},z_2)V(\Psi_{\lambda+\mu},z_4)
|0\rangle=z_2^{-(\lambda+\mu)^2}
\langle\Psi_{\lambda+\mu}|e^{L_1/z_2}e^{z_4L_{-1}}|
\Psi_{\lambda+\mu}\rangle\,,\eqno(6.50)$$
by (2.43) and the creation property (2.3) (noting that the states
$\Psi_{\lambda}$ are (quasi-)primary). Applying (2.29) then gives us
$$\eqalignno{
\langle 0|V(\Psi_{-\mu},z_1)V(\Psi_{-\lambda},z_2)&V(\Psi_\lambda,z_3)
V(\Psi_\mu,z_4)|0\rangle\sim\epsilon(-\mu,-\lambda)\epsilon(
\lambda,\mu)\cr
&\cdot
(z_1-z_2)^{\lambda\cdot\mu}
(z_3-z_4)^{\lambda\cdot\mu}(z_2-z_4)^{-(\lambda+\mu)^2}
\,.&(6.51)}$$
Hence $G(x)$ must be regular at $x=0$.

Similarly, the third limit gives us that $G(x)$ is regular at $x=1$.
Hence, by Liouville's theorem, $G(x)$ must be a constant, in fact
$G(x)\equiv 1$ from (6.48).
So we deduce that
$$\eqalignno{
\langle 0|V(\Psi_{-\mu},z_1)&V(\Psi_{-\lambda},z_2)V(\Psi_\lambda,z_3)
V(\Psi_\mu,z_4)|0\rangle=(z_1-z_4)^{-\mu^2}(z_2-z_3)^{-\lambda^2}\cr
&\cdot (z_1-z_2)^{\lambda\cdot\mu}
(z_1-z_3)^{-\lambda\cdot\mu}
(z_2-z_4)^{-\lambda\cdot\mu}
(z_3-z_4)^{\lambda\cdot\mu}
\,,
&(6.52)}$$
and
$$\epsilon(-\mu,-\lambda)\epsilon(
\lambda,\mu)=1\,.\eqno(6.53)$$
However, the OPE (6.45) together with
the hermitian property (2.43) tell us that $\epsilon(\lambda,\mu)^\ast
=\epsilon(-\mu,-\lambda)$, and so by (6.53) we see that $\epsilon(
\lambda,\mu)$ is of unit modulus. The locality property applied to (6.45)
gives immediately that
$$\epsilon(\lambda,\mu)=(-1)^{\lambda\cdot\mu}\epsilon(\mu,\lambda)\,,
\eqno(6.54)$$
and associativity implies
$$\epsilon(\lambda,\mu)\epsilon(\lambda+\mu,\nu)=\epsilon(\lambda,
\mu+\nu)\epsilon(\mu,\nu)\,.\eqno(6.55)$$
We thus see, from [11], that we may make a ``gauge transformation" of the
cocycles for $\H(\Lambda)$ if necessary
so that the symmetry factors are given by the
above set of $\epsilon(\lambda,\mu)$. Hence, our proposed isomorphism
$u:\H\rightarrow\H(\Lambda)$ is given by
$$ua_n^ju^{-1}=\hat a_n^j\,;\qquad u|\Psi_{\lambda}\rangle
=|\hat\lambda\rangle
\,,\eqno(6.56)$$
using the notation of section 5.2 for $\H(\Lambda)$
(where we use a hatted notation to distinguish the theory
$\H(\Lambda)$).
All that we are now required to show to complete the proof is that
$$uV(\psi,z)u^{-1}=\hat V(u\psi,z)\,,\eqno(6.57)$$
for all $\psi\in\H$.
{}From the definition of the $P^j(z)$, we see that this is true for the
states $a_{-1}^j|0\rangle$. Further, we may act with the modes of the $P^j(z)$
({\it i.e.} the $a_n^j$) on the states $\Psi_\lambda$ to generate
all states in $\H$ by use of the duality relation ({\it c.f.} the
argument given in [11]
to simplify the final locality relation), which
holds also in $\H(\Lambda)$. Hence, we need only verify (6.57)
for $\psi=\Psi_\lambda$, $\lambda\in\Lambda$.
Further, we see that it is only necessary to verify (6.57) for
such states acting on states $|\mu\rangle$, $\mu\in\Lambda$, since
we may act on the left with suitable combinations of the operators
$\hat V(\hat a_{-1}^j|0\rangle,\zeta)$ and move them to the
right by locality
to act on $|\mu\rangle$ and raise it to an arbitrary state in
$\H(\Lambda)$. So, we need only check
$$uV(\Psi_\lambda,z)|\Psi_\mu\rangle=\hat V(|\hat\lambda
\rangle,z)|\hat\mu
\rangle\,,\eqno(6.58)$$
for all $\lambda$, $\mu\in\Lambda$.

Let us verify (6.58) by induction on the modes. From (6.45) we have
$$V(\Psi_\lambda,z)|\Psi_\mu\rangle=z^{\lambda\cdot\mu}
\epsilon(\lambda,\mu)|\Psi_{\lambda+\mu}\rangle+\ldots\,,\eqno(6.59)
$$
{\it i.e.} the first non-zero term in the expansion is
$$V_{-{1\over 2}(\lambda+\mu)^2+{1\over 2}\mu^2}(\Psi_\lambda)|
\Psi_\mu\rangle=\epsilon(\lambda,\mu)|\Psi_{\lambda+\mu}\rangle
\,.\eqno(6.60)$$
Similarly for the right hand side of (6.58). We have chosen the
$\epsilon(\lambda,\mu)$ of $\H(\Lambda)$ such that (6.58) holds for
this mode. Now, from (2.51),
$$[L_{-1},V_n(\Psi_\lambda)]=(1-n-\hhalf\lambda^2)V_{n-1}(
\Psi_\lambda)\,.\eqno(6.61)$$
Hence
$$\eqalignno{
L_{-1}V_n(\Psi_\lambda)|\Psi_\mu\rangle&=(1-n-\hhalf\lambda^2)
V_{n-1}(\Psi_\lambda)|\Psi_\mu\rangle+
V_n(\Psi_\lambda)\mu\cdot a_{-1}|\Psi_\mu\rangle\cr
&=(1-n-\hhalf\lambda^2-\lambda\cdot\mu)V_{n-1}(\Psi_\lambda)|\Psi_\mu\rangle+
\mu\cdot a_{-1}V_n(\Psi_\lambda)|\Psi_\mu\rangle\,,
&(6.62)}$$
using $\left[a_m^j,V_n(\Psi_\lambda)\right]=
\lambda^jV_{m+n}(\Psi_\lambda)$
(which follows from the OPE (6.41)),
and similarly for $V_n(\lambda)|\mu\rangle$. So, if we have
$(1-n-{1\over 2}\lambda^2-\lambda\cdot\mu)\neq 0$ for all relevant $n$ then the
required result follows by induction. Thus, we require the coefficient
to be positive for the first non-zero term, {\it i.e.}
for $n=-{1\over 2}(\lambda+\mu)^2+{1\over 2}\mu^2$, and so
we have the desired conclusion. $\sqre$

In [20] the twist invariant subalgebras for the theories $\H(\Lambda)$ were
evaluated, {\it i.e.} the
algebras generated by the weight one states in
$\H^+(\Lambda)$, which survive into the twisted theory
$\THX(\Lambda)$,
{\it i.e.} the states $|\lambda\rangle+|-\lambda\rangle$ for
$\lambda\in\Lambda(2)$. For dimensions of 24 or greater, the
minimal conformal
weight in the twisted sector is at least 2, and so the algebra
$g_{\THX(
\Lambda)}$
is just the twist invariant subalgebra of $g_{\H(\Lambda)}$. But in
8 and 16 dimensions, the twist invariant subalgebras become ``enhanced''. There
are twisted states with conformal weight one in these dimensions, and
including
the corresponding vertex operators in the operator algebra extends the twist
invariant subalgebra. $\H(E_8)$ as we have seen has $g_{\H(E_8)}=E_8$, and the
twist invariant subalgebra $D_8$ is enhanced to $E_8$. By Proposition
6.4,
we see that $\THX(E_8)\cong\H(E_8)$, in complete analogy with
the result for the constructions of lattices from codes. Similarly, and again
mirroring the results for codes, it is found that $\THX(E_8\oplus
E_8)\cong\H(D_{16})$ and $\THX(D_{16})\cong\H(E_8\oplus E_8)$. The results
for the 24-dimensional even self-dual lattices are shown in table 2. The
lattices $\Lambda$ (except for the Leech lattice $\Lambda_{24}$) are denoted by
the algebra whose root system $\Lambda(2)$ forms,
while the straight and twisted theories $\H(\Lambda)$
and $\THX(\Lambda)$ are labelled by the algebra corresponding to the weight
one fields (except for $\THX(\Lambda_{24})\equiv V^\natural$, the natural
Monster module).

We see that there are 9 coincidences between a twisted theory and a straight
theory (Proposition 6.4 shows that the theories are indeed isomorphic)and the
other 15 twisted theories are distinct, as their algebras are
distinct.
This gives us a total of 39 self-dual bosonic theories with $c=24$. We also
remember that there are 9 doubly-even self-dual binary codes of length 24. By
comparing with the results from section 5.1 given
\noindent
in figures 1, 2 and 3 we see
that these coincidences between the straight and twisted theories are
\vskip10pt
\hbox to\hsize{\hfil
\vbox{\offinterlineskip
\halign{\vrule#&
\strut\quad\hfil $# $\hfil\quad&\vrule#&
\strut\quad\hfil $# $\hfil\quad&\vrule#&
\strut\quad\hfil $# $\hfil\quad&\vrule#\cr
\noalign{\hrule}
height4pt&\omit&&\omit&&\omit&\cr
&\Lambda&&
\H(\Lambda)&&\THX(\Lambda)&\cr
height4pt&\omit&&\omit&&\omit&\cr
\noalign{\hrule}
height4pt&\omit&&\omit&&\omit&\cr
&{E_8}^3&&{E_8}^3&&{D_8}^3&\cr
height4pt&\omit&&\omit&&\omit&\cr
&D_{16}E_8&&D_{16}E_8&&{D_8}^3&\cr
height4pt&\omit&&\omit&&\omit&\cr
&{D_8}^3&&{D_8}^3&&{D_4}^6&\cr
height4pt&\omit&&\omit&&\omit&\cr
&{D_4}^6&&{D_4}^6&&{A_1}^{24}&\cr
height4pt&\omit&&\omit&&\omit&\cr
&{A_1}^{24}&&{A_1}^{24}&&\Lambda_{24}&\cr
height4pt&\omit&&\omit&&\omit&\cr
&\Lambda_{24}&&U(1)^{24}&&V^\natural&\cr
height4pt&\omit&&\omit&&\omit&\cr
&D_{24}&&D_{24}&&{D_{12}}^2&\cr
height4pt&\omit&&\omit&&\omit&\cr
&{D_{12}}^2&&{D_{12}}^2&&{D_6}^4&\cr
height4pt&\omit&&\omit&&\omit&\cr
&{D_6}^4&&{D_6}^4&&{A_3}^8&\cr
height4pt&\omit&&\omit&&\omit&\cr
&{A_3}^8&&{A_3}^8&&{A_1}^{16}&\cr
height4pt&\omit&&\omit&&\omit&\cr
&D_{10}{E_7}^2&&D_{10}{E_7}^2&&{D_5}^2{A_7}^2&\cr
height4pt&\omit&&\omit&&\omit&\cr
&{D_5}^2{A_7}^2&&{D_5}^2{A_7}^2&&{D_4}^2{B_2}^4&\cr
height4pt&\omit&&\omit&&\omit&\cr
&A_{24}&&A_{24}&&B_{12}&\cr
height4pt&\omit&&\omit&&\omit&\cr
&A_{17}E_7&&A_{17}E_7&&D_9A_7&\cr
height4pt&\omit&&\omit&&\omit&\cr
&A_{15}D_9&&A_{15}D_9&&D_8{B_4}^2&\cr
height4pt&\omit&&\omit&&\omit&\cr
&{A_{12}}^2&&{A_{12}}^2&&{B_6}^2&\cr
height4pt&\omit&&\omit&&\omit&\cr
&A_{11}D_7E_6&&A_{11}D_7E_6&&D_6{B_3}^2C_4&\cr
height4pt&\omit&&\omit&&\omit&\cr
&{A_9}^2D_6&&{A_9}^2D_6&&{D_5}^2{A_3}^2&\cr
height4pt&\omit&&\omit&&\omit&\cr
&{A_8}^3&&{A_8}^3&&{B_4}^3&\cr
height4pt&\omit&&\omit&&\omit&\cr
&{A_6}^4&&{A_6}^4&&{B_3}^4&\cr
height4pt&\omit&&\omit&&\omit&\cr
&{A_5}^4D_4&&{A_5}^4D_4&&{A_3}^4{A_1}^4&\cr
height4pt&\omit&&\omit&&\omit&\cr
&{A_4}^6&&{A_4}^6&&{B_2}^6&\cr
height4pt&\omit&&\omit&&\omit&\cr
&{A_2}^{12}&&{A_2}^{12}&&{A_1}^{12}&\cr
height4pt&\omit&&\omit&&\omit&\cr
&{E_6}^4&&{E_6}^4&&{C_4}^4&\cr
height6pt&\omit&&\omit&&\omit&\cr
\noalign{\hrule\vskip8pt}
\multispan{6}{Table 2. Straight and twisted constructions from
even}\hfil\cr
\noalign{\vskip2pt}
\multispan{6}{self-dual lattices in 24 dimensions}
\hfil\cr
}}\hfil}
\vskip6pt
\noindent
such that
the conformal field theory given by the straight construction from the lattice
obtained by the twisted construction applied to a doubly-even self-dual binary
code is
\vskip6pt
\hbox to\hsize{\hfil
\vbox{\openup2\jot
\halign{
\strut\quad\quad\hfil $# $\hfil\qquad\qquad&
\strut\quad\quad\hfil $# $\hfil\quad\quad&
\strut\quad\quad\hfil $# $\hfil\quad\quad&
\strut\quad\quad\hfil $# $\hfil\quad\quad&
\strut\qquad\qquad\hfil $# $\hfil\quad\quad\cr
|\C_4|&\hbox{code}&\hbox{lattice}&\hbox{cft}&\dim g\cr
&&&&\cr
&&&{E_8}^3&744\cr
&&{E_8}^3&D_{16}E_8&744\cr
42&{e_8}^3&D_{16}E_8&&\cr
42&d_{16}e_8&&{D_8}^3&360\cr
&&{D_8}^3&&\cr
18&{d_8}^3&&{D_4}^6&168\cr
&&{D_4}^6&&\cr
6&{d_4}^6&&{A_1}^{24}&72\cr
&&{A_1}^{24}&&\cr
0&g_{24}&&\Lambda_{24}&24\cr
&&\Lambda_{24}&&\cr
&&&V^\natural&0\cr
&&{\rm Fig.\,\,4}&&\cr
&&&&\cr
&&&D_{24}&1128\cr
&&D_{24}&&\cr
66&d_{24}&&{D_{12}}^2&552\cr
&&{D_{12}}^2&&\cr
30&{d_{12}}^2&&{D_6}^4&264\cr
&&{D_6}^4&&\cr
12&{d_6}^4&&{A_3}^8&120\cr
&&{A_3}^8&&\cr
&&&{A_1}^{16}&48\cr
&&{\rm Fig.\,\,5}&&\cr
}}\hfil}
\vskip6pt
\noindent
isomorphic to the theory given by the twisted construction from the
lattice obtained by the straight construction applied to the same code. We may
thus extend figures 1-3 to give figures 4-6, where there is included on the
right hand side the dimension of the appropriate Lie algebra. Again, wavy
arrows denote the twisted construction and straight arrows the
straight
construction. ($\dim g_{\H(\Lambda)}=|\Lambda(2)|+24$ and
$\dim g_{\THX(\Lambda)}=\half |\Lambda(2)|$, in at least 24 dimensions,
so that, from our discussion in section 5.1, $\dim
g_{\H(\Lambda_\C)}=16|\C_4|+72$, $\dim g_{\THX(\Lambda_\C)}=
\dim g_{\H(\widetilde\Lambda_\C)}=8|\C_4|+24$ and $\dim g_{\THX(
\widetilde\Lambda_\C)}=4|\C_4|$ in 24 dimensions).
Also, figure 7 shows the results in 16 dimensions. (Note that figure 7 may be
extended indefinitely by adjoining copies of itself, a property which
is not
shared by the graphs in 24 dimensions, since in that case $\dim g$ decreases as
one descends the graph.)
\vskip6pt
\hbox to \hsize{\hfil
\vbox{\openup2\jot
\halign{
\strut\quad\quad\hfil $# $\hfil\qquad\qquad&
\strut\quad\quad\hfil $# $\hfil\quad\quad&
\strut\quad\quad\hfil $# $\hfil\quad\quad&
\strut\quad\quad\hfil $# $\hfil\quad\quad&
\strut\qquad\qquad\hfil $# $\hfil\quad\quad\cr
|\C_4|&\hbox{code}&\hbox{lattice}&\hbox{cft}&\dim g\cr
&&&&\cr
&&&D_{10}{E_7}^2&456\cr
&&D_{10}{E_7}^2&&\cr
24&d_{10}{e_7}^2&&{D_5}^2{A_7}^2&216\cr
&&{D_5}^2{A_7}^2&&\cr
&&&{D_4}^2{B_2}^4&96\cr
&&{\rm Fig.\,\,6}&&\cr
&&&&\cr
&&&{E_8}^2&496\cr
&&{E_8}^2&&\cr
28&{e_8}^2&&D_{16}&496\cr
&&D_{16}&&\cr
28&d_{16}&&{E_8}^2&496\cr
&&{e_8}^2&&\cr
&&&D_{16}&496\cr
&&{\rm Fig.\,\,7}&&\cr
}}\hfil}
\vskip6pt
We thus see that the connection with codes is more than just an analogy. Codes
can be used to understand the structure and symmetries of conformal field
theories. In section 7, we prove that $\H(\widetilde\Lambda_\C)\cong\THX(
\Lambda_\C)$ for any doubly-even self-dual binary code $\C$ (of any length - a
multiple of 8). Here, we show

\noindent
{\it {\bf Proposition 6.5}
Any coincidence between a twisted theory
and a straight theory must be due to the existence of a doubly-even self-dual
binary code.}

First we need

\noindent
{\it {\bf Lemma 6.6}
Let $\H$ be a self-dual bosonic conformal field theory with
central charge $c=d$. Then $\H$ is isomorphic to $\H(\Lambda_\C)$ for $\C$ some
doubly-even self-dual binary code of length $d$ if and only if $g_\H\supset
su(2)^d$.}

\noindent
{\it Proof.}
The proof in one direction is immediate (see (7.1-3)). Conversely,
if $g_\H\supset
su(2)^d$ then $\rank g_\H=d$.
So, from Proposition 6.4,
we see that $\H\cong\H(\Lambda)$ for some even lattice
$\Lambda$. $\Lambda$ is a weight lattice for $g_\H\supset
su(2)^d$, and so is contained between the root lattice of $su(2)^d$ and its
dual, {\it i.e.}
$$\sqrt 2\ze^d\subset\Lambda\subset{1\over{\sqrt 2}}\ze^d\,.\eqno(6.63)$$
So we can write
$$\Lambda={\C\over{\sqrt 2}}+\sqrt 2\ze^d\eqno(6.64)$$
for some doubly-even linear binary code $\C$ (these properties of $\C$ follow
since $\Lambda$ is an even lattice). Since $\H$ is self-dual,
$\chi_\H(-1/\tau)=\chi_\H(\tau)$, and so we see from (5.31) that $\Lambda$, and
hence $\C$, must be self-dual. Note that this also implies that $d$ must be a
multiple of 8 from the lattice theory discussed in section 5.1. $\sqre$

\noindent
{\it Proof of Proposition 6.5}
If a twisted theory $\THX(\Lambda)$ in $d$-dimensions
coincides with a straight theory $\H(\Lambda')$ then the corresponding algebra
must have rank $d$ (since $\H(\Lambda')$ contains the states
$a_{-1}^j|0\rangle$, $1\leq j\leq d$, for which the corresponding fields
commute). The weight one states in $\THX(\Lambda)$ are of the form
$|\lambda_+\rangle\equiv|\lambda\rangle+|-\lambda\rangle$ for $\lambda\in
\Lambda(2)$, and these satisfy the commutation relations
$$[V_n(\lambda_+),V_m(\mu_+)]=\cases{0&$\lambda\cdot\mu=0$\cr
\epsilon(\lambda,\mp\mu)V_{n+m}((\lambda\mp\mu)_+)&$\lambda\cdot\mu=\pm 1$,}
\eqno(6.65)$$
from (6.28). Hence, since the algebra has rank $d$, we must have $d$
orthogonal vectors in $\Lambda(2)$ (since for $\lambda\cdot\mu=\pm 2$,
$\lambda=\pm\mu$ and the states $\lambda_+$ and $\mu_+$ are not independent).
So, denoting these by
$\sqrt 2e_j$ for $1\leq j\leq d$, we see from (7.1-3) that
$g_{\H(\Lambda)}\supset su(2)^d$, and, since $\H(\Lambda)$ is self-dual due
to the fact that $\THX(\Lambda)$ is, the above result tells us that
$\Lambda=\Lambda_\C$ for some doubly-even self-dual binary code $\C$. But, as
is shown in section 7, $\THX(\Lambda_\C)\cong\H(\widetilde\Lambda_\C)$. So we
deduce also that $\Lambda'$ is equivalent to $\widetilde\Lambda_\C$. Therefore,
all
coincidences are labelled by codes in the manner noted from the results above.
$\sqre$

We also note those twisted theories which are distinct from untwisted theories
must have rank strictly less than $d$, from Proposition 6.4, and this
is consistent with table 2, since all of the 15 new twisted theories have
algebras of rank less than 24.

For $\C$ a doubly-even self-dual binary code, we have the decomposition
$\Lambda_\C=\Lambda_0(\C)\cup\Lambda_1(\C)$,
$\widetilde\Lambda_\C=\Lambda_0(\C)\cup\Lambda_3(\C)$.
So we can divide $\H^\pm(\Lambda_\C)$ and $\H^\pm(\widetilde\Lambda_\C)$ into
two
subspaces each according to whether the momentum is in $\Lambda_0$ or not,
{\it i.e.} we define $\V^\pm_a=\H^\pm(\Lambda_a)$ to be the subspace generated
by the Heisenberg algebra from the states $|\lambda\rangle$ with
$\lambda\in\Lambda_a$ and with $\theta=\pm 1$,  $0\leq a\leq 3$ (note that we
also include the lattice $\Lambda_2$). Similarly we can define twisted spaces
$\T_a^\pm$ (for more details see the next section). Then we have the
decompositions
\vskip6pt
\line{\hfil\hbox{
\vbox{\openup2\jot
\halign{
\strut\hfil $#$\hfil&
\strut\quad\hfil $#$\hfil&
\strut\quad\hfil $#$\hfil&
\strut\quad\hfil $#$\hfil&
\strut\quad\hfil $#$\hfil&
\strut\quad\hfil $#$\hfil&
\strut\quad\hfil $#$\hfil&
\strut\quad\hfil $#$\hfil&
\strut\quad\hfil $#$\hfil\cr
\H(\Lambda_\C)&=&\V^+_0&\oplus&\V^-_0&\oplus&\V^+_1&\oplus&\V^-_1\cr
\H(\widetilde\Lambda_\C)&=&\V^+_0&\oplus&\V^-_0&\oplus&\V^+_3&\oplus&\V^-_3\cr
\THX(\Lambda_\C)&=&\V^+_0&\oplus&\V^+_1&\oplus&\T^+_0&\oplus&\T^+_1\cr
\THX(\widetilde\Lambda_\C)&=&\V^+_0&\oplus&\V^+_3&\oplus&\T^+_0&\oplus&\T^+_3\cr
}}}\hfil(6.66)}
\vskip6pt
The triality structure of FLM is an involution acting on
$\THX(\widetilde\Lambda_\C)$ which mixes the straight and
twisted spaces, {\it e.g.} maps $\V_3^+\leftrightarrow\T_0^+$,
$\V_0^+\rightarrow\V_0^+$, $\T_3^+\rightarrow\T_3^+$. So, we might postulate
that an extension of this operator to the whole structure (6.66) would provide
the isomorphism between $\H(\widetilde\Lambda_\C)$ and $\THX(\Lambda_\C)$ and
also an automorphism of $\H(\Lambda_\C)$. In the next section, we show that
this is true, but give rather a converse argument to this, {\it i.e.} we
observe that there is an obvious triality structure of $\H(\Lambda_\C)$ which,
using the isomorphism $\H(\widetilde\Lambda_\C)\cong\THX(\Lambda_\C)$ which we
prove directly, can be extended to the first three rows of (6.66) and then
finally to $\THX(\widetilde\Lambda_\C)$, providing a simple construction of the
triality structure of FLM, which serves to generate the Monster, and also
generalising this structure beyond the particular case associated with the
Golay code. In section 8, we define further involutions to give a cubic group
of automorphisms. This strategy is summarised in figure 8.
\vskip6pt
\hbox to\hsize{\hfil
\vbox{
\halign{
\strut$# $\hfil\quad\quad&
\strut\quad\quad\hfil $# $\hfil\quad\quad&
\strut\quad\quad $# $\hfil&
\quad\quad\hfil {\sl #} \hfil\cr
&&\H(\Lambda_\C)&evident cubic\cr
&&&symmetry\cr
&\Lambda_\C&&\cr
&&&\cr
\C&&\H(\widetilde\Lambda_\C)\cong\THX(\Lambda_\C)&induced\cr
&&&isomorphism\cr
&\widetilde\Lambda_\C&&\cr
&&&\cr
&&\THX(\widetilde\Lambda_\C)&induced cubic\cr
&&&symmetry\cr
&&&\cr
}}\hfil}
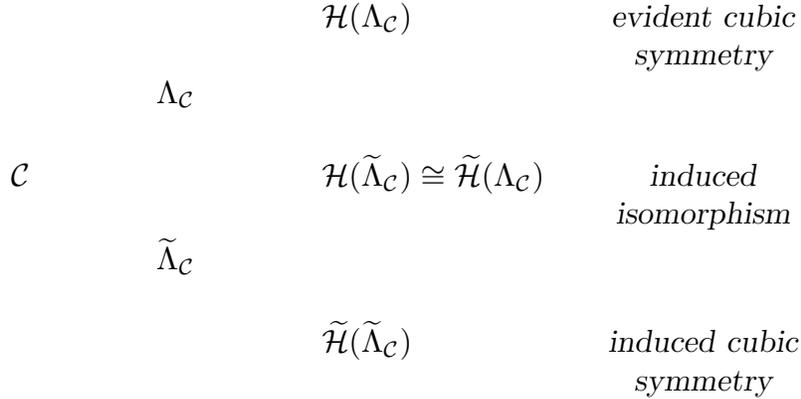
\centerline{Fig. 8}
\vskip6pt
[Note that if we
let $g_0$ be the Lie algebra corresponding to the sub-conformal
field theory $\V_0^+$, and $g_a$ for $a=1$, 2, 3 the algebras corresponding to
$\V_0^-$, $\V_1^+$ and $\V_1^-$ respectively, then we have a sort of
``elaborated symmetric space structure''
$$\eqalignno{
&[g_0,g_0]\subset g_0\qquad [g_0,g_a]\subset g_a\cr
&[g_a,g_a]\subset g_0\qquad [g_a,g_b]\subset g_c\,,&(6.67)}$$
where $(a,b,c)$ is a permutation of (1,2,3). $g=\bigoplus_{a=0}^3g_a$ can be
divided into a symmetric space in three isomorphic ways, {\it i.e.}
$g/g_0\oplus g_a$.]
\vskip12pt
\centerline{\bf 7. Construction of the triality operator}
\nobreak
\vskip3pt
\nobreak
For $\C$ a doubly-even self-dual binary code of length d, we defined the spaces
$\V_a^\pm$ for $a=0$, 1, 2, 3 in the previous section. (Note that we suppress
the $\C$ dependence for ease of notation.) We may also define 8 corresponding
twisted spaces starting from an irreducible representation $\X\equiv\X
(\Lambda_0^\ast)$ of the gamma matrix algebra
$\Gamma\equiv\Gamma(\Lambda_0^\ast)=\{\gamma_\lambda:\lambda\in\Lambda_0^\ast
\}$ (noting that $\Lambda_0^\ast=\Lambda_0\cup\Lambda_1\cup\Lambda_2\cup
\Lambda_3$), which is described in the appendix. $\X$ is of dimension
$2^{1+d/2}$
and splits into four irreducible representations $\X_a$, $0\leq a\leq 3$, of
$\Gamma_0=\Gamma(\Lambda_0)$, with $\X_0\oplus\X_a$ an irreducible
representation of dimension $2^{d/2}$ of $\Gamma(\Lambda_0\cup\Lambda_a)$ for
$a=1$, 2, 3. (Note that the lattice is only even for $a=1$, 3). Define
$\theta=(-1)^{d/8}$on $\X_a$ for $a=0$, 1, 3 and $\theta=-(-1)^{d/8}$ on
$\X_2$. Set $\T_a^\pm=\H_T^\pm(\Lambda_a)$ for $0\leq a\leq 3$, the subspace
with $\theta=\pm 1$ generated by the $c$-oscillators from $\X_a$.

Define the weight one states
$$\zeta_1^j={1\over{\sqrt 2}}a_{-1}^j|0\rangle\in\V_0^-\eqno(7.1)$$
$$\zeta_2^j=-{1\over 2}(|\sqrt 2e_j\rangle
+|-\sqrt 2e_j\rangle)\in\V_1^+\eqno(7.2)$$
$$\zeta_3^j={i\over 2}(|\sqrt 2e_j\rangle
-|-\sqrt 2e_j\rangle)\in\V_1^-\,,\eqno(7.3)$$
$1\leq j\leq d$, where the $e_j$ are the unit vectors in the direction of the
axes (which we defined in section 5.1). Set $J^{ja}(z)=V(\zeta^j_a,z)$, $1\leq
a\leq 3$, $1\leq j\leq d$. Then these currents define an affine $su(2)^d$
algebra. This follows from the relations (6.28), which give
$$[J^{ja}_m,J^{kb}_n]=i\epsilon^{abc}J^{kc}_{m+n}\delta^{jk}+\half
m\delta_{m,-n}\delta^{ab}\delta^{jk}\,,\eqno(7.4)$$
where there is an implicit sum over $c$, $1\leq c\leq 3$. For each $su(2)$,
{\it i.e.} for each j, we can define a rotation in that $su(2)$,
$$\sigma^j=\exp\left\{{i\pi\over{\sqrt 2}}(J^{j1}_0+J^{j2}_0)\right\}\,,
\eqno(7.5)$$
which rotates by $\pi$ about the axis equally inclined to the first and second
axes. Set
$$\sigma=\prod_{j=1}^d\sigma^j\,.\eqno(7.6)$$
Then we have, since the distinct $su(2)$'s commute,
$$\sigma J_m^{j1}\sigma^{-1}=J_m^{j2}\,,\,\,
\sigma J_m^{j2}\sigma^{-1}=J_m^{j1}\,,\,\,
\sigma J_m^{j3}\sigma^{-1}=-J_m^{j3}\,,\eqno(7.7)$$
and $\sigma$ defines an automorphism of $\H(\Lambda_\C)$, provided the cocycles
are chosen appropriately, which has order 2 ($\sigma^2=1$), {\it i.e.}
$\sigma$ is an involution of $\H(\Lambda_\C)$.

When $d$ is an odd multiple of 8, we shall modify the definition of $\sigma$
given above slightly, by redefining for some $l$, $1\leq l\leq d$,
$$\sigma^l=\exp\left\{{3\pi i\over{\sqrt 2}}(J^{l1}_0+J^{l2}_0)\right\}\,,
\eqno(7.8)$$
$\sigma$ still being given by (7.6). This still gives $\sigma^2=1$. [Although
each individual $su(2)$, with generators $J_0^{ja}$ for some $j$, has
half-integral spins on $\H(\Lambda_\C)$, the diagonal group, with generators
$\sum_jJ_0^{ja}$, has only integral spins, due to the way in which the
occurrence of the half-integral spins is correlated by the codewords. The
redefinition (7.8) for $d$ an odd multiple of 8 changes $\sigma$ by a factor of
$-1$ on states with half-integral spin with respect to the $su(2)$ labelled by
$l$
(but leaves it unchanged on states with integral spin with respect to this
$su(2)$).]

\noindent
{\it {\bf Proposition 7.1}
The spaces $\V_a^\pm$ and $\T_a^\pm$ for $0\leq a\leq
3$ are irreducible as representation spaces for $\V_0^+$.}\ref{7}

\noindent
{\it Proof.}
Consider initially the
space $\V_a^P$ for some $a$, $0\leq a\leq 3$, and some parity $P=\pm$. Suppose
that $U$ is an irreducible representation space for $\V_0^+$ contained in
$\V_a^P$.
Set
$$\psi^j=\half a_{-1}^ja_{-1}^j\vac\,,\,\,L^j(z)\equiv V(\psi^j,z)=\half
:P^j(z)P^j(z):\,,\eqno(7.9)$$
where there is no sum over $j$ (we drop the summation convention unless
otherwise stated) and
$$P^j(z)=i{d\over{dz}}X^j(z)\,,\eqno(7.10)$$
from (5.18). Also, for $j\neq k$, set
$$\psi^{jk}=a_{-1}^ja_{-1}^k\vac\,,\,\,L^{jk}(z)\equiv
V(\psi^{jk},z)=P^j(z)P^k(z)\,,\eqno(7.11)$$
(for $j\neq k$, $P^j$ and $P^k$ commute, so no normal ordering is necessary).
Then, since $U$ is irreducible, $L^j(z)$ and $L^{jk}(z)$ map $U\rightarrow U$,
since $\psi^j$, $\psi^{jk}\in\V_0^+$. Denoting the modes of the vertex
operators
for $\psi^j$ and $\psi^{jk}$ similarly, we have $[L_0^j,L_0^k]=0$ and so we may
write $U$ as the direct sum of simultaneous eigenspaces of the $L_0^j$. If
$\phi\in U$ is such a state, say $L_0^j\phi=\nu^j\phi$, $1\leq j\leq d$. Then
$$L_M^{jk}\phi=\sum_na_n^ja^k_{M-n}\phi\eqno(7.12)$$
($j\neq k$) must be a state in $U$, since $L^{jk}(z):U\rightarrow U$.
But
$$L_0^ja_n^ja^k_{M-n}\phi=(\nu^j-n)a_n^ja^k_{M-n}\phi\,.\eqno(7.13)$$
The projection of the state (7.12) onto the simultaneous eigenspaces must be
in $U$, so we see from (7.13) that $a_n^ja^k_{M-n}\phi\in U$, {\it i.e.}
$a_m^ja_n^k$ maps $U\rightarrow U$ for $j\neq k$. Thus, so does
$$[a_m^ja^k_{-M},a_n^ja^k_M]=-Ma_n^ja_m^j+m\delta_{m,-n}a^k_{-M}a^k_M\,,
\eqno(7.14)$$
for $j\neq k$. Hence $a^j_na^j_m$ maps $U\rightarrow U$ for $m\neq -n$. Putting
$m=-n$ in (7.14) we have that
$$ma^k_{-M}a^k_M-Ma^j_{-m}a^j_m\eqno(7.15)$$
maps $U\rightarrow U$ for $j\neq k$. Since $\psi_L\in\V_0^+$, then we can
decompose $U$ into eigenspaces of $L_0:U\rightarrow U$. Applying (7.15) to a
state $\phi\in U$ of conformal weight $h_\phi$, we see that $a^k_M$ must
annihilate $\phi$ for sufficiently large $M$, {\it i.e.} for $M>h_\phi$. Thus
$a^j_{-m}a^j_m$ maps $U\rightarrow U$, and we have that $a^j_ma^k_n$ maps
$U\rightarrow U$ for all $1\leq j$, $k\leq d$ and $m$, $n\in\ze$.

Therefore, since $S^{jk}\equiv a_0^ja_0^k$ maps $U\rightarrow U$ and the
operators $S^{jk}$ commute, we may decompose $U$ into the direct sum of the
simultaneous eigenspaces of these operators. By application of $a_m^ja_n^k$
with $m$, $n\geq 0$ to a state in such an eigenspace, we see that each such
eigenspace contains a state of the form $|\lambda_a\rangle+P|-\lambda_a
\rangle$, where $\lambda_a\in\Lambda_a(\C)$. The vectors $\sqrt 2e_j\pm\sqrt
2e_k$ and $\fred c$ for $1\leq j<k\leq d$, $c\in\C$, generate the lattice
$\Lambda_0(\C)$, and generate from $\lambda_a$ the lattice $\Lambda_a(\C)$. Set
$$\zeta^\pm_{jk}=|\sqrt 2e_j\pm\sqrt 2e_k\rangle+|-(\sqrt 2e_j\pm\sqrt
2e_k)\rangle\,,\qquad\zeta_c=|\fred c\rangle+|-\fred c\rangle\,.\eqno(7.16)$$
Then $V(\zeta^\pm_{jk},z)$ and $V(\zeta_c,z)$ map $U\rightarrow U$ (since the
states are in $\V_0^+$) and, projecting onto the simultaneous eigenspaces of
the operators $S^{jk}$, we see that application of these vertex operators takes
us from the eigenspace containing $|\lambda_a\rangle+P|-\lambda_a
\rangle$ to all eigenspaces $|\lambda\rangle+P|-\lambda
\rangle$ with $\lambda\in\Lambda_a$. Applying $a_m^ja_n^k$ with $m$, $n\leq 0$
generates $\V_a^P$ from these states ({\it e.g.} taking $m=0$, $n<0$ maps us to
a state with an odd number of creation operators and zero mode piece
$|\lambda\rangle-P|-\lambda
\rangle$), {\it i.e.} $U=\V_a^P$, and so the spaces $\V_a^\pm$ are irreducible
as representation spaces for $\V_0^+$.

The argument works in a similar way for the spaces $\T_a^\pm$. For $U$ an
irreducible representation space for $\V_0^+$ contained in $T_a^P$,
$c^j_rc^k_s$ maps $U\rightarrow U$, as above. Then by application of
$c^j_rc^k_s$ with $r$, $s>0$ we see that $U$ contains a state $\chi$ for
$P=P_0$ or a state $c_{-\half}^j\chi$ for $P=-P_0$, where $\chi\in\X_a$ and
$P_0=(-1)^{d/8}$. Acting with the vertex operators $V(\zeta_\lambda,z)$ for
$\zeta_\lambda=|\lambda\rangle+|-\lambda\rangle\in\V_0^+$ in the first case
shows that the set of all such $\chi$ appearing in $U$ must form an irreducible
representation space for the gamma matrix algebra $\Gamma_0$, {\it i.e.} $U$
must contain $\X_a$, and then acting with $c_r^jc_s^k$ for $r$, $s<0$ we
deduce that $U=\T_a^P$. In the second case, acting with $c_{-\half}^kc_\half^j$
shows that $U$ contains all
the states $c^k_{-\half}\chi$, $1\leq k\leq d$.
We may, as above, deduce that $\chi$ ranges over all of $\X_a$ (act with
$V(\zeta_\lambda,z)$ for $\lambda\cdot e_k=0$) and hence that $U=\T_a^P$.
$\sqre$

\noindent
{\it {\bf Proposition 7.2}
$\sigma$ maps $\V_0^+$ to itself. Further, $\sigma:\V_0^-\rightarrow\V_1^+$,
$\V_1^+\rightarrow\V_0^-$ and $\V_1^-
\rightarrow\V_1^-$.}

\noindent
{\it Proof.}
{}From the above argument,
$a_m^ja_n^k$, $V(\zeta_{jk}^\pm,z)$ and $V(\zeta_c,z)$, $1\leq j<k\leq d$, $m$,
$n\in\ze$, $c\in\C$, generate $\V_0^+$ from $\vac$. So, since it is clear that
$\sigma\vac=\vac$, it is only necessary to show that $\sigma$ transforms these
operators into operators which map $\V_0^+$ into itself in order to establish
the result. (7.7) gives
$$\sigma V(\zeta_1^j,z)\sigma^{-1}=V(\zeta_2^j,z)\,,\quad
\sigma V(\zeta_2^j,z)\sigma^{-1}=V(\zeta_1^j,z)\,,\quad
\sigma V(\zeta_3^j,z)\sigma^{-1}=-V(\zeta_3^j,z)\,.\quad\eqno(7.17)$$
Now $V(\zeta_a^j,z)V(\zeta_a^k,\zeta)$ for $a=1$, 2, 3, $1\leq j$, $k\leq d$,
maps $\V_0^+$ to itself, since $\V_0^+$ is defined as a subspace of
$\H(\Lambda_\C)$ by the conditions $\theta=1$ and $\theta_1=1$, where
$$\theta_1=\exp\left\{ i{\pi\over{\sqrt 2}}\one\cdot p\right\}\,,\eqno(7.18)$$
and each of these products commutes with both $\theta$ and $\theta_1$.
($\theta_1=1$ fixes the momentum to lie in $\Lambda_0(\C)$.)
Taking moments of the product with $a=1$ shows that $a_m^ja_n^k$ maps $\V_0^+$
to itself, and from
$$V(\zeta_{jk}^\pm,z)=2V(\zeta_2^j,z)V(\zeta_2^k,z)\mp 2
V(\zeta_3^j,z)V(\zeta_3^k,z)\eqno(7.19)$$
we see that $V(\zeta_{jk}^\pm,z)$ does also. Finally, we must consider
$V(\zeta_c,z)$. Set $\Psi_c=|\fred c\rangle$. Then for $j$ such that $e_j\cdot
c=1$
$$[J_0^{j1},V(\Psi_c,z)]=\half V(\Psi_c,z)\,,\qquad
[J_0^{j1},V(\Psi_{c'},z)]=-\half V(\Psi_{c'},z)\,,\eqno(7.20a)$$
$$[J_0^{j2},V(\Psi_c,z)]={i\over 2}\epsilon V(\Psi_{c'},z)\,,\qquad
[J_0^{j2},V(\Psi_{c'},z)]=-{i\over 2}\epsilon V(\Psi_c,z)\,,\eqno(7.20b)$$
where $c'=c-2e_j$ and $\epsilon=\epsilon(-\sqrt2 e_j,\fred c)=-
\epsilon(\sqrt 2e_j,\fred c')$. Therefore
$$\eqalignno{
\sigma^j\pmatrix{V(\Psi_c,z)\cr V(\Psi_{c'},z)}{\sigma_j}^{-1}&=
\exp\left\{{i\pi\over{2\sqrt 2}}\pmatrix{1&i\epsilon\cr -i\epsilon&-1}
\right\}\pmatrix{V(\Psi_c,z)\cr V(\Psi_{c'},z)}\cr
&={i\over{\sqrt 2}}\pmatrix{1&i\epsilon\cr -i\epsilon&-1}
\pmatrix{V(\Psi_c,z)\cr V(\Psi_{c'},z)}\,.&(7.21)}$$
Set $\Delta(c)=\{c':c'_k=\pm c_k\,,\,1\leq k\leq d\}$, {\it i.e.} the set of
all $d$-tuples which can be obtained from $c$ by application of (7.20) for
various $j$. We see that
$$\sigma V(\Psi_c,z)\sigma^{-1}=2^{-\half |c|}\sum_{c'\in\Delta(c)}
i^{n(c,c')}\eta(c,c')V(\Psi_{c'},z)\,,\eqno(7.22)$$
where $n(c,c')$ is the number of $j$, $1\leq j\leq d$, such that $c'_j\neq c_j$
and $\eta(c,c')$ is given by
$$\left(\prod_{c'_j=-1}\gamma_{\sqrt 2e_j}\right)\gamma_{c\over{\sqrt 2}}=
\eta(c,c')\gamma_{c'\over{\sqrt 2}}\,.\eqno(7.23)$$
Similarly, we find
$$\sigma V(\Psi_{-c},z)\sigma^{-1}=2^{-\half |c|}\sum_{c'\in\Delta(c)}
(-i)^{n(c,c')}\eta(c,c')V(\Psi_{-c'},z)\,.\eqno(7.24)$$
But $\C$ is doubly-even, so that $n(c,c')+n(c,-c')=|c|\in 4\ze$, {\it i.e.}
$(-i)^{n(c,-c')}=i^{n(c,c')}$. From the appendix,
$$\eta(c,c')=(-1)^{n(c,c')}\eta(c,-c')\,,\eqno(7.25)$$
and so (7.22) and (7.24) may be added to give
$$\sigma V(\zeta_c,z)\sigma^{-1}=2^{-\half |c|}\sum_{c'\in\Delta_+(c)}
i^{n(c,c')}\eta(c,c')V(\zeta_{c'},z)\,,\eqno(7.26)$$
where $\Delta_+(c)=\{c'\in\Delta(c):n(c,c')\in 2\ze\}$. When $n(c,c')$ is even,
$\zeta_{c'}\in\V_0^+$, since ${c'\over{\sqrt 2}}={c\over{\sqrt 2}}+\lambda$
where $\lambda\in\sqrt 2\ze_+^d$. Therefore, $V(\zeta_c,z)$ maps to operators
which map $V_0^+$ to itself. Thus, $\sigma:\V_0^+\rightarrow\V_0^+$ as
required.

Since $\V_0^-$, $\V_1^+$ and $\V_1^-$ are irreducible as representation spaces
for $\V_0^+$ and $\sigma$ maps $\V_0^+$ to itself, it is only necessary to
check the transformation of one state in each space to show that
$\sigma:\V_0^-\rightarrow\V_1^+$, $\V_1^+\rightarrow\V_0^-$ and $\V_1^-
\rightarrow\V_1^-$. Since $\sigma\zeta_1^j=\zeta_2^j$,
$\sigma\zeta_2^j=\zeta_1^j$ and $\sigma\zeta_3^j=-\zeta_3^j$, then this result
follows. $\sqre$

Hence, $\sigma$ gives an isomorphism between $\V_0^+\oplus\V_0^-$ and
$\V_0^+\oplus\V_1^+$ (note that these are conformal field theories
($\psi_L\in\V_0^+$)).

As stated in the previous section, we shall now show

\noindent
{\it {\bf Proposition 7.3}
$\H(\widetilde\Lambda_\C)\cong\THX(\Lambda_\C)$, for $\C$ a doubly-even
self-dual binary code.}

\noindent
{\it Proof.}
$\sqrt 2J_0^{j1}=p_0^j$,
$1\leq j\leq d$, provides a Cartan subalgebra for $\H(\widetilde\Lambda_\C)$,
and
we see from (6.65) that $\sqrt 2J_0^{j2}$, $1\leq j\leq d$, provides a Cartan
subalgebra for $\THX(\Lambda_\C)$ ({\it i.e.} both
$g_{\H(\widetilde\Lambda_\C)}$ and $g_{\THX(\Lambda_\C)}$  have rank $d$). The
corresponding lattice of eigenvalues in the first case
on $\V_0^+$ is $\Lambda_0(\C)$, {\it
i.e.} is $d$-dimensional. So, by (7.7), the lattice of eigenvalues of
$\sqrt 2J_0^{j2}$ in $\THX(\Lambda_\C)$ is also $d$-dimensional and
contains $\Lambda_0(\C)$. By Proposition 6.4, $\THX(\Lambda_\C)$
is isomorphic to a theory $\H(\Lambda)$ for some $d$-dimensional even lattice
$\Lambda$. $\Lambda$ contains $\Lambda_0(\C)$, and since it is even it must be
integral, and hence $\Lambda\subset\Lambda_0(\C)^\ast=\Lambda_0(\C)\cup
\Lambda_1(\C)\cup\Lambda_2(\C)\cup\Lambda_3(\C)$. Note that $\Lambda_1(\C)$,
$\Lambda_2(\C)$ and $\Lambda_3(\C)$ are all shifts of $\Lambda_0(\C)$
so that, since $\Lambda\supset\Lambda_0(\C)$, if $\Lambda$ contains any element
of $\Lambda_i(\C)$ then $\Lambda\supset\Lambda_i(\C)$ for $i=1$, 2, 3.
Thus,
there are three possibilities for $\Lambda$ even, {\it i.e.}
$\Lambda=\Lambda_0(\C)$, $\Lambda=\Lambda_0(\C)\cup\Lambda_1(\C)=\Lambda_\C$
and $\Lambda=\Lambda_0(\C)\cup\Lambda_3(\C)=\widetilde\Lambda_\C$.
For $d\geq 24$, these three possibilities can be easily distinguished by
considering their partition functions. $\dim g_{\H(\Lambda_\C)}=16|\C_4|+3d$
while $\dim g_{\H(\widetilde\Lambda_\C)}=\dim g_{\H(\Lambda_0(\C))}=
8|\C_4|+d$ from earlier discussions. Also, since $\Lambda_0(\C)$ is strictly
contained in $\widetilde\Lambda_\C$, the partition functions for
$\H(\widetilde\Lambda_\C)$ and $\H(\Lambda_0(\C))$ are distinct. Hence, all
three
cases have distinct partition functions, and so to complete the proof in $d\geq
24$ it is only necessary to verify that $\THX(\Lambda_\C)$ and
$\H(\widetilde\Lambda_\C)$ have equal partition functions. (Note that since
$\dim g_{\THX(\Lambda_\C)}=8|\C_4|+d$ we can immediately exclude the
possibility $\Lambda=\Lambda_\C$.) Since $\V_0^+\oplus\V_0^-\cong\V_0^+\oplus
\V_1^+$ from the above, then the partition functions for these parts coincide,
and it remains to consider $\V_3^+\oplus\V_3^-$ and $\T_0^+\oplus\T_1^+$. The
corresponding partition functions are
$$ \hhalf 2^{d\over 2} q^{{d\over 16}-1}\prod_{n=1}^\infty
(1-q^n)^d\left(\sum_{m=-\infty}^\infty q^{\half m+m^2}
+(-1)^{d\over 8}\sum_{m=-\infty}^\infty (-q)^{\half m+m^2}\right)
\eqno(7.27)$$
and
$$ \hhalf 2^{d\over 2} q^{{d\over 16}-1}\left(\prod_{r=\half}^\infty
(1-q^r)^{-d} + (-1)^{d\over 8}
\prod_{r=\half}^\infty (1+q^r)^{-d}\right)\,,\eqno(7.28)$$
and these are equal by virtue of the identity
$$\sum_{m=-\infty}^\infty q^{\half m+m^2}= \prod_{n=1}^\infty
(1-q^{2n})\prod_{r=\half}^\infty (1+q^r)\,.\eqno(7.29)$$
Alternatively, we may see the equality immediately since, as we have argued
previously, $\chi_{\THX(\Lambda_\C)}$ and $\chi_{\H(\widetilde\Lambda_\C)}$ are
modular invariant, and so equal up to a constant, from the discussion of
section 6.1. The constant terms coincide, both being equal to $8|\C_4|+d$, the
dimension of the corresponding Lie algebra.
The cases $d=8$ and $d=16$ can be considered separately. There is only one code
to consider in the first case and two in the second, and the results quoted in
section 6.2 show that $\THX(\Lambda_\C)\cong\H(\widetilde\Lambda_\C)$ here
also.
So we have $\THX(\Lambda_\C)\cong\H(\widetilde\Lambda_\C)$ as required.
$\sqre$

\noindent
{\it {\bf Proposition 7.4} $\sigma$ can be extended to an isomorphism
$\THX(\Lambda_\C)\cong\H(\widetilde\Lambda_\C)$.}

\noindent
{\it Proof.}
{}From Proposition 7.3,
we have an isomorphism $\sigma_0:\H(\widetilde\Lambda_\C)\rightarrow
\THX(\Lambda_\C)$ with $\sigma_0J^{j1}(z)\sigma_0^{-1}=J^{j2}(z)$.
Restricted to $\V_0^+\oplus\V_0^-$, $\sigma=\sigma_0u$, where $u$ is an
automorphism of $\V_0^+\oplus\V_0^-$ commuting with $J^{j1}(z)$, $1\leq j\leq
d$, {\it i.e.} $u$ preserves the eigenspaces of the $p^j$. Thus
$u|\lambda\rangle=v(\lambda)|\lambda\rangle$ for $\lambda\in\Lambda_0(\C)$,
where $|v(\lambda)|=1$. From $uV(|\mu\rangle,z)u^{-1}=V(u|\mu\rangle,z)$
applied
to $|\lambda\rangle$, we see that
$$v(\lambda+\mu)=v(\lambda)v(\mu)\eqno(7.30)$$
is required. $u$ can be extended to an automorphism of
$\H(\widetilde\Lambda_\C)$
by choosing $\mu\in\Lambda_3(\C)$ and $v(\mu)$ such that $v(\mu)^2=v(2\mu)$
(since $2\mu\in\Lambda_0(\C)$ this is known). This ensures that (7.30) holds.
Then replacing $\sigma_0$ by $\sigma_0u$ gives an isomorphism
$\H(\widetilde\Lambda_\C)\rightarrow\THX(\Lambda_\C)$ which coincides with
$\sigma$ on $\V_0^+\oplus\V_0^-$. Thus, $\sigma$ can be extended to the whole
of $\H(\widetilde\Lambda_\C)$ as required. $\sqre$

Therefore $\V_3^+\oplus\V_3^-$ and
$\T_0^+\oplus\T_1^+$ are equivalent as representations of the isomorphic
conformal field theories $\V_0^+\oplus\V_0^-$ and $\V_0^+\oplus\V_1^+$. It is
shown in section 8 that $\sigma$ maps $\V_3^+\rightarrow\T_0^+$ and
$\V_3^-\rightarrow\T_1^+$ (in fact, a more generalised form of this result is
demonstrated). ($\sigma$ is unique up to the automorphism $\iota$ of
$\H(\widetilde\Lambda_\C)$ which is 1 on $\V_0^+\oplus\V_0^-$ and $-1$ on
$\V_3^+\oplus\V_3^-$.) $\sigma$ thus gives an isomorphism of $\V_0^+\oplus
\V_3^+$ and $\V_0^+\oplus\T_0^+$. For $\psi\in\V_0^+$, $\phi\in\V_3^+$
$$\sigma^{-1}V(\psi,z)\sigma\phi=V(\sigma\psi,z)\phi\,.\eqno(7.31)$$

In [24] it is shown that any irreducible hermitian real meromorphic
representation
$\cal K$ of $\H(\Lambda)_+$ is equivalent either to $\H(\Lambda)_\pm$ or
$\H_T(\Lambda)_+$, a twisted analogue of Proposition 6.4. The argument
used there proceeds by showing that any representation $U$ of a
conformal field theory $\H$ can be characterised by a state in $\H$;
to do this the expectation value in a state $\chi$ of the representation
of a product of the operators $U(\psi_j,z_j)$, representing states
$\psi_j\in\H$,
is rewritten using (3.8) and locality as the scalar product of a suitable state
of $\H$
on the product of vertex operators of $\H$ acting on one of the $\psi_j$`s.
The essential representation property (3.1) can be translated into
properties of this state. This argument can be reversed to define a
representation
by a suitable state of $\H$ having these properties.
This procedure can be applied to associate a state to any
meromorphic representation of $\H(\Lambda)_+$.
The properties required for this state to
define a representation are such that the state defines also a
representation of $\H(\Lambda)$. This representation is an extension
of the initial representation of $\H(\Lambda)_+$, and must restrict to
this, in particular restricting to a meromorphic representation
of $\H(\Lambda)_+$. The (non-meromorphic) representations of
$\H(\Lambda)$ are easily classified (see {\it e.g.} [25]), and those
that restrict to a meromorphic representation of $\H(\Lambda)_+$ are
simply $\H(\Lambda)$ itself and $\H_T(\Lambda)$ (modulo inequivalent
ground state representations of the twisted cocycles). The possible cases
quoted then follow.
These
cases are clearly distinguished by a simple count of the number of
states of conformal weight one. (Note that the uniqueness of the
twisted representation has been demonstrated previously in the case of
the Leech lattice in [26] using specific features of this model, so
allowing a demonstration of triality for the Monster theory. The
arguments of [24] allow the result to be extended to encompass all
even lattices. This is in
the spirit of this paper of extending results for the Monster to a
broader class of theories by abstracting the general properties required.)
Thus, we deduce that $\T_0^+\oplus\T_3^+$ and $\V_3^+\oplus\T_3^+$ are
equivalent as representations of $\V_0^+\oplus\V_3^+$ and
$\V_0^+\oplus\T_0^+$ respectively, identified by $\sigma$. Since
$\sigma(\T_0^+)=\V_3^+$, then $\T_3^+$ corresponds to $\T_3^+$. So $\sigma$ can
be extended by a map $\rho$ from $\T_0^+\rightarrow\V_3^+$ and $\T_3^+
\rightarrow\T_3^+$ into an automorphism of $\THX(\widetilde\Lambda_\C)$, by the
arguments at the end of section 4. For $\psi\in\V_0^+$, $\phi\in\V_3^+$
$$\rho V(\psi,z)\rho^{-1}\phi=V(\sigma\psi,z)\phi\,,\eqno(7.32)$$
Considering (7.31) and (7.32) together, we deduce that acting on $\T_0^+$
$$\beta V(\psi,z)\beta^{-1}=V(\psi,z)\,,\eqno(7.33)$$
where $\beta=\sigma\rho$. But we know from the above arguments that the action
of $\V_0^+$ on $\T_0^+$ is irreducible, and so it must follow from a Schur's
lemma type argument that $\beta=\kappa 1$ for some $\kappa\in\ce$. From the
arguments at the end of section 4, we see that we must have $\rho=\pm\sigma^{
-1}$ since both $\rho$ and $\sigma$ are compatible with the conjugation and
this fixes the map up to a sign. Therefore, we may choose $\rho=\sigma^{-1}$ on
$\T_0^+$ and rename $\rho:\T_0^+\rightarrow\V_3^+$ as $\sigma$. $\sigma$
squares to 1 on $\V_0^+$, $\V_3^+$ and $\T_0^+$, and so to demonstrate that it
is an involution, we must check that $\sigma^2=1$ on $\T_3^+$.
For $\psi\in\V_3^+$, $\phi\in\T_0^+$
$$\eqalignno{
\sigma^2V(\psi,z)\phi&=\sigma V(\sigma\psi,z)\sigma\phi\cr
&=V(\sigma^2\psi,z)\sigma^2\phi\cr
&=V(\psi,z)\phi\,,&(7.34)}$$
and so, since $\V_0^+$ acts irreducibly on $\T_3^+$, (7.34) is equivalent to
saying that $\sigma^2=1$ on $\T_3^+$. This establishes that $\sigma$ is the
triality operator postulated, {\it i.e.} an automorphism of order 2 of
$\THX(\widetilde\Lambda_\C)$ such that it maps $\V_0^+$ and $\T_3^+$ into
themselves, and interchanges $\V_3^+$ and $\T_0^+$. It is defined up to an
automorphism $\alpha$ (induced by $\iota$ on $\H(\widetilde\Lambda_\C)$) equal
to
1 on $\V_0^+\oplus\T_3^+$ and $-1$ on $\V_3^+\oplus\T_0^+$, with which it
commutes.
\vskip12pt
\centerline{\bf 8. Extension to a cubic group}
\nobreak
\vskip3pt
\nobreak
Renaming the triality operator $\sigma$ constructed in the previous section as
$\sigma_3$, we may also construct, by permuting $\zeta^j_1$, $\zeta^j_2$ and
$\zeta^j_3$, automorphisms $\sigma_1$ and $\sigma_2$ of $\H(\Lambda_\C)$. Then
$1$, $\sigma_a$, $\sigma_1\sigma_2$, $\sigma_1\sigma_3$ $\iota_b$, $1\leq a$,
$b\leq 3$, (where $\iota_1=1$ on $\V_0^+\oplus\V_0^-$ and $-1$ on $\V_1^+\oplus
\V_1^-$ and $\iota_2$ and $\iota_3$ are defined by cyclically permuting
$\V_0^-$, $\V_1^+$ and $\V_1^-$) generate a group isomorphic to the symmetry
group of the cube, $S_4$. For $(a,b,c)$ a permutation of $(1,2,3)$,
$$\sigma_a\sigma_b\sigma_a=\iota_c\sigma_c\,,\quad\iota_a\iota_b=\iota_c\,,\quad
\sigma_a\iota_a=\iota_a\sigma_a\,,\quad\sigma_a\iota_b=\iota_c\sigma_a\,.
\eqno(8.1)$$
There exist subgroups isomorphic to $S_3$, {\it e.g.} $\{1,\sigma_1,\sigma_2
\iota_2,\sigma_3,\sigma_1\sigma_2\iota_2,\sigma_1\sigma_3\}$.

We therefore extend the spaces which were considered in the previous section to
give the following ``magic square'' diagram:
\vskip6pt
\line{\hfil\hbox{
\vbox{\openup2\jot
\halign{
\strut $#$\hfil\qquad&
\strut\quad\hfil $#$\hfil&
\strut\quad\hfil $#$\hfil&
\strut\quad\hfil $#$\hfil&
\strut\quad\hfil $#$\hfil&
\strut\quad\hfil $#$\hfil&
\strut\quad\hfil $#$\hfil&
\strut\quad\hfil $#$\hfil&
\strut\quad\hfil $#$\hfil&
\strut\quad\hfil $#$\hfil\cr
&&&&&(1)&&(2)&&(3)\cr
&&&&&\H(\Lambda_\C)&&\THX(\widetilde\Lambda_\C)&&\THX'(\widetilde\Lambda_\C)\cr
&&&&&\|&&\|&&\|\cr
&&&&&\V^+_0&&\V^+_0&&\V^+_0\cr
&&&&&\oplus&&\oplus&&\oplus\cr
(1)&\H(\widetilde\Lambda_\C)&=&\V^+_0&\oplus&\V^-_0&\oplus&\V^+_3&\oplus&\V^-_3\cr
&&&&&\oplus&&\oplus&&\oplus\cr
(2)&\THX(\Lambda_\C)&=&\V^+_0&\oplus&\V^+_1&\oplus&\T^+_0&\oplus&\T^+_1\cr
&&&&&\oplus&&\oplus&&\oplus\cr
(3)&\THX'(\Lambda_\C)&=&\V^+_0&\oplus&\V^-_1&\oplus&\T^+_3&\oplus&\T^-_2\cr
}}}\hfil(8.2)}
\vskip6pt
Let $\V_{ab}$ be the space in row $a$ and column $b$ and set $\V_{ab}=\V_0^+$
if $a\cdot b=0$, for $0\leq a,b\leq 3$. For $(a,b,c)$ a permutation of
$(1,2,3)$, $\sigma_a:\V_{01}\rightarrow\V_{01}$, $\V_{a1}\rightarrow\V_{a1}$,
$\V_{b1}\leftrightarrow\V_{c1}$, {\it i.e.} it defines an automorphism of
$\V_{a0}\oplus\V_{a1}$ and an isomorphism $\V_{b0}\oplus\V_{b1}\rightarrow
\V_{c0}\oplus\V_{c1}$. Define $\H_a$ to be the space given by row $a$ and
$\H^b$ to be the space given by column $b$. Then $\sigma_a$ induces an
automorphism of $\H_a$ and an isomorphism $\H_b\rightarrow\H_c$. The
generalisation of the result that $\sigma:\V_3^+\rightarrow\T_0^+$ which was
required in the last section is that $\sigma_a$ preserves the columns, in the
sense that $\sigma_a:\V_{a\alpha}\rightarrow\V_{a\alpha}$ and
$\sigma_a:\V_{b\alpha}\leftrightarrow\V_{c\alpha}$ for $\alpha=1,2$. This
induces an automorphism of the space $\H^2=\THX(\widetilde\Lambda_\C)$. In a
similar way, an automorphism of the third column is induced. Note that
$\sigma_a$, as an automorphism of the second column, is defined up to the
involution $\iota_a$, which is 1 on $\V_{02}\oplus\V_{a2}$ and $-1$ on
$\V_{b2}\oplus\V_{c2}$. Thus $\iota_a$ commutes with $\sigma_a$. Either
$\sigma_1\sigma_3\sigma_1$ is equal to $\sigma_2$ or it is equal to $\iota_2
\sigma_2$, depending on the sign choices made in the definitions. Whichever
choice is made though, $\sigma_1^2=\sigma_3^2=(\sigma_1\sigma_3\sigma_1)^2=1$,
so that $\{1,\sigma_1,\sigma_3,\sigma_1\sigma_3\sigma_1,\sigma_1\sigma_3,
\sigma_3\sigma_1\}$ forms a group isomorphic to $S_3$. Thus, we obtain a
triality group of automorphisms of $\THX(\widetilde\Lambda_\C)$, explaining the
origin of the term triality operator which has been used so far. In the case
$\C=g_{24}$, the Golay code, the triality group, or just $\sigma_3$, then, as
explained in section 6.1, together with the extension of Conway's group defined
there generates the Monster.

Note that if we reverse the situation of the previous section, and use the
modified definition (7.8) if and only if $d$ is an {\it even} multiple of 8,
then the induced maps, say $\tilde\sigma_a$, interchange columns 2 and 3 rather
than preserve them, {\it i.e.} the $\tilde\sigma_a$ induce isomorphisms
$\H^2\rightarrow\H^3$.

Finally in this section, we verify that $\sigma_a$ does in fact preserve the
columns as stated. In other words, we wish to show

\noindent
{\it {\bf Proposition 8.1}
For $\alpha=1,2$, and
$(a,b,c)$ a permutation of $(1,2,3)$,}
$$\sigma_a:\V_{b\alpha}\rightarrow\V_{c\alpha}\,,\qquad
\sigma_a:\V_{a\alpha}\rightarrow\V_{a\alpha}\,.\eqno(8.3)$$

\noindent
{\it Proof.}
{}From the arguments of the previous section, we see that for $\sigma_3$ (and
analogously for all the $\sigma_a$) that it either preserves or interchanges
the columns, {\it i.e.} it satisfies either (8.3) or
$$\sigma_a:\V_{b\alpha}\rightarrow\V_{c\beta}\,,\qquad
\sigma_a:\V_{a\alpha}\rightarrow\V_{a\beta}\,,\eqno(8.4)$$
where $\alpha=2$ and $\beta=3$ or vice versa. Since, on $\H(\Lambda_\C)$ we
have $\sigma_a\sigma_b\sigma_a=\iota_a\sigma_c$ then we see that on
$\V_{d2}\oplus\V_{d3}$, for $1\leq d\leq 3$, we must have $\sigma_a
\sigma_b\sigma_a=\pm\sigma_c$, thus showing that if either one of $\sigma_b$ or
$\sigma_c$ preserves the columns, then both do, since with two applications of
$\sigma_a$ whether it preserves or interchanges the columns is of no
consequence. Thus, we can deduce that $\sigma_1$, $\sigma_2$ and $\sigma_3$
preserve the columns if we can show that any one of them does. $\sigma_1$ is
the simplest of the three to consider.

By cyclic permutation of the corresponding property for $\sigma\equiv
\sigma_3$ from section 7, we have the relations
$$\sigma_1\zeta_1^j=-\zeta_1^j\,,\qquad\sigma_1\zeta_2^j=\zeta_3^j\,,\qquad
\sigma_1\zeta_3^j=\zeta_2^j\,,\eqno(8.5)$$
or
$$\sigma_1 a_n^j\sigma_1^{-1}=-a_n^j\,,\qquad
\sigma_1|\sqrt 2e_j\rangle=i|-\sqrt 2e_j\rangle\,,\qquad
\sigma_1|-\sqrt 2e_j\rangle=-i|\sqrt 2e_j\rangle\,.\eqno(8.6)$$
Its action on $V(\Psi_c,z)$ is given by an argument similar to that given in
section 7 in showing that $\sigma$ mapped $\V_0^+$ to itself, {\it i.e.} for
$e_j\cdot c=1$, we use
$$[J_0^{j3},V(\Psi_c,z)]=-\half\epsilon V(\Psi_{c'},z)\,,\qquad
[J_0^{j3},V(\Psi_{c'},z)]=-\half\epsilon V(\Psi_c,z)\,\eqno(8.7)$$
together with (7.20) to give
$$\eqalignno{
\sigma^j_1\pmatrix{V(\Psi_c,z)\cr V(\Psi_{c'},z)}&=
\exp\left\{{i\pi\over{2\sqrt 2}}\pmatrix{0&(i-1)\epsilon\cr -(i+1)\epsilon&0}
\right\}\pmatrix{V(\Psi_c,z)\cr V(\Psi_{c'},z)}\cr
&={i\over{\sqrt 2}}\pmatrix{0&-(1+i)\epsilon\cr (1-i)\epsilon&0}
\pmatrix{V(\Psi_c,z)\cr V(\Psi_{c'},z)}\,.&(8.8)}$$
where
$$\sigma_1^j=\exp\left\{{i\pi\over{\sqrt 2}}(J^{j2}_0+J^{j3}_0)\right\}\,,
\eqno(8.9)$$
({\it c.f.} (7.5)). So it follows by successive application of (8.8) that
$$\sigma_1V(\Psi_c,z)\sigma_1^{-1}=(-1)^{{1\over
4}|c|}V(\Psi_{-c},z)\,.\eqno(8.10)$$
Together with (8.6), this gives
$$\sigma_1=\theta\exp\{i\pi w\cdot p\}\,,\eqno(8.11)$$
where $w={1\over{2\sqrt 2}}\one$, which is in $\Lambda_3(\C)$ for d a multiple
of 16. This defines $\sigma_1$ on $\H(\Lambda_\C)$, although it can also be
used to define $\sigma_1$ on $\V_3^\pm$, giving an automorphism of $\H_1=
\H(\widetilde\Lambda_\C)$ which preserves $\V_0^\pm$, $\V_3^\pm$. On the
twisted
spaces $\T_0^+$, $\T_1^+$, $\T_2^-$ and $\T_3^+$ $\sigma_1$ may be defined by
the analogous relation
$$\sigma_1=\theta\gamma_w\,.\eqno(8.12)$$
This provides an isomorphism $\H_2\rightarrow\H_3$ which preserves the columns,
and the required result follows. When $d$ is an odd multiple of 8, if we use
the same definition of $\sigma_1$, then we have that $w\in\Lambda_2(\C)$, and
$\sigma_1$ (and hence $\sigma_2$ and $\sigma_3$) interchanges the columns, as
noted above.
Otherwise, we redefine $\sigma_1^l$ for some $l$ by analogy with (7.8), which
serves to modify $w$ to ${1\over{2\sqrt 2}}\one+\sqrt 2e_l$, which is once more
an element of $\Lambda_3(\C)$, and hence $\sigma_1$ preserves the columns.
(Making this modification when $d$ is a multiple of 16 gives a $\sigma_1$ which
interchanges the columns.) $\sqre$
\vskip12pt
\centerline{\bf 9. Conclusions}
\nobreak
\vskip3pt
\nobreak
The main result of this paper is the demonstration that the remarkable
results of Frenkel, Lepowsky and Meurman on the construction of the
natural representation of the Monster group as a conformal field
theory generalise to a wider class of theories. This generalization
exhibits the features which lead to the existence of the ``triality''
structure more clearly, and specific features of particular models are
not required.

Following in this spirit, the discussion of the structure and
representations of chiral bosonic meromorphic conformal field theories
and the construction of orbifolds illustrates a general progam and
approach which it is hoped to take further.

The nature of what were previously thought of as merely useful
analogies of conformal field theory with the theories of lattices and
codes has been extended to deeper links between their structures, and
it will prove interesting in the future to extend the depth and, more
importantly, the
understanding of such connections.
\vskip12pt
\centerline{\bf Acknowledgement}
\nobreak
\vskip3pt
\nobreak
Paul Montague is grateful to the S.E.R.C. for a Research Studentship,
and to Gonville and Caius College for a Research Fellowship. The
authors are also grateful to Klaus Lucke for useful comments.
\vskip12pt
\centerline{\bf Appendix}
\nobreak
\vskip3pt
\nobreak
In this appendix we wish to define the gamma matrix algebra $\Gamma\equiv
\Gamma(\Lambda_0^\ast)=\{\pm\gamma_\lambda:\lambda\in\Lambda_0^\ast\}$ required
for the extension of the triality structure in section 7. In order to do this,
we must specify the symmetry factor $S(\lambda,\mu)$\ref{22} in
$$\gamma_\lambda\gamma_\mu=S(\lambda,\mu)\gamma_\mu\gamma_\lambda\,,
\eqno(\h{A}.1)$$
such that
$$S(\lambda,\nu)S(\mu,\nu)=S(\lambda+\mu,\nu)\,,\quad
S(\lambda,\mu)=1/S(\mu,\lambda)\,,\quad S(\lambda,\lambda)=1\,,
\eqno(\h{A}.2)$$
for $\lambda$, $\mu$, $\nu\in\Lambda_0^\ast$. The definition of [11]
requires modification, since the lattice is not even in this case. Following
[11], $S(\lambda,\mu)=(-1)^{\lambda\cdot\mu}$ for $\lambda$, $\mu\in
\Lambda_\C=\Lambda_0\cup\Lambda_1$ or $\widetilde\Lambda_\C=\Lambda_0\cup
\Lambda_3$. Making the choices $S(\sqrt 2e_j,{1\over{2\sqrt 2}}\one)=i=-
S({1\over{2\sqrt 2}}\one,\sqrt 2e_j)$ for $1\leq j\leq d$ gives, from (A.2),
$S(\lambda,\mu)$ as follows:
\vskip6pt
\line{\hfil\hbox{
\vbox{\openup2\jot
\halign{
\strut\quad\hfil $#$\hfil\quad&
\strut\quad\hfil $#$\hfil\quad&
\strut\quad\hfil $#$\hfil\quad&
\strut\quad\hfil $#$\hfil\quad&
\strut\quad\hfil $#$\hfil\quad\cr
&\mu\in\Lambda_0&\mu\in\Lambda_1&\mu\in\Lambda_2&\mu\in\Lambda_3\cr
\lambda\in\Lambda_0&(-1)^{\lambda\dt\mu}&(-1)^{\lambda\dt\mu}&
(-1)^{\lambda\dt\mu}&(-1)^{\lambda\dt\mu}\cr
\lambda\in\Lambda_1&(-1)^{\lambda\dt\mu}&(-1)^{\lambda\dt\mu}&
-e^{i\lambda\dt\mu\pi}&-e^{i\lambda\dt\mu\pi}\cr
\lambda\in\Lambda_2&(-1)^{\lambda\dt\mu}&e^{i\lambda\dt\mu\pi}&
(-1)^{\lambda\dt\mu+1}&-e^{i\lambda\dt\mu\pi}\cr
\lambda\in\Lambda_3&(-1)^{\lambda\dt\mu}&e^{i\lambda\dt\mu\pi}&
e^{i\lambda\dt\mu\pi}&(-1)^{\lambda\dt\mu}\cr
}}}\hfil(\h{A}.3)}
\vskip6pt
If $\X$ is an irreducible representation of $\Gamma$, then we consider the
division of $\X$ into irreducible representations of $\Gamma(\Lambda_\C)$.
Since $\Lambda_\C$ is an even self-dual lattice, we see from appendix C of [11]
that
such a representation $\X(\Lambda_\C)\subset\Gamma$ is of dimension $2^{\half
d}$. Then $\X'(\Lambda_\C)=\gamma_w\X(\Lambda_\C)$ for $w={1\over{2\sqrt
2}}\one$ is also an irreducible representation space for $\X(\Lambda_\C)$.
Noting that $\gamma_w\X'(\Lambda_\C)=\X(\Lambda_\C)$ (since $2w\in\Lambda_\C$
as $\one\in\C$ for $\C$ doubly-even) and that $\gamma_w$ and
$\Gamma(\Lambda_\C)$ generate $\Gamma$, gives $\X=\X(\Lambda_\C)\oplus
\X'(\Lambda_\C)$. In other words, $\X$ is of dimension $2^{\half d+1}$. Also,
from appendix C of [11],
an irreducible representation of $\Gamma(\Lambda_0)$ has
dimension $2^{\half d-1}$, so similarly $\X(\Lambda_\C)=\X_0(\Lambda_0)
\oplus\X_1(\Lambda_0)$, where $\X_0(\Lambda_0)$ and $\X_1(\Lambda_0)$ are
irreducible representations of $\Gamma(\Lambda_0)$. Similarly, $\X'(\Lambda_\C)
=\X_2(\Lambda_0)\oplus\X_3(\Lambda_0)$. Also, we have a decomposition of $\X$
into irreducible representations of $\X(\widetilde\Lambda_\C)$, which must be
different sums of the $\X_j(\Lambda_0)$ (because $\Gamma(\Lambda_\C)$ and
$\Gamma(\widetilde\Lambda_\C)$ together generate $\Gamma$). Thus, we take
$\X_0\oplus\X_3$ and $\X_1\oplus\X_2$ (dropping the explicit $\Lambda_0$
dependence) to be such sums.

The gauge choice $\gamma_\lambda=\gamma_{-\lambda}$ made in [11] cannot
be extended to the whole of $\Gamma$. We have $\gamma_\lambda\gamma_{-\lambda}=
\epsilon(\lambda,-\lambda)\gamma_0=\epsilon(\lambda,-\lambda)1$. So
$\gamma_\lambda=\gamma_{-\lambda}$ implies that ${\gamma_\lambda}^2=\pm 1$, and
so is a central element of $\Gamma$. From the table (A.3) we can see that this
can only be true for $\lambda\in\Lambda_0$. With the definition of $\theta$
given in section 7 however, {\it i.e.} inserting an additional $-1$ on $\X_2$,
then $\theta\gamma_\lambda\theta\gamma_\lambda$ is central, and so, by
irreducibility and Schur's lemma, must be a multiple of the identity. Since
$\gamma_{-\lambda}=\pm{\gamma_\lambda}^{-1}$, from the above, then
$\gamma_\lambda=\pm\theta\gamma_{-\lambda}\theta^{-1}$ ($\theta=
\theta^{-1}$ as $\theta$ is an involution). By a choice of gauge, we take
$$\theta\gamma_{-\lambda}\theta^{-1} = \gamma_\lambda\,.\eqno(\h{A}.4)$$
This gives the involution
$$\theta V(\psi,z)\theta^{-1} = V(\theta\psi,z)\,,\eqno(\h{A}.5)$$
for $\psi\in\H(\Lambda_a)$, $a=0$, 1, 3, acting on the full twisted space
$\T_0^+\oplus\T_1^+\oplus\T_2^+\oplus\T_3^+$, explaining the reasoning behind
the apparently unnatural definition of $\theta$. This still gives
$\gamma_\lambda=\gamma_{-\lambda}$ restricted to $\X_0\oplus\X_j$ for $j=1$, 3
and $\lambda\in\Lambda_0\cup\Lambda_j$, as we required for such self-dual
lattices in appendix B of [11].

To make the above analysis perhaps a bit clearer, an explicit construction is
now given of $\Gamma\equiv\Gamma(\Lambda_0^\ast)$ in terms of
$\Gamma(\Lambda_0)$. Suppose $\Gamma(\Lambda_0)=\{\pm s_\lambda:\lambda\in
\Lambda_0\}$. $s_{2\nu}$ is central in $\Gamma(\Lambda_0)$, for
$\nu\in\Lambda_\C$, as $2\nu\cdot\lambda$ is even for all
$\lambda\in\Lambda_0$. So $s_\lambda$ is proportional to $s_{\lambda+2\nu}$,
and we arrange the gauge so that they coincide, {\it i.e.}
$$ s_\lambda s_\mu = (-1)^{\lambda\cdot\mu} s_\mu s_\lambda\,,\qquad
s_\lambda^2=(-1)^{\hhalf\lambda^2}\,,\qquad
s_\lambda= s_{\lambda+2\nu} =s_{-\lambda}\,,\eqno(\h{A}.6)$$
for $\lambda$, $\mu\in\Lambda_0$. Choose $\kappa\in\Lambda_1$ and
$\rho\in\Lambda_3$ with $\kappa\cdot\rho=\half$. Then
$S(\kappa,\rho)=-S(\rho,\kappa)=i$, from the table. We can take $s_{2\rho}=1$
as it is central in $\Gamma(\Lambda_0)$ (since $2\rho\cdot\lambda$ is even for
all $\lambda\in\Lambda_0$). Then we have a representation of
$\Gamma(\Lambda_\C)=\{\pm\beta_\nu:\nu\in\Lambda_\C\}$:
$$\beta_\lambda=\pmatrix{s_\lambda &0\cr 0&\varepsilon_\lambda^\kappa
s_\lambda}\,,\qquad
\beta_{\lambda+\kappa}=\pmatrix{0& \varepsilon_\lambda^\kappa
s_\lambda \cr
 s_\lambda & 0}\,,\eqno(\h{A}.7)$$
for $\lambda\in\Lambda_0$,
$\varepsilon^\kappa_\lambda=(-1)^{\kappa\cdot\lambda}$. This satisfies
$\beta_\lambda=\beta_{-\lambda}$ and ${\beta_\lambda}^2=(-1)^{\half
\lambda^2}$, as required. A representation of $\Gamma(\Lambda_0^\ast)$ is then
given by defining
$$\gamma_\lambda=\pmatrix{
s_\lambda &0&0&0\cr
0&\varepsilon_\lambda^\kappa s_\lambda &0&0\cr
0&0&\varepsilon_\lambda^{\kappa+\omega} s_\lambda &0\cr
0&0&0&\varepsilon_\lambda^\omega s_\lambda\cr}\,,\qquad
\gamma_{\lambda +\kappa}=\pmatrix{
0&\varepsilon_\lambda^\kappa s_\lambda &0&0\cr
s_\lambda &0&0&0\cr
0&0&0&i\varepsilon_\lambda^\omega s_\lambda\cr
0&0&i\varepsilon_\lambda^{\kappa+\omega} s_\lambda &0\cr}\,,$$
$$\gamma_{\lambda +\omega}=\pmatrix{
0&0&0&\varepsilon_\lambda^\omega s_\lambda\cr
0&0&-\varepsilon_\lambda^{\kappa+\omega} s_\lambda &0\cr
0&\varepsilon_\lambda^\kappa s_\lambda &0&0\cr
s_\lambda &0&0&0\cr}\,,$$
$$\gamma_{\lambda +\kappa+\omega}=\pmatrix{
0&0&i\varepsilon_\lambda^{\kappa+\omega} s_\lambda &0\cr
0&0&0&-i\varepsilon_\lambda^\omega s_\lambda\cr
s_\lambda &0&0&0\cr
0&-\varepsilon_\lambda^\kappa s_\lambda &0&0\cr}\,,\eqno(\h{A}.8)$$
with $\varepsilon_\lambda^\omega=(-1)^{\omega\cdot\lambda}$ and
$\varepsilon_\lambda^{\kappa+\omega}=\varepsilon^\kappa_\lambda
\varepsilon_\lambda^\omega$.
This gives $\theta\gamma_{-\lambda}\theta^{-1}=-\gamma_\lambda$ for
$\lambda\in\Lambda_2$, although (A.4) holds for the remaining sectors. This
could be corrected by a change of gauge, although $\H(\Lambda_2)$ need not be
considered anyway, since it corresponds to a fermionic conformal field theory.

Finally, we consider the cocycles for $\Lambda_\C$ necessary for defining the
triality operator $\sigma$. Set
$$\eta_c=\prod_{j:c_j=1}\gamma_{\sqrt 2e_j}\,,\eqno(\h{A}.9)$$
for $c\in\C$. Then $\eta_c=\pm 1$, since it is proportional to $\gamma_{
\sqrt 2c}$, which is central. We wish to choose the gauge such that $\eta_c=1$
for all $c\in\C$, and also preserve the gauge choice $\gamma_\lambda=
\gamma_{-\lambda}$. This means that we may change $\gamma_{\sqrt 2e_j}$ by a
factor $\epsilon_j$ only if we change $\gamma_{-\sqrt 2e_j}$ by the same
factor. Let $\C_1=\{c\in\C:\eta_c=1\}$. This is a sub-code of $\C$, since
$\eta_{c+c'}=\eta_c\eta_{c'}$, where $c+c'$ is performed modulo 2 ({\it i.e.}
inside $\C$) (since the $\gamma_{\sqrt 2e_j}$ commute). Then $\C=\C_1$ or
$\C=\C_1\cup(\C_1+c_0)$, where $\eta_{c_0}=-1$. In the case $\C_1\neq\C$,
$\C_1^\ast\supset\C$. Choose $c_2\in\C_1^\ast$ such that $c_2\notin\C$. Then
$c_2\cdot c\in 2\ze$ for all $c\in\C_1$ and $c_2\cdot c_0\in 2\ze+1$ (otherwise
$c_2\in\C^\ast=\C$). Set $\epsilon_j=-1$ for $(c_2)_j=1$ and $\epsilon_j=1$
otherwise. Then $\eta_c=1$ for all $c\in\C$ as required.
If $\eta(c,c')$ is defined as in (5.23), then
$$\left(\prod_{c_j'= 1} \gamma_{\sq e_j}\right)\gamma_{c\over\sq}
=\eta(c,-c')\gamma_{c'\over\sq}\,.\eqno(\h{A}.10)$$
Noting $\gamma_{{c\over{\sqrt 2}}}^2=\gamma_{{c'\over{\sqrt 2}}}^2=
(-1)^{{1\over 4}|c|}$, multiply (5.23) and (A.10) to give
$$\eqalignno{
\eta(c,c')\eta(c,-c') &=
\left(\prod_{c_j\ne c_j'} \gamma_{\sq e_j}\right)\gamma_{c\over\sq}
\left(\prod_{c_j= c_j'} \gamma_{\sq e_j}\right)\gamma_{c\over\sq}
(-1)^{\quar|c|}
\cr
&=\eta_c (-1)^{n(c,c')}\,,&(\h{A}.11)}$$
and (5.25) follows.
\vskip12pt
\centerline{\bf References}
\nobreak
\vskip3pt
\nobreak
\item{ 1.} L. Dolan, P. Goddard and P. Montague, Phys. Lett. B 236 (1990) 165.
\item{ 2.} E. Verlinde, Nucl. Phys. B300 (1988) 360.
\item{ 3.} P. Ginsparg, {\it Applied Conformal Field Theory} (1988) preprint
HUTP-88/A054.
\item{4.} G. Moore and N. Seiberg, {\it Lectures on RCFT}, preprint RU-89-32,
YCTP-P13-89 (1989).
\item{ 5.} I. Frenkel, J. Lepowsky and A. Meurman, Proc. Natl. Acad. Sci.
USA 81 (1984) 3256.
\item{ 6.} I. Frenkel, J. Lepowsky and A. Meurman in {\it Vertex Operators in
Mathematics and Physics, Proc. 1983 MSRI Conf.}, eds. J. Lepowsky
{\it et al.} (Springer, Berlin, 1985) p. 231.
\item{ 7.} I. Frenkel, J. Lepowsky and A. Meurman, {\it Vertex Operator
Algebras
and the Monster}
(Academic Press, New York, 1988).
\item{ 8.} R. Griess, Invent. Math. 69 (1982) 1.
\item{ 9.} J.H. Conway and N.J.A. Sloane, {\it Sphere Packings, Lattices and
Groups} (Springer-Verlag, New York, 1988).
\item{ 10.} J. H. Conway, R. T. Curtis, S. P. Norton, R. A. Parker, R. A.
Wilson,
${\Bbb A}{\Bbb T}{\Bbb L}{\Bbb A}{\Bbb S}$ {\it of Finite Groups:
Maximal Subgroups and Ordinary
Characters for Simple Groups} (Clarendon Press, Oxford, 1985).
\item{ 11.} L. Dolan, P. Goddard and P. Montague, Nucl. Phys. B338 (1990) 529.
\item{12.} P. Goddard, `Meromorphic Conformal Field Theory'
in {\it Infinite dimensional Lie algebras and Lie groups:
Proceedings of the CIRM--Luminy Conference, 1988}
(World Scientific, Singapore, 1989) 556.
\item{13.} R.E. Borcherds, Proc. Natl. Acad. Sci. USA 83 (1986) 3068.
\item{14.} I.B. Frenkel, Y.-Z. Huang and J. Lepowsky, {\it On axiomatic
approaches to vertex operator algebras and modules} (1989).
\item{15.} P. Goddard, A. Kent and D. Olive, Phys. Lett. 152B (1985) 88;
Commun. Math. Phys. 103 (1986) 105.
\item{16.} J.H. Conway and V. Pless, Journal of Combinatorial Theory  Series A
28 (1980) 26.
\item{17.} B.B. Venkov, Trudy Matematicheskogo Instituta imeni V. A. Steklova
148 (1978) 65; Proceedings of the Steklov Institute of Mathematics 4 (1980) 63.
\item{18.} J.G. Thompson, Bull. London Math. Soc. 11 (1979) 352.
\item{19.} D. Bruce, E. Corrigan and D. Olive, Nucl. Phys.
B95 (1975) 427.
\item{20.} T.J. Hollowood, {\it Twisted Strings, Vertex Operators and Algebras}
(Durham University Ph.D. Thesis, 1988).
\item{21.} J. Tits {\it R\'esum\'e de Cours, Annuaire du Coll\`ege de France}
1982-1983  89; Invent. Math. 78 (1984) 491.
\item{22.} P. Goddard, W. Nahm, D. Olive and A. Schwimmer,
Commun. Math. Phys. 107 (1986) 179.
\item{23.} P, Montague, {\it Codes, Lattices and Conformal Field Theories}
(Cambridge University Ph.D. Thesis, 1992).
\item{24.} P. Montague, {\it On Representations of Conformal Field
Theories and the Construction of Orbifolds} (to appear, 1994).
\item{25.} C. Dong, {\it Twisted Modules for Vertex Algebras
Associated with Even Lattices} (Santa Cruz preprint, 1992).
\item{26.} C. Dong, {\it Representations of the Moonshine Module
Vertex Operator Algebra} (Santa Cruz preprint, 1992).
\item{27.} P. Montague, {\it Continuous Symmetries of Lattice
Conformal Field Theories and their $\ze_2$-Orbifolds} (to appear, 1994).
\bye